\documentclass[10pt,twoside,aps,prd,showpacs,nofootinbib,preprintnumbers,twocolumn,longbibliography]{revtex4-1}
\makeatletter

\def\l@subsubsection#1#2{}
\makeatother
\usepackage{amssymb}
\usepackage{amsmath}
\usepackage{bm}
\usepackage{bbm}
\usepackage{fancyhdr}
\usepackage{natbib}
\usepackage[dvipsnames]{xcolor}
\usepackage{graphicx}
\usepackage[utf8]{inputenc}
\usepackage{dsfont}
\usepackage{wrapfig}
\usepackage{float} 
\usepackage{enumitem}
\usepackage[section]{placeins} 
\usepackage{xspace}
\usepackage{adjustbox}
\usepackage[absolute,overlay]{textpos}
\usepackage{multirow}
\newcommand{\Ca}[2]{$^{#1}$Ca${}^{#2}$}
\newcommand{\CaStoD}[1]{$4^{2}S_{1/2} \rightarrow 3^{2}D_{#1}$\xspace}
\newcommand{\CaDtoD}{$3^{2}D_{3/2} \rightarrow 3^{2}D_{5/2}$\xspace}

\newcommand{\CaStoP}{$4s^2S_{1/2}\rightarrow 4 p^2 P_{1/2}$\xspace}
\newcommand{\CaDtoP}{$3d^2D_{3/2}\rightarrow 4 p^2 P_{1/2}$\xspace}

\newcommand{\castodfive}{$\nu_{729}$\xspace}
\newcommand{\castodthree}{$\nu_{732}$\xspace}
\newcommand{\castop}{$\nu_{397}$\xspace}
\newcommand{\cadfine}{$\nu_{\mathrm{DD}}$\xspace}

\newcommand{\YbStoDfive}{$^2S_{1/2}\rightarrow^2D_{5/2}$\xspace}
\newcommand{\YbStoDthree}{$^2S_{1/2}\rightarrow^2D_{3/2}$\xspace}
\newcommand{\YbStoDtwo}{$^1S_{0}\rightarrow^1D_{2}$\xspace}

\newcommand{\dvec}{\boldsymbol{d}}
\newcommand{\dveca}[1]{\boldsymbol{d}^{\,{#1}}}

\newcommand{\dai}[2]{d_{#2}^{#1}}
\newcommand{\dorigai}[2]{\hat{d}_{#2}^{#1}}
\newcommand{\Deltai}[2]{\Delta_{#2}^{#1}}
\newcommand{\Deltaveca}[1]{\boldsymbol{\Delta}^{#1}}
\newcommand{\deltaveca}[1]{\boldsymbol{\delta}^{#1}}
\newcommand{\sdeltaveca}[1]{\boldsymbol{\sigma[\delta^{#1}]}}
\newcommand{\deltai}[2]{\delta^{#1}_{#2}}
\newcommand{\sdeltai}[2]{\sigma[\deltai{#1}{#2}]}
\newcommand{\covd}{\boldsymbol{\Sigma}_{\dvec}}
\newcommand{\covdab}[1]{\Sigma_{\dvec}^{#1}}
\newcommand{\covtheta}{\boldsymbol{\Sigma}_{K\phi}}
\newcommand{\covthetaij}[1]{\covtheta^{(#1)}}

\newcommand{\gammavec}{\boldsymbol{\gamma}}
\newcommand{\hvec}{\boldsymbol{\tilde{\gamma}}}
\newcommand{\gammabarvec}{\boldsymbol{\bar{\gamma}}}
\newcommand{\ha}[1]{\tilde{\gamma}^{#1}}

\newcommand{\eFvec}{\boldsymbol{\hat{e}\!_\mathcal{F}}}

\newcommand{\Fijvec}{\boldsymbol{\mathcal{F}}}
\newcommand{\phij}[1]{\phi_{#1}}

\newcommand{\Kijvec}{\boldsymbol{\mathcal{K}}}
\newcommand{\Kij}[1]{K_{#1}}
\newcommand{\Xijvec}{\boldsymbol{\mathcal{X}}}
\newcommand{\Kperp}{\Kijvec^\bot}
\newcommand{\Kperpij}[1]{K_{#1}^\bot}

\newcommand{\ma}[1]{m^{#1}}
\newcommand{\sma}[1]{\sigma[\ma{#1}]}
\newcommand{\map}[1]{m^{#1^\prime}}
\newcommand{\smap}[1]{\sigma[\map{#1}]}
\newcommand{\mua}[1]{\mu^{#1}}

\newcommand{\muvec}{\boldsymbol{\mu}}
\newcommand{\mmuvec}{\boldsymbol{1}}
\newcommand{\mmusqvec}{\boldsymbol{\tilde{\mu}^{(2)}}}
\newcommand{\mmua}[1]{\tilde{\mu}^{#1}}
\newcommand{\nuveci}[1]{\boldsymbol{{\nu}}_{#1}}
\newcommand{\mnuveci}[1]{\boldsymbol{\tilde{\nu}}_{#1}}
\newcommand{\mnuveca}[1]{\boldsymbol{\tilde{\nu}}^{#1}}

\newcommand{\mnuailin}[2]{\mnuai{#1}{#2}\vert_\mathrm{lin.}}
\newcommand{\mnuvecalin}[1]{\mnuveca{#1}\vert_\mathrm{lin.}}
\newcommand{\mnuaipred}[2]{\mnuai{#1}{#2}\vert_\mathrm{pred.}}
\newcommand{\mnuvecapred}[1]{\mnuveca{#1}\vert_\mathrm{pred.}}
\newcommand{\mnuvecaexp}[1]{\mnuveca{#1}}

\newcommand{\fnuveci}[1]{\boldsymbol{\bar{\nu}}_{#1}}
\newcommand{\fmuvec}{\boldsymbol{\bar{\mu}}}

\newcommand{\nuai}[2]{\nu_{#2}^{#1}}
\newcommand{\mnuai}[2]{\tilde{\nu}_{#2}^{#1}}
\newcommand{\snuai}[2]{\sigma[\nuai{#1}{#2}]}
\newcommand{\smnuai}[2]{\sigma [\mnuai{#1}{#2}]}

\newcommand{\drsq}{\delta \langle r^2\rangle}
\newcommand{\drsqvec}{\boldsymbol{\drsq}}
\newcommand{\drsqa}[1]{\drsq ^{#1}}
\newcommand{\mdrsqa}[1]{\delta \langle \widetilde{r^2}\rangle^{#1}}
\newcommand{\mdrsqvec}{\boldsymbol{\delta\langle \widetilde{r^2}\rangle}}

\newcommand{\alphaEM}{\alpha_\mathrm{EM}}
\newcommand{\alphaNP}{\alpha_\mathrm{NP}}
\newcommand{\sigalphaNP}{\sigma[\alphaNP]}
\newcommand{\bestalpha}{\alpha_*}
\newcommand{\bestalphaexp}[1]{\alpha_*^{#1}}
\newcommand{\alphaup}{\alpha_\uparrow}
\newcommand{\alphaupexp}[1]{\alpha_\uparrow^{#1}}

\newcommand{\alphadown}{\alpha_\downarrow}
\newcommand{\alphadownexp}[1]{\alpha_\downarrow^{#1}}
\newcommand{\alphadownblock}[1]{\alpha_\downarrow^{(#1)}}
\newcommand{\alphaNPdEM}{\frac{\alphaNP}{\alphaEM}}

\newcommand{\Vdat}{\mathrm{V}_\mathrm{dat}}
\newcommand{\Vpred}{\mathrm{V}_\mathrm{pred}}

\newcommand{\kifit}{\texttt{kifit}\xspace}

\newcommand{\Lambdavec}[1]{\boldsymbol{\Lambda}^{{#1}}}
\newcommand{\lambdacoordi}[2]{\lambda_{#2}^{{#1}}}

\renewcommand{\arraystretch}{2} 

\newcommand{\etatildvec}{\boldsymbol{\tilde{\eta}}}
\newcommand{\etatilda}[1]{\tilde{\eta}^{{#1}}}

\usepackage{listings}
\definecolor{codegreen}{rgb}{0,0.6,0}
\definecolor{codegray}{rgb}{0.5,0.5,0.5}
\definecolor{codepurple}{rgb}{0.58,0,0.82}
\definecolor{backcolour}{rgb}{0.95,0.95,0.92}

\lstdefinestyle{mystyle}{
    backgroundcolor=\color{backcolour},   
    commentstyle=\color{codegreen},
    keywordstyle=\color{magenta},
    numberstyle=\tiny\color{codegray},
    stringstyle=\color{codepurple},
    basicstyle=\ttfamily\footnotesize,
    breakatwhitespace=false,         
    breaklines=true,                 
    captionpos=b,                    
    keepspaces=true,                 
    numbers=left,                    
    numbersep=5pt,                  
    showspaces=false,                
    showstringspaces=false,
    showtabs=false,                  
    tabsize=2
}

\definecolor{teal}{rgb}{0.21, 0.46, 0.53}
\definecolor{persianred}{rgb}{0.8, 0.2, 0.2}
\definecolor{darklavender}{rgb}{0.45, 0.31, 0.59}
\definecolor{deepcarmine}{rgb}{0.66, 0.13, 0.24}
\usepackage[normalem]{ulem}
\usepackage[colorlinks=true,citecolor=teal,linkcolor=teal, urlcolor=teal]{hyperref}
\fancypagestyle{firststyle}{
    \fancyhead[HR]{TIF-UNIMI-2025-13, DESY-25-080}
}
\begin{document}
\preprint{ }

\title{Towards a Global Search for New Physics with Isotope Shifts}
\author{Elina Fuchs}
\email{elina.fuchs@itp.uni-hannover.de}
\affiliation{Deutsches Elektronen-Synchrotron DESY, Notkestr.~85, 22607 Hamburg, Germany}
\affiliation{Institut für Theoretische Physik, Leibniz Universität Hannover, Appelstraße 2, 30167 Hannover, Germany}
\affiliation{Physikalisch-Technische Bundesanstalt, Bundesallee 100, 38116 Braunschweig, Germany}

\author{Fiona Kirk}
\email{fiona.kirk@itp.uni-hannover.de}
\affiliation{Physikalisch-Technische Bundesanstalt, Bundesallee 100, 38116 Braunschweig, Germany}
\affiliation{Institut für Theoretische Physik, Leibniz Universität Hannover, Appelstraße 2, 30167 Hannover, Germany}
		
\author{Agnese Mariotti}
\email{agnese.mariotti@itp.uni-hannover.de}
\affiliation{Institut für Theoretische Physik, Leibniz Universität Hannover, Appelstraße 2, 30167 Hannover, Germany}

\author{Jan Richter}
\email{jan.richter@ptb.de}
\affiliation{Physikalisch-Technische Bundesanstalt, Bundesallee 100, 38116 Braunschweig, Germany}

\author{Matteo Robbiati}
\email{matteo.robbiati@cern.ch}
\affiliation{TIF Lab, Dipartimento di Fisica, Università degli Studi di Milano, Italy}
\affiliation{European Organization for Nuclear Research (CERN), Geneva 1211, Switzerland}

\begin{abstract}
Isotope shifts have emerged as a sensitive probe of new bosons that couple to electrons and neutrons, and of nuclear structure. 
The recent Hz- or even sub-Hz-level isotope shift measurements across different elements call for a global assessment of all available data. 
In this work, we present the fit framework \kifit that for the first time enables a combined analysis of isotope shift data from several elements, taking into account correlations. 
We provide a thorough comparison of analytical methods and the fit to analyse linear and nonlinear King plots and quantify their uncertainties. Finally, we provide recommendations for future measurements that could enhance the sensitivity to new physics and offer new insights into nuclear structure.
\end{abstract}


\maketitle
\thispagestyle{firststyle}
\tableofcontents

\newpage
\section{Introduction}
Atomic precision spectroscopy has emerged as a powerful tool for probing the Standard Model of particle physics (SM) and its extensions, which address long-standing questions such as the properties of dark matter~\cite{Safronova:2017xyt,Kozlov:2018mbp,Antypas:2022asj}. 
The groundbreaking developments of laser
cooling~\cite{HANSCH197568,WinelandDehmelt1975,RevModPhys.58.699,METCALF1994203}, optical frequency
combs~\cite{RevModPhys.78.1297,RevModPhys.78.1279}, optical tweezers~\cite{Ashkin:1970mb,Ashkin:1986emb,Chu:1986zz}, optical lattices~\cite{Letokhov:1968,Balykin:1988,mauri1992two,grimm1999opticaldipoletrapsneutral} and quantum logic
spectroscopy~\cite{Schmidt:2005_Science_QuantumLogicSpectroscopy, PhysRevA.42.2977,PhysRevLett.74.4091,Wineland:1997mg}
have increased the relative precision of optical atomic clocks to the level of $10^{-18}$~\cite{Ye:2024_10-19precision_PhysRevLett.133.023401, Hausser:2025_InYbCoulombCrystalClock}. As a result of this unparalleled precision, clock comparisons and other differential spectroscopic measurements probe relevant parts of the parameter space of light dark matter and dark portal models.

A prime example for such developments is isotope-shift spectroscopy, a
well-established technique commonly used to determine nuclear charge
radii~\cite{king1963comments,kingbook,ANGELI201369,Gebert:2015_Ca100kHz,shi2018unexpectedly}, but more
recently proposed as a method to search
for hypothetical new bosons mediating an additional interaction between neutrons
and electrons\cite{Delaunay:2016brc,Berengut:2017zuo}. Isotope shifts can probe bosons in the eV to MeV mass range,
allowing them to bridge the gap between 
fifth force searches via the
Casimir effect~\cite{BordagBook}, beam dump experiments~\cite{Essig:2009nc,Essig:2010xa,Batell:2009di} and searches for exotic meson decays at colliders~\cite{BaBar:2013npw,E949:2008btt}. 
New vector or scalar bosons, such as the ones probed by isotope shift spectroscopy~\cite{Frugiuele:2016rii}, are predicted
by a wide range of extensions of the SM. These include models 
that gauge baryon
minus lepton number ($B-L$), which result in a new vector boson $Z'$, or so-called portal models that introduce new light scalar or vector mediators between the SM and dark matter~\cite{Knapen:2017xzo}.

The new physics search proposed in
Refs.~\cite{Delaunay:2016brc,Berengut:2017zuo} is based on the so-called King
plot method~\cite{king1963comments,kingbook}, which at leading order predicts a linear relation between 
isotope shifts in different electronic transitions. 
Using King plots reduces the reliance on atomic structure calculations, which are limited in precision by non-perturbative and many-body effects, particularly within the nucleus. This approach allows new physics signals to be constrained using isotope shift data.

Given the long history of remarkably linear King plots, King plot searches for new physics attracted a
lot of attention and significant progress was made in improving the precision of
the isotope shift and nuclear mass measurements, with notable advancements in
neutral Ca~\cite{kramida2020atomic,Roser:2024vxa},
Ca${}^+$~\cite{Gebert:2015_Ca100kHz,Knollmann:2019gcc,Solaro:2020dxz,Chang:2023teh,Wilzewski:2024wap}
and Ca${}^{14+}$~\cite{Wilzewski:2024wap}, as well as in neutral
Yb~\cite{clark1979optical,Bowers:1999_Yb0,Kleinert:2016_Yb0,Figueroa:2022mtm} and Yb$^+$~\cite{Counts:2020aws,Hur:2022gof,Door:2024qqz}. The first observation of a
nonlinear King plot was reported in Ref.~\cite{Counts:2020aws}, which employed
the ${}^2S_{1/2}\to {}^2D_{3/2}$ vs. ${}^2S_{1/2}\to {}^2D_{5/2}$ transitions in Yb$^+$, measured at
a precision of about 300 Hz. This so-called ``King nonlinearity'' was confirmed
by subsequent measurements~\cite{Hur:2022gof,Door:2024qqz} and is currently
found to be at the level of 20.17(2) kHz~\cite{Door:2024qqz}. Recently,
nonlinear isotope shifts have also been observed in the ${}^3P_0\to {}^3 P_1$ transition,
measured at sub-Hz precision in Ca${}^{14+}$, and in the ${}^2S_{1/2}\to {}^2D_{5/2}$ transition,
measured in Ca${}^{+}$, combined with nuclear mass ratios with relative
uncertainties below $4\times 10^{-11}$~\cite{Wilzewski:2024wap}.

Although the presence of nonlinearities complicates King plot searches for new
physics, it is by no means a show-stopper:
Progress in atomic and nuclear structure theory has facilitated the identification of the
leading higher-order effects in Yb~\cite{Allehabi2021,Hur:2022gof,Door:2024qqz}
and in Ca King plots~\cite{Viatkina:2023qop,Wilzewski:2024wap}, whereas
the development of the Generalised King Plot~\cite{Berengut:2020itu} showed that
the data-driven King plot approach can be generalised so as to provide
constraints on new physics, even in the presence of higher-order nuclear effects.
For a recent review of the relevance of (nonlinear) King plots for the search
for new (nuclear) physics, see Ref.~\cite{Berengut:2025nxp}.

The King plot analysis is often restricted to even isotopes to avoid the effects of
the nuclear spin, namely hyperfine interactions, which are expected to introduce
additional nonlinear effects in the King plot~\cite{palmer1982theory}. Ref.~\cite{Berengut:2024gxl} discusses the challenges associated with using odd Yb isotopes in King plot analyses. Another example is provided by Refs.~\cite{Hofsass:2022koa, Roeser:2024_Zn_hyperfine}: although the King plots for Cd
and Zn are linear, the hyperfine interactions would potentially be observable at
Hz precision.

With a wealth of high-precision isotope shift data now available for different
elements and charged states, a systematic approach is needed to assess the
constraints on new physics contributions, which are expected to be governed by
the same couplings to electrons and to neutrons, irrespective of the element under
consideration. 
In this work we first review the (generalised) King
plot~\cite{Berengut:2017zuo,Berengut:2020itu}, the no-mass (generalised) King
plot~\cite{Berengut:2020itu} and the projection~\cite{Solaro:2020dxz} methods,
which follow a purely algebraic approach (Section \ref{sec:alg}), before presenting our code \kifit, which provides for the first time a framework to combine all linear King plots across
elements in one global constraint on new physics (Section \ref{sec:kifit}). The framework provided by
\kifit is based on the fit presented in Ref.~\cite{Frugiuele:2016rii}, but was
significantly extended to handle contemporary high-precision isotope shift data.
\kifit performs a fit to linear King plots plus
new physics, while the incorporation of
higher-order nuclear effects, that would be necessary to analyse nonlinear King
plots, is left to future work. Nonetheless, \kifit presents an important step
towards a global view of isotope shift data. 
In Section~\ref{sec:fitvsalg} we compare \kifit with the
algebraic methods introduced in Section~\ref{sec:alg}, before concluding in Section~\ref{sec:conclusions}. In the appendices we provide further details on the electronic structure calculations with \texttt{AMBiT}~\cite{kahl2019ambit} (Appendix~\ref{sec:ambit}), how the choice of transitions used in King plots affects the sensitivity to new physics (Appendix~\ref{sec:sensitivity_NP}), the impact of experimental uncertainties on the margin for new physics (Appendix~\ref{sec:uncertainty_projections}), a short manual for our code \kifit and a list of implemented validation checks (Appendix~\ref{sec:kifit_package}),
an analysis of the impact of data sparsity on the fit (Appendix~\ref{sec:sparsity}), 
as well as a summary of the available state-of-the-art isotope shift data and isotope masses (Appendix~\ref{sec:data}).

\section{Searching for New Physics with King Plots}
\label{sec:alg}

An isotope shift $\nu_i^{AA'}\equiv \nu_i^A-\nu_i^{A'}$ corresponds to the
frequency difference between the electronic transition $i$ measured in isotopes
$A$ and $A'$.
Isotope shifts are dominated by two effects, both of which can be described by a
product of an electronic coefficient ($K_i$, $F_i$) and a nuclear quantity
($\mua{AA'}$, $\delta\langle r^2\rangle^{AA'}$),
respectively~\cite{Breit:1958_IS-theory,King:1963,Stacey:1966IS_expth,Heilig:1974},
    \begin{align}
        \nu_i^{AA'}\approx K_i\mua{AA'}+ F_i \delta\langle r^2\rangle^{AA'}\,.
        \label{eq:linIS}
    \end{align}
The first term in Eq.~\eqref{eq:linIS}, which is proportional to the difference of
inverse nuclear masses~\footnote{
        The measured mass $\ma{A^0}$ of a neutral atom $A^0$ can be translated into a
        nuclear mass $\ma{A}$ by subtracting the masses $m_e$ and binding energies
        $E^b_i$ of the $N_e$ electrons~\cite{Solaro:2020dxz}:
                    \begin{align*}
                        \ma{A}=\ma{A^0}-N_e m_e+\sum_{i=1}^{N_e}E^b_i\,.
                    \end{align*}
        },
$\mua{AA'} \equiv \frac{1}{\ma{A}}-\frac{1}{\ma{A'}}$, is known as the
first-order \textit{mass shift} (MS) and describes the nuclear-recoil correction
to the electron kinetic energy~\cite{palmer1987reformulation}. The second term,
referred to as the first-order \textit{field shift} (FS), accounts for the
nuclear charge radius variance
$\delta \langle r^2\rangle ^{AA'}\equiv \langle r^2\rangle ^{A}-\langle
r^2\rangle ^{A'}$ and describes the energy shift due to changes in the nuclear
charge distribution between the different isotopes~\cite{Blundell:1987}.

Since we will mostly be dealing with isotope pairs in the following, we
introduce the isotope pair index $a=AA'$.

Thanks to the factorisation of the isotope shifts into electronic and nuclear
quantities at leading order, we can combine isotope shift measurements for two different transitions to eliminate the nuclear charge radius, which is neither experimentally nor theoretically precisely determined~\cite{King:1963,AME2020II}. We obtain a linear relation between the
isotope shifts of the two transitions labelled $1$ and $2$~\cite{King:1963},~\footnote{
         Instead of the mass-normalised isotope shifts in Eq.~\eqref{eq:linKPrel},
         it is possible to normalise the isotope shifts by those of the reference
         transition $\nuai{a}{1}$. The King relation for
         \textit{frequency-normalised} isotope shifts takes the form
             \begin{align}
                 \bar{\nu}_2^{a} \equiv \frac{\nuai{a}{2}}{\nuai{a}{1}} 
                 = F_{21}+K_{21}\bar{\mu}^{a}\,,
             \end{align}
         where $\bar{\mu}^{a} \equiv \mu^{a}/\nuai{a}{1}$, while the electronic
         coefficients appear in the same combinations as in
         Eq.~\eqref{eq:FijKijdef}.
        }
    \begin{align}
        \mnuailin{a}{2} = K_{21}+F_{21}\mnuai{a}{1}
        \label{eq:linKPrel}
    \end{align}
where we defined the so-called mass-normalised isotope
shifts~\cite{Hansen:1967,Heilig:1974,Palmer:1984_CaIS} $\mnuai{a}{i} \equiv
\frac{\nuai{a}{i}}{\mua{a}}$ and introduced the electronic coefficients
    \begin{align} 
        F_{21}\equiv\frac{F_2}{F_1} \,, \qquad 
        K_{21} \equiv K_2-F_{21}K_1\,.
        \label{eq:FijKijdef}
    \end{align}
The subscript ``lin.'' indicates that Eq.~\eqref{eq:linKPrel} only describes the leading linear behaviour.

\begin{figure}[t]
    \centering    \includegraphics[width=1\linewidth]{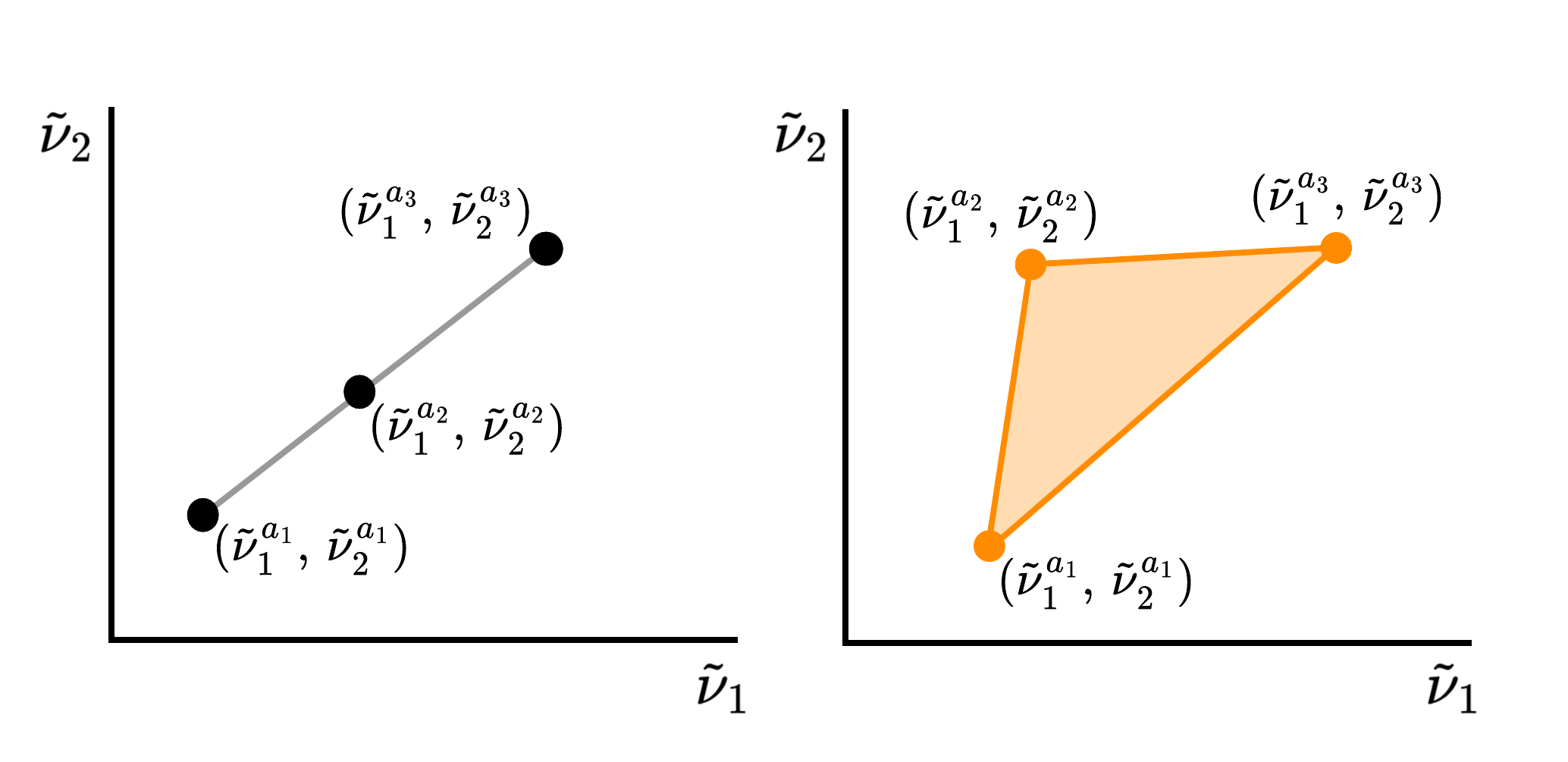}
    \caption{
    Left: Linear 2-dimensional King plot. The isotope shifts follow the relation given in Eq.~\eqref{eq:linIS}.
    Right: In the presence of higher-order nuclear or new physics contributions to the isotope shifts, the data points deviate from the King line and define a non-zero volume.
    }\label{fig:alg_sketch}
\includegraphics[width=0.5\linewidth]{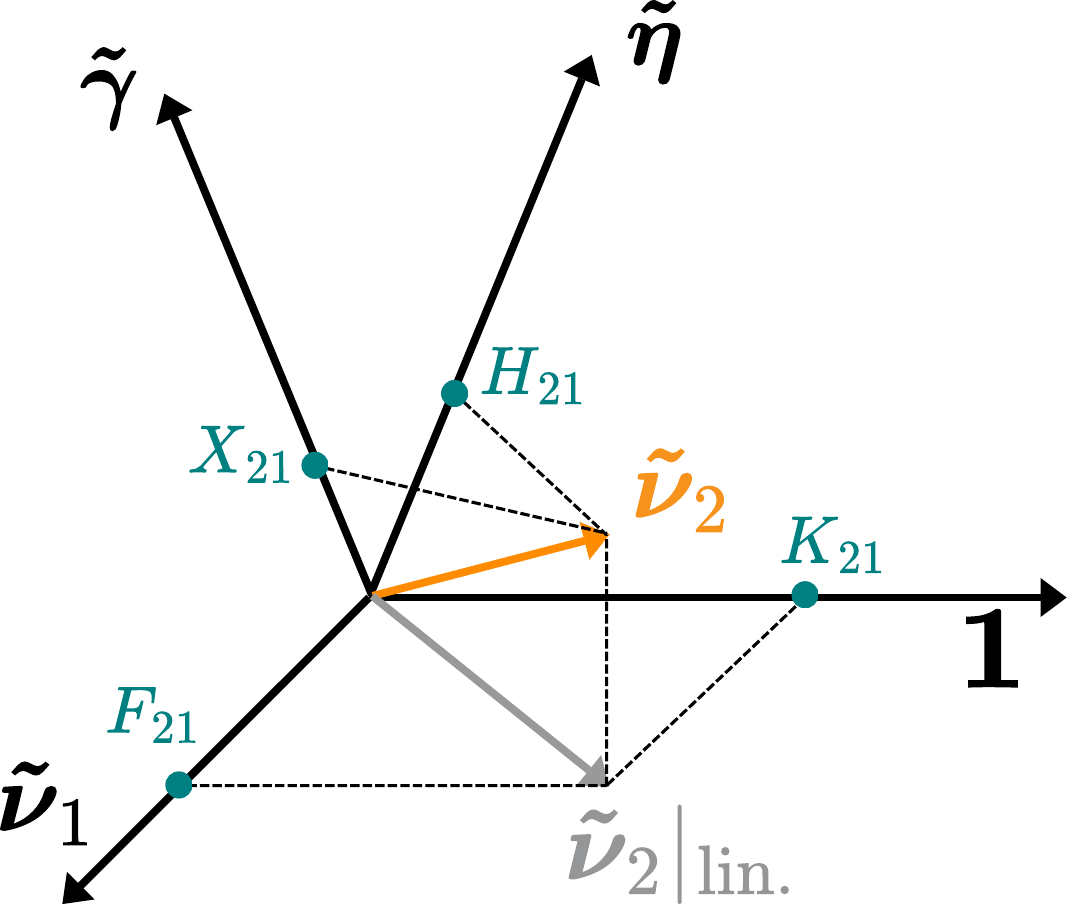}
    \caption{Illustration of the ``King plane'' spanned by the vectors 
    $\nuveci{1}$ and $\mathbf{1}$ in isotope
    pair space. If Eq.~\eqref{eq:linvecKP} holds, $\nuveci{2}$ lies in this plane, while a nonlinearity in the King plot would lead to an
    out-of-plane component for $\nuveci{2}$. Both new physics, here assumed to be proportional to $\hvec$, and higher-order
    SM terms, represented by the vector $\boldsymbol{\tilde{\eta}}$ (see Eq.~\eqref{eq:nonliNPIS}) can lead to this kind of effect. 
    }
    \label{fig:out-of-plane}
\end{figure}

If multiple isotope pairs ($a=1, \dots n$) can be probed, it is useful to
arrange the isotope shifts $\mnuai{a}{i}$, $i=1,2$, into vectors in isotope-pair
space. Defining the $n$-vectors $\mnuveci{i}=(\mnuai{1}{i},\dots,\mnuai{n}{i})$
and $\mmuvec=(1,\dots,1)$, one obtains
    \begin{align}
        \mnuveci{2}\vert _{\mathrm{lin.}} = K_{21}\mmuvec+F_{21}\mnuveci{1}\,.
        \label{eq:linvecKP}
    \end{align}
Eq.~\eqref{eq:linvecKP} can be visualised in a so-called King
plot~\cite{King:1963}, illustrated in the left half of Fig.~\ref{fig:alg_sketch}. 
The
electronic coefficients $K_{21}$ and $F_{21}$, corresponding to the intercept and slope of the \emph{King line},
can be determined via a linear fit to the isotope shift data. 
Alternatively, the isotope shift data can be arranged in isotope pair space. Fig. \ref{fig:out-of-plane} shows the \emph{plane of King linearity}, which is spanned by the vectors $\mmuvec$ and $\mnuveci{1}$. If $\mnuveci{2}$ is described by Eq.~\eqref{eq:linvecKP}, it will lie in this plane.

\subsection{Algebraic Methods for Linear King Plots}
\label{sec:alglin}
In the following, we will distinguish linear and nonlinear King plots. As King nonlinearities we understand any deviation from the linear King relation in Eq.~\eqref{eq:linvecKP}. As long as these are smaller or comparable to the experimental uncertainties, the King plot is considered linear.

Nonlinearities may be caused by higher-order nuclear or electronic corrections, or by new physics
contributions. As long as the higher-order SM contributions are negligible 
or can be determined by additional experimental and theoretical input, King plots can be used to place constraints on the mass and couplings of a new boson $\phi$ inducing
a new interaction between the bound electrons and the neutrons inside the
nucleus~\cite{Berengut:2017zuo}. 
Assuming $\phi$ couples linearly to neutrons and electrons, this effect can be described by a Yukawa potential~\cite{Haber:1978jt,Delaunay:2016brc}\\
    \begin{align}
        V_{\mathrm{NP}}(r,m_\phi)= - \alpha_\mathrm{NP} (A-Z)\frac{e^{-m_\phi r}}{r}\,,
        \label{eq:YukawaPotential}
    \end{align}
expressed here in relativistic units ($c=\hbar=1$).
Here, $\alpha_\mathrm{NP} = (-1)^s \frac{y_e y_n}{4\pi}$ is defined as the
product of the new physics couplings to electrons ($y_e$) and to neutrons
($y_n$), while $s=0,1,2$ denotes the spin, $m_\phi$ corresponds to the mass of the
new light boson $\phi$ and $A$ and $Z$ are respectively the mass and atomic
numbers of the considered isotope.
This introduces a new term in the isotope shifts~\cite{Berengut:2017zuo}
    \begin{align}
        \nu_i^{a}\vert_\text{pred.} 
        = K_i\mua{a}+ F_i \delta\langle r^2\rangle^{a}
        +\alphaNPdEM X_i\gamma^{a}\,
        \label{eq:liNPIS}
    \end{align}
where we introduced the subscript ``pred'' to indicate the predicted isotope shift in the presence of new physics. Assuming a linear King plot, we can use Eq.~\eqref{eq:liNPIS}
to set a bound on the new physics coupling $\alphaNP$, expressed in units
of the fine structure constant $\alpha_\mathrm{EM}$~\footnote{
Note that alternative conventions exist in the literature. 
An equivalent relation in terms of $y_e y_n$
    rather than $\alpha_\mathrm{NP}/\alpha_{\mathrm{EM}}$ can be obtained by
    rescaling the $X$ coefficients (see Appendix~\ref{sec:ambit}).
    }. The factor $\gamma^a\equiv A-A'$ is the 
difference in neutron number between the isotopes $A$ and $A'$, while the electronic coefficients $X_i$ quantify the sensitivity of the transition
frequency $\nu_i$ to shifts induced by the new physics potential in
Eq.~\eqref{eq:YukawaPotential}.
The leading behaviour can be understood in a perturbative approach, in which the
radial wave functions $\Psi_i^{\mathrm{init,fin}}(r)$ of the states involved in
the transition $i$
are assumed to be unchanged compared to the SM. In this case, the coefficient
$X_i$ corresponds to the difference between the overlap of the new physics potential (Eq.~\eqref{eq:YukawaPotential}) with the final- and initial-state wave functions:
    \begin{align}
        X_i(m_\phi) = \int \frac{e^{-m_\phi r}}{r}
        \left[|\Psi^{\mathrm{fin}}_i(r)|^2-|\Psi^{\mathrm{init}}_i(r)|^2\right]\mathrm{d}r\,.
        \label{eq:Xdef}
    \end{align}
The overlap with the new physics potential is particularly pronounced for S-state wave functions. Consequently narrow transitions involving S-states are favourable for isotope shift-based new physics searches.

As discussed in Appendix~\ref{sec:ambit}, for the numerical calculation of the
new physics electronic coefficients, we follow Ref.~\cite{Berengut:2017zuo} and
use the public code \texttt{AMBiT}~\cite{kahl2019ambit}, which applies a
non-perturbative finite field method that adds the new physics potential
directly to the Hamiltonian in the Dirac equation. \\

Combining Eq.~\eqref{eq:liNPIS} for the transitions $i=1,2$, we obtain the modified King relation 
    \begin{align}
        \mnuveci{2}\vert_{\text{pred.}} = K_{21} \mmuvec + F_{21} \mnuveci{1} + \alphaNPdEM X_{21} \hvec\,, 
        \label{eq:liNPKP}
    \end{align}
where we defined $X_{21}\equiv X_2- F_{21}X_1$ and $\hvec$ with elements
$\ha{a}\equiv \frac{\gamma^{a}}{\mua{a}}$.
Eq.~\eqref{eq:liNPKP} implies that $\mnuveci{2}\vert_{\text{pred.}}$ has a component along $\hvec$, which pushes it out of the King plane, as visualised in Fig.~\ref{fig:out-of-plane}.

If $n=3$ isotope pairs are probed, the system of isotope shift equations Eq.~\eqref{eq:liNPKP} can be solved for the new physics coupling~\cite{Berengut:2017zuo}:
    \begin{equation}
        \alphaNPdEM =\frac{
            \mathrm{det}\left(\mnuveci{1},\mnuveci{2},\mmuvec\right)
        }{
            \varepsilon_{ij}\mathrm{det}\left(X_i \hvec, \mnuveci{j},\mmuvec\right)
        }
        =\frac{\Vdat}{\Vpred}\,,
        \label{eq:ANP3x2}
     \end{equation}
where $\varepsilon_{ij}$ is the 2-dimensional Levi-Civita symbol and summation over repeated indices is implied. $\det (\mathbf{u}, \mathbf{v}, \mathbf{w})$ denotes the determinant of the square matrix whose columns are the vectors $\mathbf{u}$, $\mathbf{v}$ and $\mathbf{w}$.
In the following, we will refer to Eq.~\eqref{eq:ANP3x2} as the \emph{Minimal King Plot (KP)} formula~\footnote{
        In the \kifit code,
        Eq.~\eqref{eq:ANP3x2} corresponds to the 3-dimensional case of the generalised
        King plot formula, presented in Sec.~\ref{sec:generalised_algebraic}. Here 3 refers to the number of isotope pairs, i.e. to the
        number of data points in the King plot.
        }. 
Note that the normalisation of the vectors entering Eq.~\eqref{eq:ANP3x2} does not affect the result, meaning that $(\mnuveci{1}, \mnuveci{2}, \mmuvec)$ can be replaced by
$(\nuveci{1}, \nuveci{2}, \muvec)$ or by frequency-normalised counterparts.
Geometrically, $\Vdat$ can be visualised either as the volume spanned by data points in the King plot (see Fig.~\ref{fig:alg_sketch}) or as the volume of the parallelepiped spanned by $\mnuveci{1}$, $\mnuveci{2}$ and $\mathbf{1}$ (see Fig.~\ref{fig:out-of-plane}). Similarly, $\Vpred$ measures the predicted volume spanned by the data vectors in the presence of new physics contributions.

It is instructive to split $\Vdat$ and $\Vpred$ into their respective electronic and
nuclear contributions. Assuming the isotope shifts satisfy
Eq.~\eqref{eq:liNPIS} and defining
\begin{align}
        \mathcal{M} =
            \begin{pmatrix}
                X_1 & F_1\\
                X_2 & F_2
            \end{pmatrix}\,, \quad
        \mathcal{N} =
            \begin{pmatrix}
                \ha{1} & \mdrsqa{1} & 1 \\
                \ha{2} & \mdrsqa{2} & 1 \\
                \ha{3} & \mdrsqa{3} & 1 \\
            \end{pmatrix}\,,
            \label{eq:ElNuclMatKP}
    \end{align}
where $\mdrsqa{a} \equiv  \delta \langle r^2 \rangle ^a /\mua{a}$ denotes the 
mass-normalised charge radius variance, we can rewrite $\Vpred$ as
    \begin{align}
        \Vpred 
        =&  \varepsilon_{ij}\mathrm{det}\left(X_i \hvec, \mnuveci{j},\mmuvec\right)
        =  \mathrm{det}\left(\mathcal{M}\right)
        \mathrm{det}\left(\mathcal{N}\right) \notag\\
        =&  (F_1 X_2-F_2 X_1)\mathrm{det}\left(\hvec,\mdrsqvec,\mmuvec\right)\,.
        \label{eq:VpredefKP}
    \end{align}
This equation explicitly shows that King plots are sensitive to the new physics
coupling when the nuclear and electronic quantities simultaneously ``open up''
new dimensions in isotope-pair space and in transition space. \\

The solution of $\alphaNP$ in Eq.~\eqref{eq:ANP3x2} corresponds to the
value of the coupling required to reproduce the central
values of the King plot points $\mnuai{a}{i}$, $i=1,2$, $a=1,2,3$.
However, the King plot method cannot exclude higher-order SM contributions to the isotope shifts and thus cannot identify an observed nonlinearity as being of pure new physics origin. In other words, it is not a discovery tool, but can only be used to set bounds on the coupling $\alphaNP$.
The $N\sigma$ bounds on $\alphaNP$ can be estimated as
    \begin{align}
        \langle\alphaNP\rangle + N \sigma[\alphaNP]\,.
        \label{eq:nsigma_bound}
    \end{align}
\\
With $\langle\alphaNP\rangle$ we denote the absolute value of $\alphaNP$ obtained for the experimental central values of the input parameters (isotope shifts $\nuai{a}{i}$ and nuclear masses $\ma{A}$). The uncertainty on $\alphaNP$, $\sigalphaNP$, can
either be estimated using (linear) error propagation or with a Monte Carlo approach. The latter involves the generation of samples~\footnote{In the \kifit code, the number of
samples \texttt{num\_det\_samples} is a hyperparameter that can be set by the
user.} of the input parameters from a normal distributions fixed by the
respective experimental central values and uncertainties:
    \begin{align}
        \begin{split}
            \nuai{a}{i}\sim  \;\mathcal{N}\left(\langle \nuai{a}{i}\rangle, \snuai{a}{i}\right)\,,\quad
            & \ma{A}     \sim  \;\mathcal{N}\left(\langle \ma{A} \rangle,    \sma{A}\right)
        \end{split}
    \label{eq:elemsampling}
    \end{align}
and similarly for the reference isotope masses $\map{A}$.
Then, $\sigalphaNP$ is estimated as the standard deviation of the corresponding $\alphaNP$ values. In this case one should pay attention whether the distribution of the $\alphaNP$ values can be accurately described by a Gaussian.\\ 

As already discussed in Ref.~\cite{Berengut:2017zuo}, the $m_\phi$-range that King plots are particularly sensitive to is dictated by 
the $m_\phi$-dependence of the $X$
coefficients (see
Eqs.~\eqref{eq:YukawaPotential} and \eqref{eq:Xdef} or
Fig.~\ref{fig:X_vs_mass}):
In the \emph{``massless'' limit,} the Yukawa potential (Eq.~\eqref{eq:YukawaPotential}) takes the form $V_{\mathrm{NP}} \propto 1/r$ and the electronic coefficients $X$, and consequently the bounds on $\alphaNP$, become
independent of $m_\phi$ (see Figs.~\ref{fig:X_vs_mass} and
\ref{fig:compare_KP_lin}, respectively).
This corresponds to the case where the interaction range of the mediator $\phi$ (Compton wavelength
$\lambda_\phi =h/(m_\phi c)$) exceeds the size of the
atom or ion, and the overlap of the Yukawa
potential with the electronic wave functions saturates at maximal King plot sensitivity to new physics.

In the \emph{intermediate $m_\phi$ region}, the values of the $X$
coefficients are sensitive to the value of $m_\phi$ and
the combination of electronic coefficients that enters $\Vpred$ (see
Eq.~\eqref{eq:VpredefKP}) may cancel for specific values of $m_\phi$, leading to the characteristic peaks that are
visible e.g. in Fig.~\ref{fig:compare_KP_lin}.

In the \emph{large-$m_\phi$ limit,} the Yukawa potential can be
approximated by $V_{\mathrm{NP}} \propto \delta (r)/(m_\phi r)^2$. This limit
corresponds to the case where the Compton wavelength of $\phi$ is smaller
than the nuclear charge radius and the interaction mediated by $\phi$ becomes a
contact interaction. In this case the King
plot method is no longer able to distinguish the new physics shift from the field shift induced by changes in the nuclear radius.
The $X$ coefficients align themselves with the electronic
field shift coefficients~\footnote{
        Note that uncertainties in the $X$ coefficients, which are predicted by means of
        atomic structure calculations, generally spoil this behaviour. However, it can
        easily be restored by rescaling the $X_i$ coefficients in such a way that
        $X_i/X_j \to F_{ij}\vert_{\text{dat}}$, where $F_{ij}\vert_{\text{dat}}$ is the
        slope of best fit line to the King plot data.
        }
$F_i\propto
|\Psi_i^{\mathrm{fin}}(0)|-|\Psi_i^{\mathrm{init}}(0)|$ in such a way that
$X_i/X_j \to F_i/F_j$, and consequently the new physics parameters $\alphaNPdEM \gamma^a$ can no longer be distinguished from the charge radius variance $\drsqa{a}$. The resulting suppression of the sensitivity to new
physics can be
observed e.g. in Fig.~\ref{fig:compare_KP_lin}.

In Appendix~\ref{sec:sensitivity_NP}, a King plot analysis is applied to selected transitions in Ca$^+$ to illustrate the behaviour of the electronic part of $\Vpred$ across different transitions. Moreover, the impact of the uncertainties on the $X$ coefficients is briefly discussed therein.

\subsubsection*{Additional Algebraic Methods}
For completeness, let us briefly introduce the \emph{No-Mass King Plot} and the
\emph{Projection Method}, which provide alternatives to Eq.~\eqref{eq:ANP3x2}
that can be beneficial, respectively, in the case of large nuclear mass
uncertainties and in the case of isotope shift data sets for two transitions but
more than 3 isotope pairs.

\paragraph*{No-Mass King Plot ---}
The \emph{No-Mass King Plot (NMKP)} uses isotope shift data not only to
eliminate the charge radius variance $\drsq$ from Eq.~\eqref{eq:liNPIS}, but also the nuclear masses. This requires isotope shift measurements for an additional transition $\nuveci{3}$, yielding~\cite{Berengut:2020itu}~\footnote{
    In the \kifit code this formula is referred to as the ``dimension 3 No-Mass
    Generalised King Plot''.}:
    \begin{align}
        \alphaNPdEM = \frac{
            2\mathrm{det}\left(
            \nuveci{1},\nuveci{2},\nuveci{3}
            \right)}{
            \varepsilon_{ijk}\mathrm{det}\left(
            X_i\boldsymbol{\gamma},\nuveci{j},\nuveci{k}\right)}\,.
        \label{eq:ANP3x3}
    \end{align}
The denominator takes a particularly symmetric form in this case: 
    \begin{align}
    \Vpred = \frac{1}{2}\varepsilon_{ijk}\mathrm{det}\left(
        X_i\boldsymbol\gamma,\nuveci{j},\nuveci{k}\right) 
        = \mathrm{det}(\mathcal{M}) \mathrm{det}(\mathcal{N})
        \label{eq:VpredefNMKP}
    \end{align}
with
    \begin{align}
         \mathcal{M} = \begin{pmatrix}
         X_1 & K_1 & F_1\\
         X_2 & K_2 & F_2\\
         X_3 & K_3 & F_3
         \end{pmatrix}\,,
         \quad
         \mathcal{N} = \begin{pmatrix}
         \gamma^{1} & \mua{1} & \drsqa{1} \\
         \gamma^{2} & \mua{2} & \drsqa{2} \\
         \gamma^{3} & \mua{3} & \drsqa{3} 
         \end{pmatrix}\,.
         \label{eq:ElNuclMatNMKP}
    \end{align}
The No-Mass King Plot formula allows to mitigate uncertainties on the nuclear
masses when their effect on the uncertainty estimation in
Eq.~\eqref{eq:nsigma_bound} dominates over that of the isotope shift
uncertainties.

\paragraph*{Projection Method ---}

The \emph{Projection Method} presented in Ref.~\cite{Solaro:2020dxz}, yields a bound
on new physics using isotope shift data for 2 transitions measured in $n$
isotope pairs.
Introducing the $(n\times2)$ matrix
    \begin{align}
        D_\mathbf{w}\equiv\left(\mnuveci{1},\mathbf{w}\right)\,,
    \end{align}
where $\mathbf{w}$ is a generic $n$-vector in isotope-pair space, we can define the projection of the vector $\mmuvec$ on the plane spanned by the vectors $\mnuveci{1}$ and $\mathbf{w}$ as
    \begin{equation}
        \mathbf{p}_\mathbf{w}=D_\mathbf{w}\left(D_\mathbf{w}^\top D_\mathbf{w}\right)^{-1}D_\mathbf{w}^\top\mmuvec\,,
    \end{equation}
where $\top$ denotes the transpose. 
Then, $\alphaNP$ can be expressed as
    \begin{equation}
        \left\vert\alphaNPdEM\right\vert=\frac{
                        \mathrm{V}\left(\mmuvec,\mnuveci{1},\mnuveci{2}\right)
                        }{
                        |X_{21}| \mathrm{V}\left(\mmuvec,\mnuveci{1},\hvec\right)
                        },
        \label{eq:ANP_proj}
    \end{equation}
where
    \begin{align}
        &\mathrm{V}\left(\mmuvec,\mnuveci{1},\mathbf{w}\right)
        =\left\|\mmuvec-\mathbf{p}_\mathbf{w}\right\|\sqrt{\|\mnuveci{1}\|^2
        \|\mathbf{w}\|^2-\left(\mnuveci{1} \cdot \mathbf{w}\right)^2}\,, 
    \end{align}
is the volume of the parallelepiped spanned by the vectors $\mmuvec$, $\mnuveci{1}$ and $\mathbf{w}$.
Notice that Eq.~\eqref{eq:ANP_proj} reduces to an equation similar to the Minimal King Plot formula (Eq.~\eqref{eq:ANP3x2}), for $n=3$ isotope pairs,
however, it is insensitive to the sign of $\alphaNP$.
When sampling $\alphaNP$ (see discussion around Eq.~\eqref{eq:nsigma_bound}), this insensitivity to the sign can lead to a distortion of the distribution.

\subsection{Algebraic Methods for Nonlinear King Plots}
\label{sec:algnonlin}
In the previous section, we discussed how isotope shifts that produce linear King plots can be used to set bounds on the new physics coupling. However, the increasing sensitivity of isotope shift and nuclear mass measurements has resolved deviations from Eq.~\eqref{eq:linIS} (see e.g. samarium (Sm)~\cite{Striganov:1962_Sm,King:1963,Stacey:1966_IS_interpretation} and ytterbium (Yb)~\cite{Counts:2020aws,Hur:2022gof,Figueroa:2022mtm,ono2022observation,kawasaki2024isotope,Door:2024qqz} King plots, and the combination of singly and highly-charged calcium (Ca)~\cite{Wilzewski:2024wap}~\footnote{
            After a nonlinearity had been measured recently in cadmium (Cd)~\cite{Ohayon:2022ijv}, linearity was restored in \cite{Hofsass:2022koa}.
            }),
calling for a more detailed analysis of the SM background. Assuming factorisability of the electronic and nuclear contributions, we denote the next-to-leading SM contribution by
$H_i \eta^a$, such that
    \begin{align}
        \nuai{a}{i} = K_i \mua{a} + F_i \drsqa{a} 
                      + H_i \eta^a + \ldots
                      + \alphaNPdEM X_i \gamma^{a}\,,
        \label{eq:nonliNPIS}
    \end{align}
where the ellipsis is a placeholder for unresolved higher-order SM contributions. In the presence of King plot nonlinearities, methods beyond the ones discussed in the previous section are needed. In the following section we discuss how the main sources of
nonlinearity can be identified and eliminated, either in a data-driven way, or
using additional theoretical input, and why King plots remain a useful tool to
search for new physics.

\subsubsection*{The Nonlinearity Decomposition}
If at least four isotope pairs are probed, 
there is a clear hierarchy between the King nonlinearities, and predictions for their nuclear structure are available, the \textit{Nonlinearity
Decomposition Plot}~\cite{Counts:2020aws,Hur:2022gof} can provide useful insights into the origin of the dominant King nonlinearity.
The main idea is to project the isotope shift data onto the $n$ basis
vectors $(\mmuvec, \mnuveci{1}, \Lambdavec{1}, \ldots, \Lambdavec{n-2})$, where
$\mmuvec$ and $\mnuveci{1}$ define the plane of ``King linearity'', and the linearly independent vectors $\lbrace\Lambdavec{\ell}\rbrace_{\ell = 1}^{(n-2)}$, 
which can be chosen to be orthogonal~\footnote{The results in this section can, of course, be generalised to the case where the vectors $\Lambdavec{\ell}$ are not orthogonal to the basis vectors $\mmuvec$ and $\mnuveci{1}$. In this case the projections of the higher-order terms such as $H_i \eta^a$ onto $\mmuvec$ and $\mnuveci{1}$ need to be taken into account. 
} to $\mmuvec$ and $\mnuveci{1}$, define a
basis for the ``King nonlinearity space'':
    \begin{align}
        \mnuveci{i} = K_{i1} \mmuvec + F_{i1} \mnuveci{1} 
                      + \sum_{\ell = 1}^{n-2}\lambdacoordi{\ell}{i1} \Lambdavec{\ell}\,.
        \label{eq:lambda_dec}
    \end{align}

In the minimal case of 4 isotope pairs, as in
Refs.~\cite{Counts:2020aws,Hur:2022gof,Door:2024qqz} and
Ref.~\cite{Wilzewski:2024wap}, which considered the 5 even stable isotopes of Yb
and Ca, respectively, the basis vectors of the King nonlinearity space can be expressed
as~\cite{Hur:2022gof}
    \begin{align}
        \begin{split}
            \Lambdavec{+}  &\sim \left(\mnuai{3}{1}-\mnuai{2}{1},\mnuai{1}{1}-\mnuai{4}{1},\mnuai{4}{1}-\mnuai{1}{1},\mnuai{2}{1}-\mnuai{3}{1}\right)\,,\\
            \Lambdavec{-}  &\sim \left(\mnuai{4}{1}-\mnuai{2}{1},\mnuai{1}{1}-\mnuai{3}{1},\mnuai{2}{1}-\mnuai{4}{1},\mnuai{3}{1}-\mnuai{1}{1}\right)\,,
        \end{split}
    \label{eq:NLDvecs}
    \end{align}
and Eq.~\eqref{eq:lambda_dec} reads
    \begin{equation}
         \mnuveci{i} = K_{i1} \mmuvec + F_{i1} \mnuveci{1} + \lambdacoordi{+}{i1} \Lambdavec{+} + \lambdacoordi{-}{i1} \Lambdavec{-}\,.
         \label{eq:lambda_dec_4}
    \end{equation}

\begin{figure}
        \centering
        \includegraphics[scale=.35]{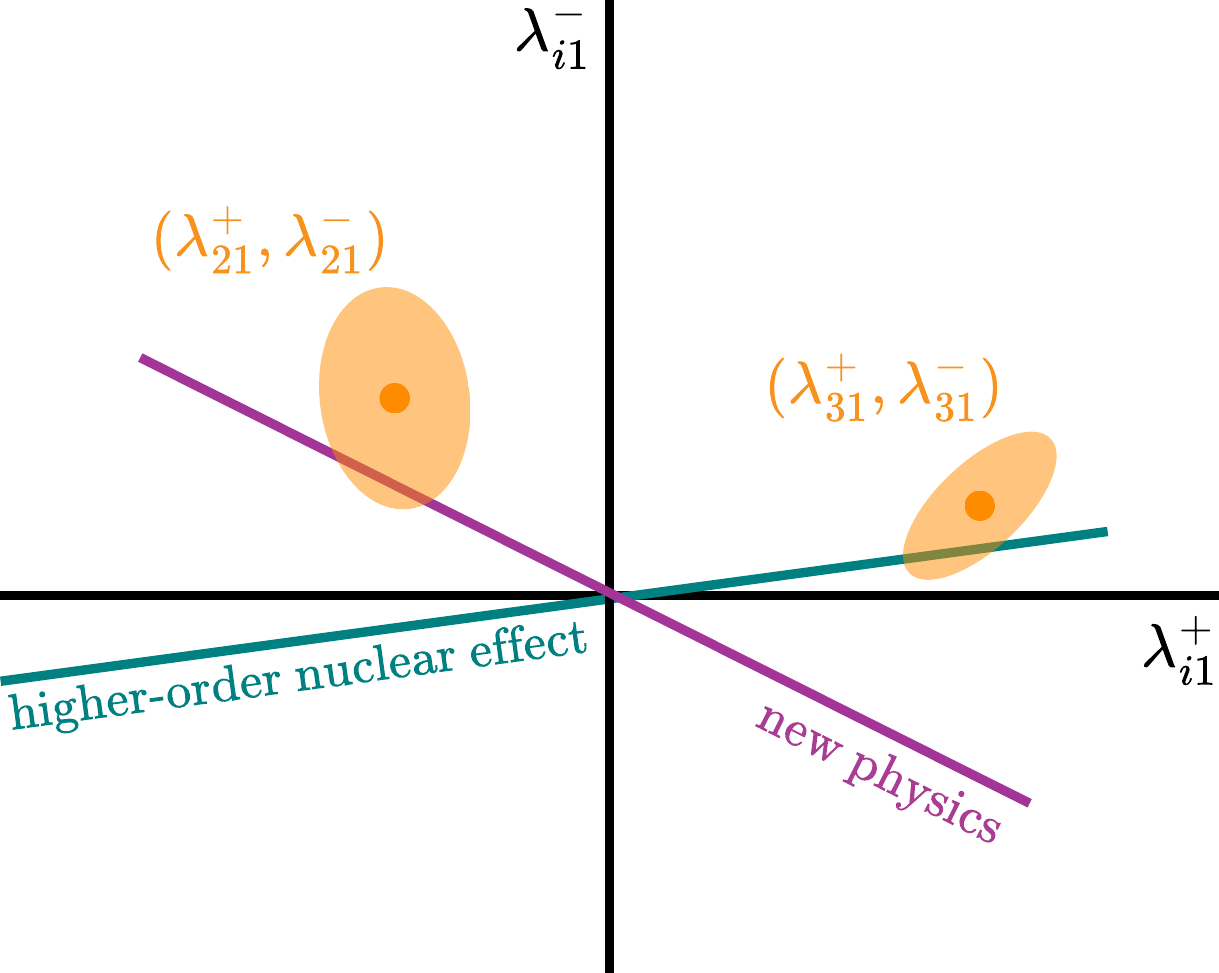}
        \caption{
        Schematic illustration of the Nonlinearity Decomposition Plot in
        the minimal case of 4 isotope pairs.
        The two orange points with uncertainty ellipses represent the nonlinearities observed in the King plots constructed from the isotope shift data for the transition pairs $(1,2)$ and $(1,3)$. The purple and teal lines show the predicted slopes of the new physics term and a higher-order SM contribution to the isotope shift, respectively.
        }
    \label{fig:lambda_dec}
    \end{figure}

The projections $\lambdacoordi{+}{i1}$ and $\lambdacoordi{-}{i1}$
are used as coordinates in the Nonlinearity Decomposition Plot, which in the case of 4 isotope pairs is 2-dimensional, as illustrated in Fig.~\ref{fig:lambda_dec}:
Each point $(\lambdacoordi{+}{i1}, \lambdacoordi{-}{i1})$, $i=2,3$ in the Nonlinearity Decomposition Plot represents the nonlinearities in the 2-dimensional King plot constructed from the transitions $i$ and $1$. If its uncertainty ellipse contains the origin of the $(\lambdacoordi{+}{i1}, \lambdacoordi{-}{i1})$ plane, the associated King plot can be considered to be linear, else additional terms need to be added to the isotope shift equations to describe the data.

The Nonlinearity Decomposition Plot is particularly useful when there is a strong hierarchy between the nonlinearities. In the case of one dominant King nonlinearity that can be factorised into electronic and nuclear contributions (as e.g. $H_i $ and $\eta^a$ in Eq.~\eqref{eq:nonliNPIS}), the data points approximately lie on one line in the Nonlinearity Decomposition Plot. 
Indeed, in this case the slope $\lambdacoordi{-}{i1}/\lambdacoordi{+}{i1}$ defined by the projections onto the space of King nonlinearity is transition independent
\begin{align}
     \frac{\lambdacoordi{-}{}}{\lambdacoordi{+}{}} \equiv
    \frac{\lambdacoordi{-}{i1}}{\lambdacoordi{+}{i1}} = \frac{
    (\boldsymbol{\tilde{\eta}}\cdot\Lambdavec{-} )
    (\Lambdavec{+}\cdot \Lambdavec{+})
    -
    (\boldsymbol{\tilde{\eta}}\cdot \Lambdavec{+} )
    (\Lambdavec{-}\cdot \Lambdavec{+})
    }{
    (\boldsymbol{\tilde{\eta}}\cdot\Lambdavec{+} )
    (\Lambdavec{-}\cdot \Lambdavec{-})
    -
    (\boldsymbol{\tilde{\eta}}\cdot \Lambdavec{-} )
    (\Lambdavec{-}\cdot \Lambdavec{+}        )
    }
\end{align}
and a comparison with predictions for different nuclear parameters $\boldsymbol{\tilde{\eta}}$ can be used to identify the origin of the dominant nonlinearity in the data.

The purple line through the origin in Fig.~\ref{fig:lambda_dec} illustrates how the new physics hypothesis introduced in Eq.~\eqref{eq:liNPIS} can be tested with the help of the Nonlinearity Decomposition Plot: If all uncertainty ellipses overlap with this purple line, the observed nonlinearities are compatible with the new physics hypothesis (or any other effect that is proportional to the neutron number).
A similar statement can be made for higher-order nuclear contributions. The slopes of the corresponding lines can either be determined via nuclear structure calculations or using complementary experimental input.

Note that more than one source of nonlinearity, as well as input for their electronic structure, is needed to describe the data that does not lie on a line through the origin of the $(\lambdacoordi{+}{i1}, \lambdacoordi{-}{i1})$ plane.\\

\paragraph*{Generalisation to $n$ isotope pairs}
If isotope shift data for $n$ isotope pairs is available, more than one nonlinearity can be identified in this data-driven way. We define 
the vector of coordinates associated to the King plot for the transitions $i$ and $1$ as
$\boldsymbol{\lambda}_{i1} = (\lambdacoordi{1}{i1},  \ldots,  \lambdacoordi{n-2}{i1})^\top$ 
and introduce the $ (n \times (n-2))$-matrix 
$\boldsymbol\Omega = (\Lambdavec{1},  \ldots , \Lambdavec{n-2} )$ of vectors that span the nonlinearity space. Eq.~\eqref{eq:lambda_dec} can then be generalised using
$    \sum_{\ell = 1}^{n-2}\lambdacoordi{\ell}{i1} \Lambdavec{\ell} 
        =  \boldsymbol\Omega \boldsymbol{\lambda}_{i1}\,.$

Using the Moore-Penrose pseudoinverse of $\boldsymbol\Omega$ to solve for the 
coordinates $\lambdacoordi{\ell}{i1}$, one obtains the generalisation of Eq.~\eqref{eq:NLDslope}
    \begin{align}        
        \frac{\lambdacoordi{\ell'}{i1}}{\lambdacoordi{\ell}{i1}} 
        = \frac{\left[(\boldsymbol\Omega^\top \boldsymbol\Omega)^{-1}
            \boldsymbol\Omega^\top\etatildvec\right]^{(\ell')}}{\left[(\boldsymbol\Omega^\top
            \boldsymbol\Omega)^{-1} \boldsymbol\Omega^\top\etatildvec\right]^{(\ell)}}\,, \quad \ell'\neq \ell\,,
        \label{eq:NLDslope}
    \end{align}
where $[\cdot]^{(\ell)}$ is to be understood as the $\ell^\mathrm{th}$ entry of $[\cdot]$. Since $\ell, \ell'$ run from 1 to $n-2$, and $\ell'\neq \ell$, this method can resolve $n-3$ sources of King plot nonlinearity, i.e. one in the case of 4 isotope pairs.

Note that while the slope of the predicted nonlinearity in the King plane is transition-independent, the coordinates $\lambdacoordi{\ell}{i1}$ are not and their magnitude
    \begin{align}
        \Vert \boldsymbol{\lambda}_{i1} \Vert  
            = \sqrt{\sum_\ell \big(\lambdacoordi{\ell}{i1}\big)^2} 
            = |H_{i1}| \sqrt{\etatildvec^\top
            \boldsymbol\Omega(\boldsymbol\Omega^\top \boldsymbol\Omega)^{-2}
            \boldsymbol\Omega^\top\etatildvec}\,,
        \label{eq:NLDrad}
    \end{align}
or 
    \begin{align}
        \lambdacoordi{+}{i1}
            &= H_i \boldsymbol{\tilde{\eta}} \cdot 
            \frac{\Lambdavec{+}(\Lambdavec{-}\cdot\Lambdavec{-})-\Lambdavec{-}(\Lambdavec{-}\cdot\Lambdavec{+})}{(\Lambdavec{+}\cdot\Lambdavec{+})(\Lambdavec{-}\cdot\Lambdavec{-})-(\Lambdavec{-}\cdot\Lambdavec{+})^2}
            \,,\notag\\            
        \lambdacoordi{-}{i1}
            &= H_i \boldsymbol{\tilde{\eta}} \cdot 
            \frac{\Lambdavec{-}(\Lambdavec{+}\cdot\Lambdavec{+})-\Lambdavec{+}(\Lambdavec{-}\cdot\Lambdavec{+})}{(\Lambdavec{+}\cdot\Lambdavec{+})(\Lambdavec{-}\cdot\Lambdavec{-})-(\Lambdavec{-}\cdot\Lambdavec{+})^2}
    \end{align}
in the case of 4 isotope pairs,
can be used to extract the electronic coefficients $H_{i1}$ from data. 

\subsubsection*{Generalised King Plots}
\label{sec:generalised_algebraic}
In the recent Yb ~\cite{Counts:2020aws,Hur:2022gof,Figueroa:2022mtm,ono2022observation,kawasaki2024isotope,Door:2024qqz} and Ca~\cite{Wilzewski:2024wap} King plots, the leading nonlinearity is not compatible with the new physics term and so-called \emph{Generalised King Plots}~\cite{Berengut:2020itu} were employed to set bounds on $\alphaNP$. These generalise Eq.~\eqref{eq:ANP3x2} and \eqref{eq:ANP3x3} to higher dimensions, i.e. they use additional measurements to eliminate higher-order nuclear contributions to
the isotope shift equations.

For concreteness, let us assume the isotope shift equations take the form of Eq.~\eqref{eq:nonliNPIS}:
As discussed in the case of the \emph{Minimal King Plot}, isotope shift data for 
two transitions, $i=1,2$, allow us to eliminate the charge radius variance, 
$\drsq$, resulting in
    \begin{align}
        \mnuai{a}{2} = K_{21} + F_{21} \mnuai{a}{1}
                      + H_{21} \etatilda{a} + \ldots 
                      + \alphaNPdEM X_{21} \ha{a}\,,
    \end{align}
where $H_{21} \equiv H_2 - F_{21} H_1$. 

If isotope shifts are measured in a third transition $i=3$, they can be used to 
eliminate the higher-order SM term $\etatilda{a}$, leading to
    \begin{align}
        \mnuai{a}{3} = K_{321} + F_{321} \mnuai{a}{1} 
                       + H_{321} \mnuai{a}{2} + \ldots
                       + \alphaNPdEM X_{321} \ha{a}\,,
        \label{eq:liNPKP43}
    \end{align}
with $H_{321} \equiv H_{31}/H_{21}$ and $P_{321}=P_{31}-H_{321}P_{21}$, with $P\in \{K,F,X\}$. Assuming the higher-order SM terms represented by the ellipsis are negligible, the isotope shift data can be used to set bounds on the new physics term (see fit in Ref.~\cite{Hur:2022gof}). 
Note, however, that any direct extraction of $\alphaNP$ from Eq.~\eqref{eq:liNPKP43} will be affected by the large theoretical uncertainties associated with the prediction for the electronic coefficient $X_{321}$, corresponding to a difference differences of the $X$ coefficients introduced in Eq.~\eqref{eq:liNPIS}.\\

One of the main strengths of the \emph{Generalised King Plot} formulas is that they directly use the $X$ coefficients introduced in Eq.~\eqref{eq:liNPIS} rather than relying on differences thereof. Assuming isotope shifts for $m$ transitions are measured in $n=m+1$ isotope pairs, solving the system of $m \times n$ isotope shift equations for $\alphaNP$ leads to the \emph{Generalised King Plot (GKP)} formula~\cite{Berengut:2020itu}~\footnote{
    In the \kifit code this formula is referred to as the ``dimension $n$
    Generalised King Plot'', where $n$ is the number of isotope pairs.\\
    \noindent
    Using frequency-normalised quantities, i.e.
    $\bar{\tau}^a\equiv \tau^a/\nuai{a}{1}$, 
    $\tau \in \lbrace \nu_{i}, \mu, \gamma\rbrace$, $i=1,\ldots,m$,
    Eq.~\eqref{eq:ANPnxm} takes the form
    \begin{align*}
        \alphaNPdEM &=\frac{(n-2)!\;
        \mathrm{det}\left(\fnuveci{1}, \ldots,\fnuveci{n-1}, \fmuvec\right)
        }{\varepsilon_{i_1\ldots i_{n-1}}
        \mathrm{det}\left(
        X_{i_1}\gammabarvec,\fnuveci{i_2},\ldots,\fnuveci{i_{n-1}},\fmuvec
        \right)}\,.
    \end{align*}
    }:
    \begin{align}
        \alphaNPdEM &= \frac{(n-2)!\;\mathrm{det}\left(
        \mnuveci{1}, \ldots, \mnuveci{n-1}, \mmuvec
        \right)}{\varepsilon_{i_1\ldots i_{n-1}}\mathrm{det}\left(
        X_{i_1} \hvec,\mnuveci{i_2},\ldots,\mnuveci{i_{n-1}},\mmuvec\right)}\,,
        \label{eq:ANPnxm}
    \end{align}

As briefly suggested in Ref.~\cite{Berengut:2020itu}, if higher-order SM contributions are present and the uncertainties on the
nuclear masses are a limiting factor, it can be advantageous to combine the benefits of the GKP and NMKP (Eq.~\eqref{eq:ANP3x3}) formulas, obtaining a \emph{No-Mass Generalised King Plot
(NMGKP)} formula:
    \begin{align}
        \alphaNPdEM = \frac{(n-1)!\;\mathrm{det}\left(
        \nuveci{1}, \nuveci{2},\ldots,\nuveci{n}\right)
        }{\varepsilon_{i_1,i_2,\ldots, i_n}\mathrm{det}\left(
        X_{i_1}\gammavec, \nuveci{i_2},\ldots,\nuveci{i_n}\right)}\,.
        \label{eq:ANPnxn}
    \end{align}
    {\setlength{\tabcolsep}{4pt}
    \renewcommand{\arraystretch}{1.1}
    \begin{table}[]
        \centering
        \begin{tabular}{l|ccccc}
        \hline
        \hline
                      & KP & NMKP & GKP & NMGKP & \kifit\\
                      & Eqs.~\eqref{eq:ANP3x2} & \eqref{eq:ANP3x3}
                      & \eqref{eq:ANPnxm} & \eqref{eq:ANPnxn} &
                      Sec.~\ref{sec:kifit}\\

        \hline
        Isotope pairs & 3 & 3 & $3 \leq n$    & $3\leq n$   & $3 \leq n$ \\
        Transitions   & 2 & 3 & $n-1$ & $n$   & $2 \leq m$ \\
        Spurions      & - & - & $n-3$ & $n-3$ & - \\
        \hline
        \hline
        \end{tabular}
        \label{tab:dof_KP}
        \caption{
        Summary of the number of transitions and isotope
        pairs that are required by the different algebraic methods presented in Section~\ref{sec:alg} and by the King plot fit \kifit introduced in
        Section~\ref{sec:kifit}. (Note that $m$ and $n$ are independent
        integers.) For the GKP and the NMGKP we also give the number of
        higher-order nuclear parameters (spurions) that are eliminated while
        setting a bound on the coupling $\alphaNP$. 
        KP and NMKP are the minimal cases of GKP and NMGKP, respectively.
        }
        \label{tab:fitvsalg_dims}
    \end{table}
    }

\subsubsection*{Subtracting SM Nonlinearities}
For lighter systems such as Ca, where the mass shift dominates over the field shift, the next-to-leading order SM contribution to 
Eq.~\eqref{eq:linKPrel} might be the second-order mass shift~\cite{Viatkina:2023qop}, 
$\nuai{a}{i}\vert_{\mathrm{MS}(2)}=K_i^{(2)} \mua{a(2)}$, where
$\mua{a(2)}\equiv 1/(\ma{A})^2-1/(\map{A})^2$ is the difference of the squared
inverse nuclear masses and $K_i^{(2)}$ is an electronic coefficient. 
Since in this case the nuclear parameters (namely, the isotope masses) can be determined experimentally at high precision, the
second-order mass shift can be kept explicit in the isotope shift equations, such that
    \begin{align}
           \mnuai{a}{2} 
            = K_{21} + F_{21} \mnuai{a}{1} +\alphaNPdEM X_{21} \ha{a}\, + K_{21}^{(2)}\mmua{a(2)}.
            \label{eq:KPrelMS2(1)}
    \end{align}
If the frequency shifts $\nuai{a}{i}$, $i=1,2$, are measured in four isotope
pairs, Eq.~\eqref{eq:KPrelMS2(1)} can then be solved for $\alphaNP$, yielding
    \begin{equation}
        \alphaNPdEM=\frac{
            \det\left( \mnuveci{1},\mnuveci{2},\mmuvec, \mmusqvec \right)
        }{
            \varepsilon_{ij}
            \det\left(X_i \hvec, \mnuveci{j}, \mmuvec, \mmusqvec\right)}\,,
        \label{eq:ANP_2nd_MS}
    \end{equation}
where all bold symbols are 4-vectors in isotope-pair space. In this case one higher-order effect can be taken into account without adding an additional transition. We will refer to this equation as the \emph{Nuclear Input King Plot (NIKP)} formula.

Alternatively, if the the electronic coefficients $K_i^{(2)}$ can be predicted with sufficient accuracy (see e.g. Ref.~\cite{Wilzewski:2024wap, Viatkina:2023qop}, where a $10\%$ uncertainty is assigned to the prediction of $K_i^{(2)}$), the second-order mass shift can be subtracted from the measured isotope shift~\cite{Wilzewski:2024wap}, yielding an isotope shift equation which
resembles Eq.~\eqref{eq:linKPrel}:
    \begin{align}
        \left(\mnuai{a}{2} - K_{21}^{(2)}\mmua{a(2)} \right) 
        = K_{21} + F_{21} \mnuai{a}{1} +\alphaNP X_{21} \ha{a}\,.
        \label{eq:KPrelMS2}
    \end{align}
Now $\alphaNP$ can be obtained using Eq.~\eqref{eq:ANP3x2} with the modification $\mnuai{a}{2} \to \left(\mnuai{a}{2}\right)'= \mnuai{a}{2} - K_{21}^{(2)}\mmua{a(2)}$. 
Since $K_i^{(2)}$ depends only on the electronic structure, it induces correlations between the objects $\lbrace\left(\mnuai{a}{2}\right)'\rbrace_{a=1}^n$, which in turn reduce the impact of the uncertainty on the coefficients $K_i^{(2)}$ on $\sigalphaNP$~\cite{Wilzewski:2024wap}.

The strategy of subtracting higher-order SM contributions until relations such as Eq.~\eqref{eq:KPrelMS2} arise, could be promising to bridge the gap between light elements such as hydrogen, deuterium or helium
~\cite{Delaunay:2017dku,Jones:2019qny,Potvliege:2023lvf,Potvliege:2024xly},
where spectroscopic bounds can be derived from a direct comparison of predictions 
and measurements, and heavy elements such as Yb, where corrections to the charge
distribution, such as nuclear deformation, are sizeable but hard to predict from
first principles~\cite{Door:2024qqz}.

\subsubsection*{Combining Data Sets}
In Section~\ref{sec:alg}, we discussed the algebraic methods summarised in
Table~\ref{tab:fitvsalg_dims}. The advantage of these methods is their
simplicity. However, they can only be applied to data sets of the specific
dimensions given in Table~\ref{tab:fitvsalg_dims}. If the available data
sets come in different shapes, we are forced to derive bounds using subsets of
the data, and how these resulting bounds are best combined is unclear.

In the following section we will follow an alternative approach consisting of a
fit to isotope shift data. This strategy is more flexible regarding the dimensions of the input data and in particular enables a fit to isotope shift data from different elements.

\section{The King Plot Fit}
\label{sec:kifit}

The King plot fit presented in this section is inspired by
Ref.~\cite{Frugiuele:2016rii} but significantly extended such that it can handle the contemporary precision of isotope shift data and potentially large
hierarchies between the levels of uncertainty of different subsets of the data. 
In the following, we describe the theoretical framework and the main
steps of the algorithm, which is structured into a \emph{build phase}, a
\emph{search phase}, an \emph{experiment phase} and a \emph{consolidation
phase}, as illustrated in Fig.~\ref{fig:kifit_algo}.
An implementation of this procedure is provided in the form of the Python
package \kifit, which is publicly available at 
\href{https://github.com/QTI-TH/kifit}{https://github.com/QTI-TH/kifit}. More instructions on how to use the package can be found in Appendix~\ref{sec:kifit}. 
The data provided by the current version of \kifit is summarised in Appendix~\ref{sec:data}. 

\subsection{Geometric Construction (Build Phase)}
\label{sec:build}
One of the advantages of the King plot fit is that it is more flexible than the algebraic methods described in Section~\ref{sec:alglin} (see
Table~\ref{tab:fitvsalg_dims} for an overview) in regard to the number of transitions and isotope pairs that it can combine.
As in Section~\ref{sec:algnonlin}, $m$ denotes the number of transitions and $n$ the number of 
isotope pairs considered for a given element.

Depending on the application, it can be advantageous to arrange the same isotope
shift data in terms of isotope pairs or transitions:
For the construction of the determinants discussed in
Section~\ref{sec:alglin}, it is beneficial to describe the data in terms of the
vectors $\mmuvec$, $\mnuveci{i}$ and $\hvec$, which are fixed by the transition
frequency and mass measurements, as well as the neutron number difference.
For the fit, which is performed directly on the level of the King plot and constrains
deviations from the King line defined by the electronic coefficients $F_{i1}$,
$K_{i1}$, $i=2,\ldots, m$, it is more natural to work in transition space.

Fixing $i=1$ as the reference transition and assuming the isotope shift data to be linear, we can 
construct $(m-1)$ relations of the form of Eq.~\eqref{eq:linKPrel}, and arrange them into $n$ linear $m$-vector equations, one for every isotope pair $a$~\cite{Frugiuele:2016rii}:
    \begin{align}
        \mnuvecalin{a}\equiv \! \!
        \begin{pmatrix}
            \mnuai{a}{1}\\[-3mm]
            \mnuailin{a}{2}\\[-3mm]
            \vdots\\[-3mm]
            \mnuailin{a}{m}
        \end{pmatrix} \! \!
        = \!
        \begin{pmatrix}
            0\\[-3mm]
            K_{21}\\[-3mm]
            \vdots\\[-3mm]        
            K_{m1}
        \end{pmatrix}
        \!
        + \mnuai{a}{1} \!
        \begin{pmatrix}
            1\\[-3mm]
            F_{21}\\[-3mm]
            \vdots\\[-3mm]
            F_{m1}
        \end{pmatrix} \!
        \equiv \Kijvec + \mnuai{a}{1} \Fijvec\,.
        \label{eq:linISfitdef}
    \end{align}
Here we introduced the $m$-dimensional electronic coefficient vectors $\Kijvec$ and $\Fijvec$. Note that the isotope shift vectors introduced in Eq.~\eqref{eq:linvecKP}
and used in Sections~\ref{sec:alglin} and \ref{sec:algnonlin} were vectors in isotope
pair space, meaning that we had one for each transition, whereas the vectors that appear in Eq.~\eqref{eq:linISfitdef} are vectors in transition
space, meaning that we have one for each isotope pair.

\begin{figure}
        \centering
        \includegraphics[scale=0.33]{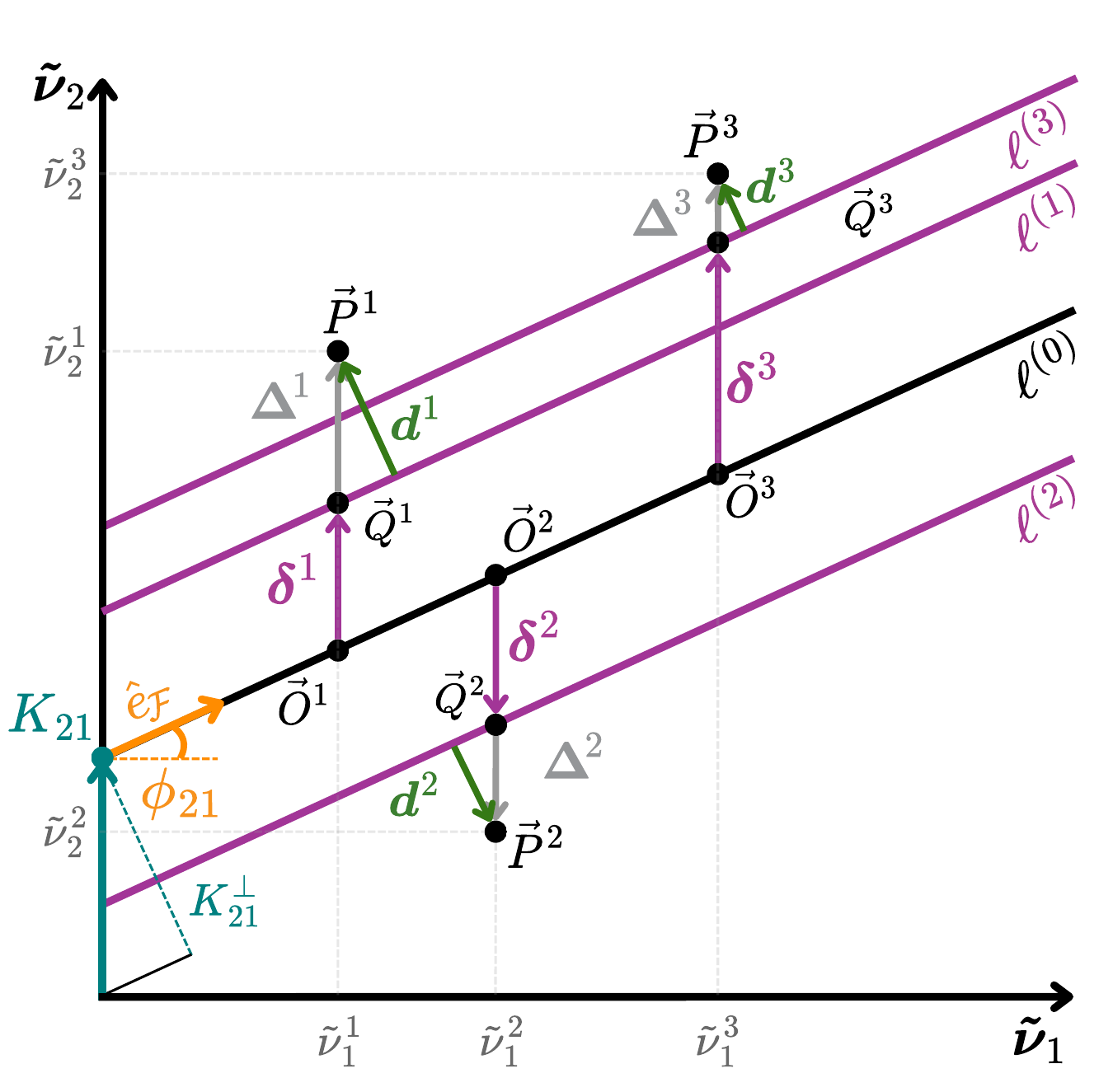}
        \caption{
        Illustration of the \kifit construction in the simplest case of $m=2$
        transitions and $n=3$ isotope pairs.
        We use the notation
        $\vec{O}^a = (\mnuai{a}{1}, \mnuailin{a}{2})$, $a=1,2,\ldots, n$ for the
        linear prediction (Eq.~\eqref{eq:linISfitdef}),
        $\vec{Q}^a = (\mnuai{a}{1}, \mnuaipred{a}{2})$ for the nonlinear
        prediction including new physics (Eq.~\eqref{eq:NPISfit}) and
        $\vec{P}^a = (\mnuai{a}{1}, \mnuai{a}{2})$ for the data points.
        The King line obtained from the initial fit to the isotope shift data is
        denoted $\ell^{(0)}$
        and described by the fit parameters $\Kperpij{21}$ and $\phij{21}$, 
        whereas the parallel line associated to the isotope pair $a$ is
        denoted $\ell^{(a)}$.
        }
        \label{fig:kifit_sketch}
\end{figure}

\begin{figure}
        \centering
        \includegraphics[scale=.2]{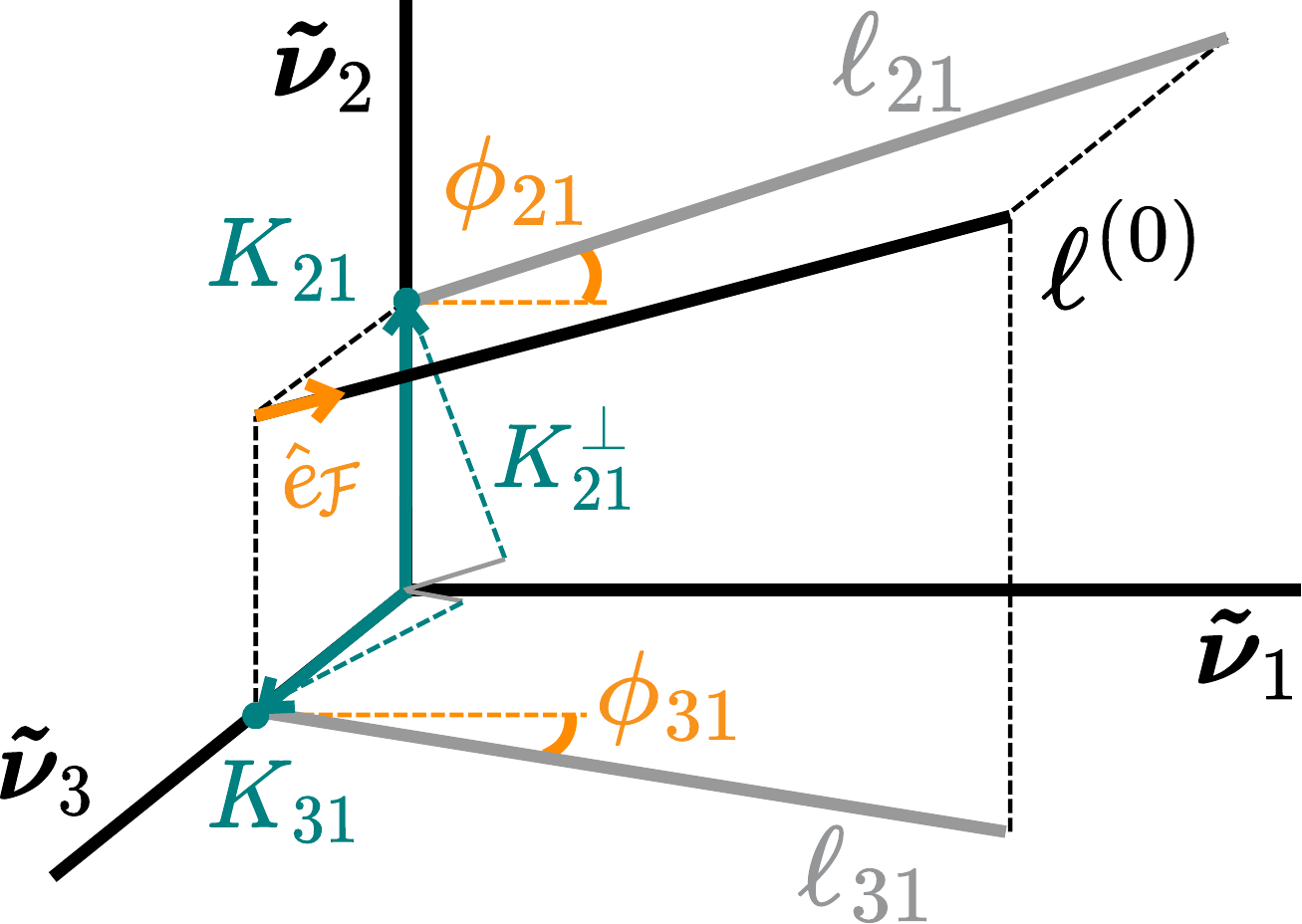}
        \caption{
        Illustration of a 3 dimensional King plot with a King line $\ell^{(0)}$ (black), fixed by the fit parameters $\phi_{j1}$ 
        (see Eq.~\eqref{eq:eFdef}) and $\Kperpij{j1}$, $j=2,3$ (see Eq.~\eqref{eq:Kperpdef}) 
        and oriented in the direction of the unit vector $\eFvec$. 
        The lines $\ell_{j1}$, $j=2,3$, represent the King lines of the 2-dimensional King plots
        constructed for the transition pairs $(1,2)$ and $(1,3)$, respectively.
        }
        \label{fig:Kperp_3d}
    \end{figure}

Fig.~\ref{fig:kifit_sketch} is a 2-dimensional illustration of Eq.~\eqref{eq:linISfitdef} for the case of $m=2$ transitions and $n=3$ isotope pairs: 
By construction, the points $\vec{O}^a = (\mnuai{a}{1}, \mnuailin{a}{2})$, $a=1,2,3$ lie on a \emph{King line} $\ell^{(0)}$ with intercept $K_{21}$ and slope $F_{21}$.
In the \kifit code, the vector $\Fijvec$ of electronic field shift coefficients
is expressed in terms of the inclination angles $\phi_{j1}$ of the King lines in
the $(\mnuveci{1}, \mnuveci{j})$ plane. This is reasonable, since only its
orientation matters.
The unit vector $\eFvec$, directed along the King line $\ell^{(0)}$
(orange vector in Fig.~\ref{fig:kifit_sketch} and Fig.~\ref{fig:Kperp_3d}), takes the form:
    \begin{align}
        \eFvec \equiv \frac{\Fijvec}{\|\Fijvec\|}
        =\frac{1}{\sqrt{1 +\sum_{j=2}^m \tan^2 \phi_{j1}}} 
        \begin{pmatrix}
        1\\[-3mm]
        \tan\phi_{21}\\[-3mm]
        \vdots\\[-3mm]
        \tan\phi_{m1}
        \end{pmatrix}
        \,.
        \label{eq:eFdef}
    \end{align}
We can define an angle $\phi$ which satisfies
\begin{align}
\cos \phi = \frac{1}{\sqrt{1 +\sum_{j=2}^m \tan^2 \phi_{j1}}}
\label{eq:phidef}
\end{align}
and corresponds to the angle between the direction 
$\boldsymbol{\hat{e}_1}=(1, 0, \ldots, 0)$ fixed by the reference transition $1$
and the King line. In the case of 2 transitions, $\phi=\phij{21}$
(see also Fig.~\ref{fig:kifit_sketch}).

Instead of taking the $m-1$ intercepts $\lbrace\Kij{j1}\rbrace_{j=2}^m$ to be
fit parameters, we introduce
    \begin{align}
        \Kperp\equiv 
        \begin{pmatrix}
            \Kperpij{11}\\[-3mm]
            \Kperpij{21}\\[-3mm]
            \vdots\\[-3mm]
            \Kperpij{m1}
        \end{pmatrix}
        = 
        \begin{pmatrix}            
            -\sum_{j=2}^m\sin \phij{j1}\Kij{j1}\\[-3mm]
            \cos \phi_{21} \Kij{21}\\[-3mm]
            \vdots\\[-3mm]
            \cos \phi_{m1} \Kij{m1}
        \end{pmatrix}\,,
        \label{eq:Kperpdef}
    \end{align}
which is orthogonal to $\eFvec$,
and fit to the the $m-1$ projected intercepts 
$\lbrace \Kperpij{j1} \rbrace_{j=2}^m$. This parametrisation is crucial for the
numerical stability of the fit in the limit where $\eFvec$ and $\Kijvec{}$
approach collinearity and $\Kperpij{j1} \ll \Kij{j1}$.

The initial guesses $\lbrace \langle \Kperpij{j1}\rangle, \langle
\phij{j1}\rangle \rbrace_{j=2}^m$ for the $2(m-1)$ fit parameters and their
covariance matrix $\covthetaij{j1}$ are determined by a set
of $(m-1)$ linear fits.
Since King plots are constructed from pairs of isotope shifts,
all of which have uncertainties, we perform orthogonal distance
regression (ODR)~\cite{odr} that is implemented in the 
\texttt{scipy.odr} method of the Scipy package~\cite{2020SciPy-NMeth}.

Once the background (i.e. the linear King plot) is fixed, we can proceed with the parametrisation of the
signal:
the new physics term introduced in Eq.~\eqref{eq:liNPKP} leads to isotope-pair
dependent shifts of the data points in transition space (i.e. in the King
plot):
    \begin{align}
        \deltaveca{a} 
        =\alphaNPdEM\ha{a}\Xijvec\,,
        \qquad 
        \Xijvec = 
        \left(
        0,  
        X_{21},  
        \ldots, 
        X_{m1}
        \right)^\top .
        \label{eq:NPvec}
    \end{align}
Here, 
$X_{j1} \equiv X_j-F_{j1}X_1 =  X_j-\tan \phi_{j1}X_1$, $j=2,\ldots,m$,
are the electronic coefficients that are constructed from the fit parameter
$\phij{j1}$ and the coefficients $X_i$, $i=1,\ldots,m$ determined by means of atomic
structure calculations (see Section~\ref{sec:alglin} and Appendix~\ref{sec:ambit}).
The predictions $\mnuvecapred{a}$ incorporating the linear isotope shift
behaviour plus the shifts predicted by the new physics term are simply given by
    \begin{align}
        \mnuvecapred{a} = & \mnuvecalin{a} + \deltaveca{a}  \,,\quad 
        a=1,\ldots, n.
        \label{eq:NPISfit}        
    \end{align}
For each isotope pair $a$, Eq.~\eqref{eq:NPISfit} defines a parallel line
$\ell^{(a)}$ to the King line $\ell^{(0)}$. In Fig.~\ref{fig:kifit_sketch}, $\vec{Q}^a = (\mnuai{a}{1}, \mnuaipred{a}{2})$, $a=1,2,3$, and the vectors $\deltaveca{a}$ with their associated lines $\ell^{(a)}$ are depicted in purple.\\

Since it is impossible to distinguish a constant shift of all points
$\mnuai{a}{i}$, $1<i\leq m$, from a variation of the mass shift coefficients $K_{i1}$, we
only include the variations $\sdeltaveca{a}$
with respect to the mean new physics-induced shift in our new physics predictions. The elements of $\sdeltaveca{a}$ can be estimated as
    \begin{align}
         \sdeltai{a}{j} = \alphaNPdEM \left(\ha{a} - \langle \ha{}\rangle^j
         \right)X_{j1}\,,
         \label{eq:sdeltaidef}
    \end{align}
where
    \begin{align}
        \langle \ha{}\rangle^j = 
        \frac{\sum_{a=1}^n \ha{a} \smnuai{a}{j}}{\sum_{b=1}^n \smnuai{b}{j}}\,,
        \quad j=2,\ldots,m
        \label{eq:asymmh}
    \end{align}
is the average value of the new physics nuclear parameter
$\ha{a}$, weighted by the uncertainties on the mass-normalised isotope shifts.
(Note that by construction $\sdeltai{a}{1}=0\; \forall a$, since $X_{11}=0$, see
Eq.~\eqref{eq:NPvec}). 
In this way the constant shift induced by the quantities $\langle \ha{}\rangle ^i$ is effectively absorbed in the fit parameters $\lbrace
\Kperpij{j1}\rbrace_{j=2}^m$, leading to a corrected mass shift vector
    \begin{align}
        \Kijvec' = \Kijvec +  \boldsymbol{\langle \ha{}\rangle}\,,
        \label{eq:Kprime}
    \end{align}
with $\boldsymbol{\langle \ha{}\rangle} = 
(\langle \ha{}\rangle^1, \, \ldots , \, \langle \ha{}\rangle^n ) ^\top$. 
Using these definitions, we obtain for the isotope shift
    \begin{align}
        \mnuvecapred{a} = & \Kijvec' + \mnuai{a}{1} \Fijvec + \sdeltaveca{a}\,,
        \quad a=1,\ldots, n.
        \label{eq:Kijp}
    \end{align}
For simplicity, we will omit the prime in $\Kijvec'$ in the following.

Note that in general Eq.~\eqref{eq:asymmh} induces a slight asymmetry under
exchange of transitions. In order to cross-check the parametrisation, we implement in the \kifit code a transition-independent version of $\sdeltaveca{a}$, with
\begin{align}
    \langle \ha{}\rangle \equiv \frac{1}{n}\sum_{a=1}^n \ha{a} 
    = \langle \ha{}\rangle^i\,,\quad \forall i\,.
    \label{eq:symmh}
\end{align}
The invariance of the \kifit construction under exchange of transitions is
discussed in more detail in Appendix~\ref{sec:kifit_validation}.

\subsubsection*{Construction of the Log-Likelihood}

The aim of the King plot fit is to minimize the distances of the experimental
data points $\mnuveca{a}$ (denoted $\vec{P}^a$ in Fig.~\ref{fig:kifit_sketch}) from the
parallel lines $\ell^{(a)}$. These distances are given by
    \begin{align}
        \dvec^{a} = &
       \Deltaveca{a}- \left(\Deltaveca{a}\cdot \eFvec \right) \eFvec \,,\quad a=1,\ldots, n\,.
       \label{eq:ddef}
    \end{align}
with $\eFvec$ as defined in Eq.~\eqref{eq:eFdef} and the vectors $\Deltaveca{a}$
that connect the predictions $\mnuvecapred{a}$ with the data points 
$\mnuvecaexp{a}$~\footnote{
    Here we are assuming that we only ever compare one data point
    $\mnuvecaexp{a}$ with a given prediction $\mnuvecapred{a}$.
    In presence of $s$ measurements of the same isotope shift $\nuai{a}{i}$, $s$
    different copies of the element $\Deltai{a}{i}$ would need to be taken into
    account. In the current version of the \kifit code, we assume these
    complementary measurements to be independent, such that they can be combined
    in a similar way to separate elements (see Eq.~\eqref{eq:llelem}).
}:
    \begin{align}
        \Deltaveca{a}\equiv & \mnuvecaexp{a}-\mnuvecapred{a}
        =\mnuvecaexp{a}-\left(\Kijvec + \mnuai{a}{1}\Fijvec +
        \sdeltaveca{a}\right)\,.
        \label{eq:Deltadef}
    \end{align}
More explicitly,
    \begin{align}
    \begin{split}
        \dai{a}{1} 
       =& -\frac{1}{\Vert \Fijvec \Vert^2}\sum_{k=2}^m\tan (\phi_{k1})
        \Deltai{a}{k1}\,,\\
     \dai{a}{j\neq 1}
       =& \Deltai{a}{j1} + \tan (\phij{j1}) \dai{a}{1}\,,
       \label{eq:d_explicit}
   \end{split}
   \end{align}
where $\Vert \Fijvec \Vert ^2 = 1+\sum_{j=2}^m \tan^2\phi_{j1}$
(see Eq.~\eqref{eq:eFdef}) and
\begin{align}
    \Deltai{a}{j1} =& \mnuai{a}{j}-\left(K_{j1}     
    + \frac{\tan\phij{j1}}{\Vert \Fijvec\Vert} \mnuai{a}{1}
    + \sdeltai{a}{j1}\right)\,.
\end{align}
Assuming that the Euclidean norms $\Vert \dveca{a}\Vert = \sqrt{\sum_{i=1}^m
\left(\dai{a}{i}\right)^2}$ approximately follow a multivariate normal
distribution, we define the negative log-likelihood 
    \begin{align}
        -\log \mathcal{L}  \propto 
        \frac{1}{2}\sum_{a=1}^n\sum_{b=1}^n\left[
        \log \Sigma^{ab}
        +\Vert \dveca{a}\Vert \left(\Sigma_{\dvec}^{ab}\right)^{-1}\Vert\dveca{b}\Vert \right]\,,
        \label{eq:ll}
    \end{align}   
where $\Sigma_{\dvec}^{ab}\equiv \mathrm{Cov}\left(\Vert\dveca{a}\Vert, \Vert
\dveca{b}\Vert \right)$ denotes the covariance of the norms
$\Vert\dveca{a}\Vert$ and $\Vert \dveca{b}\Vert$.
In vector notation,
\begin{align}
    -\log \mathcal{L} \propto \frac{1}{2}\left[
        \log \det \covd
        +\mathbf{d} \covd^{-1}\mathbf{d} \right]\,,
        \label{eq:llvec}
\end{align}
where $\mathbf{d}\equiv(\Vert\dveca{1}\Vert, \ldots,\Vert\dveca{n}\Vert)^\top$ and
$\covd$ is the associated covariance matrix. The latter can be estimated in different ways. For an order-of-magnitude estimate,
linear error propagation is sufficient. 
Assuming the frequency measurements are
independent, 
\begin{align}
    &\covdab{ab} = 
    \sum_{i=1}^m \sum_{c=1}^n \left(
    \frac{\partial \Vert \dveca{a}\Vert}{\partial  \nuai{c}{i}} \snuai{c}{i}
    \right)^2 \delta^{ab}\notag\\
    &+ \sum_{A}
    \frac{\partial \Vert \dveca{a}\Vert}{\partial  \ma{A}} \sma{A}^2
    \frac{\partial \Vert \dveca{b}\Vert}{\partial  \ma{A}}
    + \sum_{A'}
    \frac{\partial \Vert \dveca{a}\Vert}{\partial  \map{A}} \smap{A}^2
    \frac{\partial \Vert \dveca{b}\Vert}{\partial  \map{A}}\notag\\
    &+ \sum_{j=2}^m \left(\boldsymbol{\nabla_{K\phi}}^{(j1)} \Vert \dveca{a}
    \Vert\right)^\top
    \covthetaij{j1} \left(\boldsymbol{\nabla_{K\phi}}^{(j1)} \Vert \dveca{b} 
    \Vert\right)
    \,,\label{eq:covd_linprop}
\end{align}
 where the small indices run over isotope pairs and the capital ones run over isotopes.
Here we introduced the notation 
$\boldsymbol{\nabla_{K\phi}^{j1}} \equiv \big(\partial_{\Kperpij{j1}},
\partial_{\phij{j1}}\big)^\top$ and the $2\times 2$ covariance matrix
$\covthetaij{j1}$ of the fit parameters $\Kperpij{j1}$ and $\phij{j1}$, which is
determined by performing ODR on the data points $\lbrace(\mnuai{a}{1},
\mnuai{a}{j})\rbrace_{a=1}^n$, as explained in the previous paragraph.

Rather than linear error propagation, the \kifit code estimates the covariance
matrix $\covd$ by means of a simple Monte Carlo approach,
in which the input parameters $\ma{A}$, $\ma{A'}$ and $\nuai{a}{i}$, $a=1,\ldots, n$ are sampled from Gaussian distributions fixed by the
experimental central values and uncertainties (see Eq.~\eqref{eq:elemsampling}),
whereas the fit parameters $\lbrace \Kperpij{j1}, \phij{j1} \rbrace_{j=2}^m$ are
sampled from the Gaussian distribution fixed by the best fit results $\lbrace
\langle \Kperpij{j1}\rangle, \langle \phij{j1}\rangle \rbrace_{j=2}^m$ and the
covariance matrix $\boldsymbol{\Sigma_{K \phi }}^{(j1)}$:
\begin{align}
    (\Kperpij{j1}, \phij{j1})  \sim & \mathcal{N}\left( 
    \langle \Kperpij{j1}\rangle, 
    \langle \phij{j1}\rangle, 
    \boldsymbol{\Sigma_{K \phi }}^{(j1)}\right)\,.
    \label{eq:fitparamsampling}
\end{align}

For better numerical stability, the Cholesky decomposition~\cite{chol} 
$\boldsymbol{\Sigma_d}=\mathbf{LL}^\top$ into two uniquely determined, lower triangular matrices $\mathbf{L}$, is performed and 
Eq.~\eqref{eq:llvec} is reformulated as
\begin{align}
    -\log \mathcal{L}  \propto \frac{1}{2}\left[
        2 \sum_{a=1}^n \log (L_{aa})\, \mathbf{x}
        +\mathbf{x}^\top\mathbf{x}\right]\,,
        \label{eq:choll}
\end{align}
where $\mathbf{x}=\mathbf{L}^{-1}\mathbf{d}$ and $L_{aa}$ are the diagonal elements of $\mathbf{L}$.

The numerical stability can be further improved by employing a regularised
version of the covariance matrix,
$\covd^{(\lambda)} = \covd + \lambda \mathbf{I}_n$, 
where $\mathbf{I}_n$ is the identity matrix and $\lambda\ll 1$ is
a regulator with the default value $\lambda=0$. 
We explicitly checked that the spectral decomposition and the Cholesky
decomposition lead to comparable results, that these are stable and that the
associated numerical uncertainties are negligible.

Since the Fisher information $I(\alphaNP)= - \mathbb{E}\left[\frac{\partial ^2
\log \mathcal{L}}{\partial \alphaNP^2}\right]$, i.e. the curvature of the
log-likelihood with respect to $\alphaNP$, is invariant under linear rescaling
of $\mathbf{d}$, so are the confidence intervals for $\alphaNP$. This was
checked explicitly in the \kifit code by varying the normalisation of
$\mathbf{d}$.\\

Having defined the log-likelihood for one set of independent isotope shifts, the generalisation to a combined log-likelihood is straightforward: Assuming zero correlation between the data sets, the total log-likelihood
corresponds to the direct sum of the log-likelihoods $\log\mathcal{L}^{(E)}$,
associated to the chemical elements (or independent data sets) $E=1,2, \ldots$:
    \begin{align}
        -\log\mathcal{L}(\alphaNP) = -\sum_{E} \log\mathcal{L}^{(E)}(\alphaNP)\,.
        \label{eq:llelem}
    \end{align}
For a fixed value of $\alphaNP$ and a set of input and fit parameter samples (see Eqs.~\eqref{eq:elemsampling}, \eqref{eq:fitparamsampling}), the code evaluates this sum and computes the $\Delta \chi^2$ values
    \begin{align}
        \Delta \chi^2 \equiv 2 \left(x - Q\left[x, p\right]\right)\,,
        \label{eq:dchi2}
    \end{align}
where $x=-\log\mathcal{L}(\alphaNP)$ and $Q[x,p]$ denotes the $p^\text{th}$
percentile of the negative log-likelihood values of the full set of $\alphaNP$ samples.
The value of $p$ is a hyperparameter that can be fixed by the user (\texttt{min\_percentile}, in \kifit). For $p=0$,
the minimum of the negative log-likelihood,
$\min\left(-\log\mathcal{L}(\alphaNP)\right)$, is computed. The use of small
values of $p$ can reduce the sensitivity to numerical outliers but too
large $p$-values will lead to an excessive downward shift of the $\Delta \chi^2$
values and thus to an overly conservative confidence interval.

In the next section, we will illustrate how the window of favoured $\alphaNP$ values is determined in the \kifit code, while in Sec.~\ref{sec:exp} we will discuss how the confidence intervals and their uncertainties are estimated.

\begin{figure*}[t]
    \centering
\includegraphics[width=.8\linewidth]{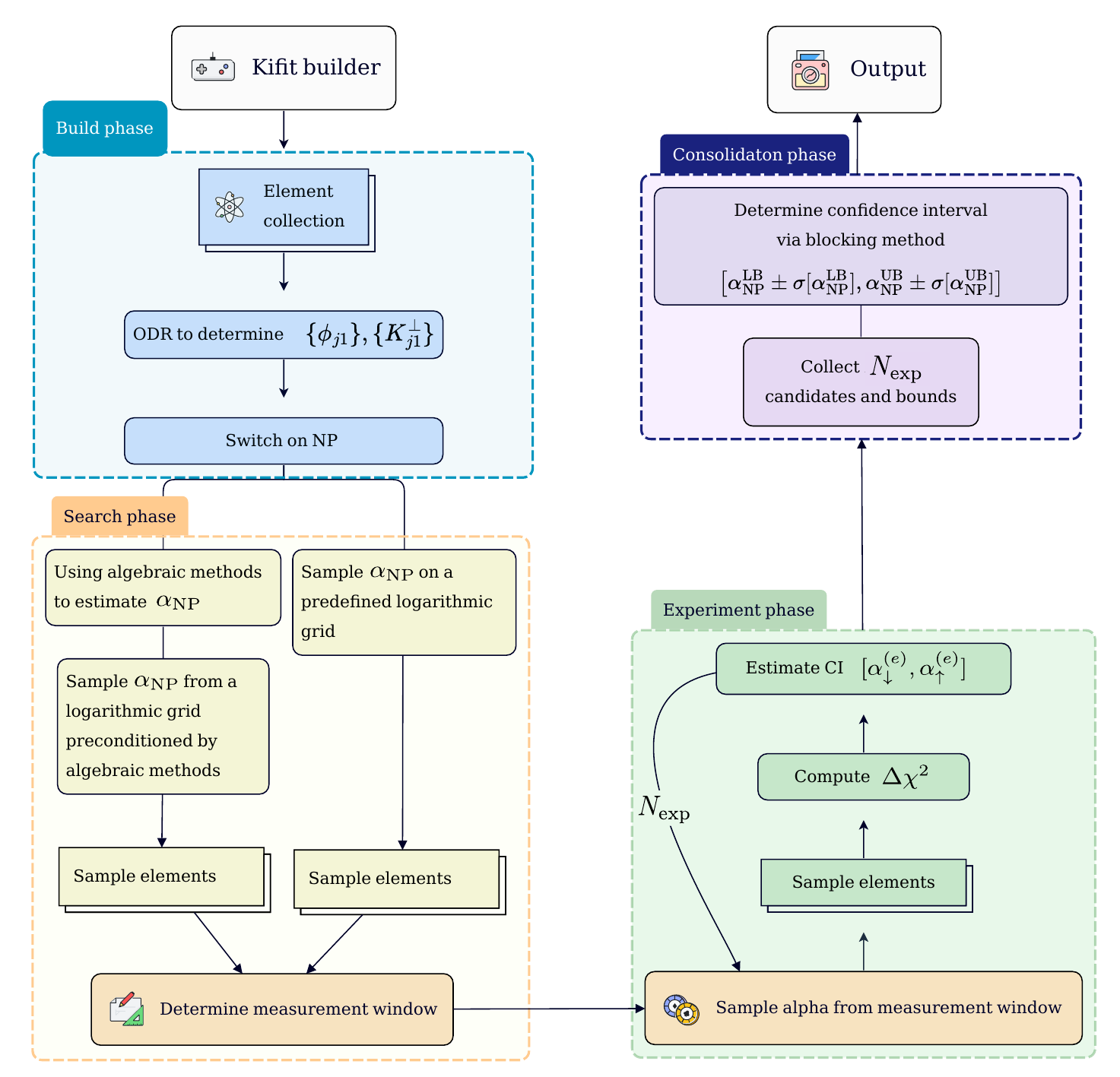}
    \caption{Illustration of the \kifit algorithm used to determine a confidence
    intervals for $\alphaNP$. The algorithm is composed of a \emph{build phase},
    a \emph{search phase}, and \emph{experiment phase} and a \emph{consolidation
    phase}. See Section~\ref{sec:kifit} for a description of the individual
    steps. 
    }
    \label{fig:kifit_algo}
\end{figure*}

The general structure of the \kifit code is illustrated in
Fig.~\ref{fig:kifit_algo}.

\begin{figure}[t]
\includegraphics[width=\linewidth]{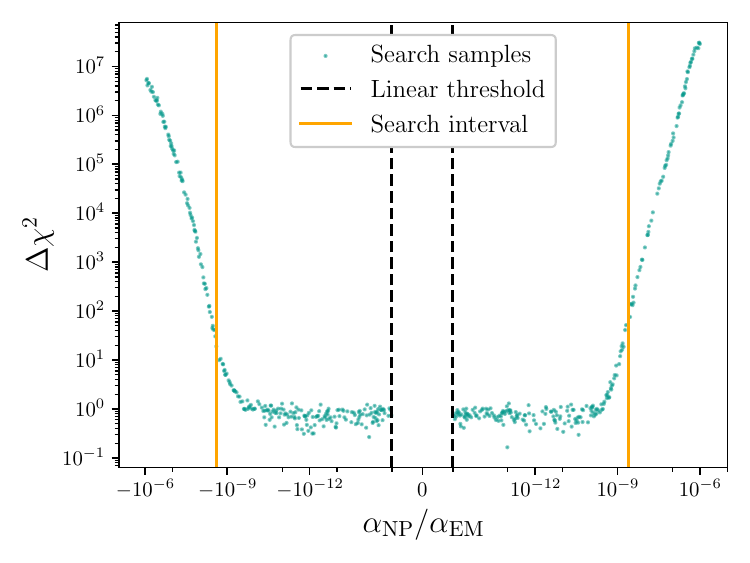}
\caption{
    Output of the \emph{search phase} for a short run of $N_\alpha=500$ $\alphaNP$ samples and only
    $N_\mathrm{exp}=10$ experiments.
    The blue scatter shows the ($\alphaNP$, $\Delta \chi^2(\alphaNP)$) pairs 
    obtained through the procedure described in Section~\ref{sec:search}.
    The upper and lower limits of the search window are indicated by vertical orange lines, the black dashed lines mark the thresholds between which the $\alphaNP$-scaling is linear.}
    \label{fig:search_phase}
\end{figure}

\begin{figure}[t]
\includegraphics[width=\linewidth]{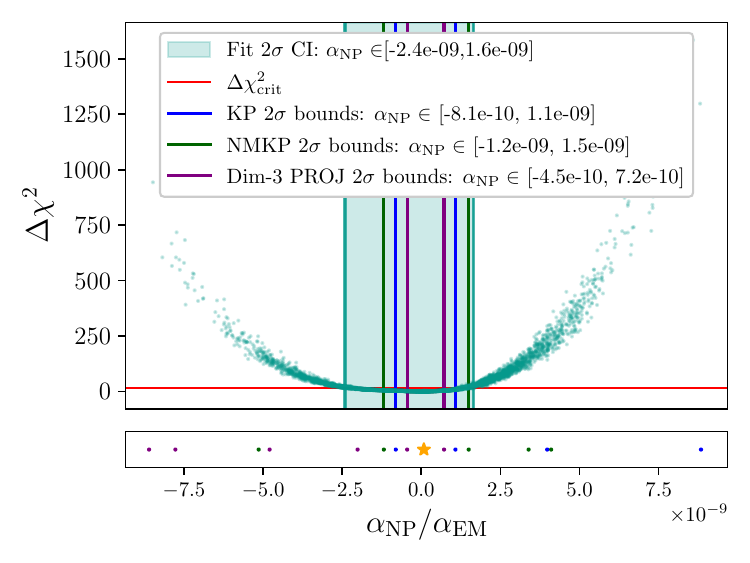}\\
\includegraphics[width=\linewidth]{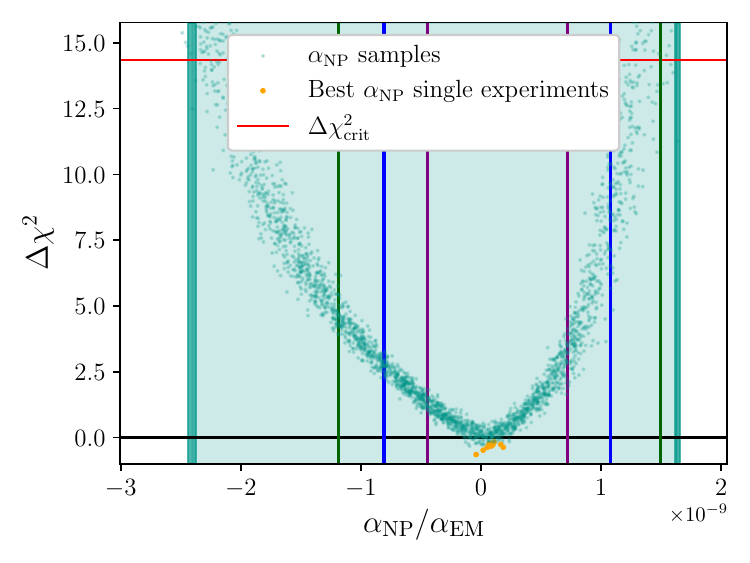}
\caption{
    Output of the \emph{experiment phase}
    for a short run of $N_\alpha=500$ $\alphaNP$ samples and only
    $N_\mathrm{exp}=10$ experiments.
    The teal scatter shows the ($\alphaNP$, $\Delta \chi^2(\alphaNP)$) pairs 
    obtained through the procedure described in Section \ref{sec:exp}.
    Some of the $\Delta\chi^2$ values are negative due to the choice of $p=1$, (see Eq.~\eqref{eq:dchi2}).
    The confidence interval, also shown in teal, is defined as the region in which there are
    $\Delta\chi^2(\alphaNP)$ values below the ``critical'' $\Delta\chi^2$ value $\Delta \chi^2 _\mathrm{crit}$, 
    indicated by the horizontal red line.
    The dark teal vertical bars show the uncertainties on
    the confidence interval. These are estimated by means of the \emph{blocking
    method} (see Section~\ref{sec:consolidation}). The best fit
    $\alphaNP$ value is represented by an orange star, its uncertainty being smaller than its width. 
    The lower plot is an enlargement of the confidence interval in the upper plot. The best $\alphaNP$ values of the $N_\mathrm{exp}$ experiments are shown in orange. 
    } 
\label{fig:experiment_phase}
\end{figure}

\subsection{Determining the Search Window (Search Phase)}
\label{sec:search}
The purpose of the \emph{search phase} is to determine a favoured range of $\alpha_\mathrm{NP}$ values. These will serve as a pool for the Monte Carlo sampling that \kifit uses to estimate the confidence interval. In the code, two different search options are implemented:
    \begin{itemize}
        \item The generally faster \texttt{detlogrid} option constructs a logarithmic grid of
        $\alpha_\mathrm{NP}$ values, with limits fixed by
            \begin{align}
                [\alphaNP\vert^\text{alg}_\text{min} &- \texttt{logrid\_frac}, \nonumber \\ &\alphaNP\vert^\text{alg}_\text{max} + \texttt{logrid\_frac}],
            \end{align}
        where $\alphaNP\vert^\text{alg}_\text{min}$ and $\alphaNP\vert^\text{alg}_\text{max}$ are the minimum and maximum values found by the algebraic methods (see Eqs.~\eqref{eq:ANP3x2}, \eqref{eq:ANP3x3}, \eqref{eq:ANPnxm} and \eqref{eq:ANPnxn}), for all possible combinations of the experimental input data, while \texttt{logrid\_frac} is a user-defined additional number of orders of magnitude used to enlarge the $\alphaNP$ scan region.
        \item The \texttt{globalogrid} option performs an even more  agnostic search on a logarithmic grid between $\alphaNP=10^{-15}$ and a maximal exponent which increases with $m_\phi$. 
    \end{itemize}

As described in the previous section, evaluating the log-likelihood for a given value of $\alphaNP$ involves sampling both the input and the fit parameters.

The number of $\alphaNP$ samples used in the search phase, and the number of input parameters and fit parameters $\lbrace \Kperpij{j1},
\phij{j1}\rbrace_{j=2}^m$ per $\alphaNP$ sample can be set by the user. The search window $[\alphadown,\alphaup]$ is then fixed by the minimum and maximum $\alphaNP$ values,
$\alphadown$ and $\alphaup$, whose $\Delta\chi^2$ values lie
below a $\Delta \chi^2$ value that is sufficiently large to capture all potentially interesting regions in $\alphaNP$ space (see Fig.~\ref{fig:search_phase}).

Although several orders of magnitude have to be scanned and the Monte Carlo
approach only gives reliable results if sufficient statistics are available, we
find that both search strategies work and give comparable results (see
Fig.~\ref{fig:Ca24}).

\subsection{Estimating the Confidence Interval (Experiment Phase)}
\label{sec:exp}

Once the \textit{search window} $[\alphadown,\alphaup]$ is fixed,
$N_{\mathrm{exp}}$ \emph{experiments} are performed with the aim of determining
a reliable confidence interval for $\alphaNP$. 
In each experiment: 
\begin{enumerate}
\item A user-defined number $N_\alpha$ of $\alpha_\mathrm{NP}$ samples are drawn from the normal distribution
    $$\alphaNP \sim
        \mathcal{N}\left(\bestalpha,\max\left(|\bestalpha-\alphadown|, |\alphaup
        - \bestalpha|\right)\right)\,,$$
where $\bestalpha$ is the $\alphaNP$ value within the search window with the lowest negative log-likelihood.
 \item For each of the $\alphaNP$ samples, the input and fit parameters are sampled as described
     in Eq.~\eqref{eq:elemsampling} and \eqref{eq:fitparamsampling}.
 \item Values of $\Delta\chi^2$ (Eq.~\eqref{eq:dchi2}) are computed using the samples determined in the previous steps. We obtain $N_\alpha N_\mathrm{exp}$ points in the ($\alphaNP$, $\Delta\chi^2$)-plane (see Fig.~\ref{fig:experiment_phase}).
 \item For each experiment $e$, a confidence interval 
     $\left[\alphadownexp{e},\alphaupexp{e} \right]$ is determined by finding the minimum and maximum $\alphaNP$ samples whose
        $\Delta \chi^2$ values are below the critical $\Delta \chi^2$ value
        associated to $2m+1$ degrees of freedom and the number $N$ of $\sigma$ ($N=2$ being the
        default value, see red horizontal lines in Fig.~\ref{fig:experiment_phase}).
 \item For each experiment $e$, a \textit{best fit value} $\alpha^e_*$ is determined, defined    as the $\alphaNP$ value with the smallest associated negative log-likelihood.
\end{enumerate}
The number of experiments, $N_\mathrm{exp}$, the number of input and fit parameter samples, the number $N_\alpha$ of $\alphaNP$ samples and the number $N$ of $\sigma$ are hyper-parameters that can be fixed by the user. For a stability analysis, see Appendix~\ref{sec:kifit_general}.\\

Finally, the results of the $N_{\mathrm{exp}}$ \emph{experiments} are collected and the final estimation of the confidence interval for $\alphaNP$ is determined in the \emph{consolidation phase}.

\subsection{Producing Results (Consolidation Phase)}
\label{sec:consolidation}

\begin{figure}
\includegraphics[width=\linewidth]{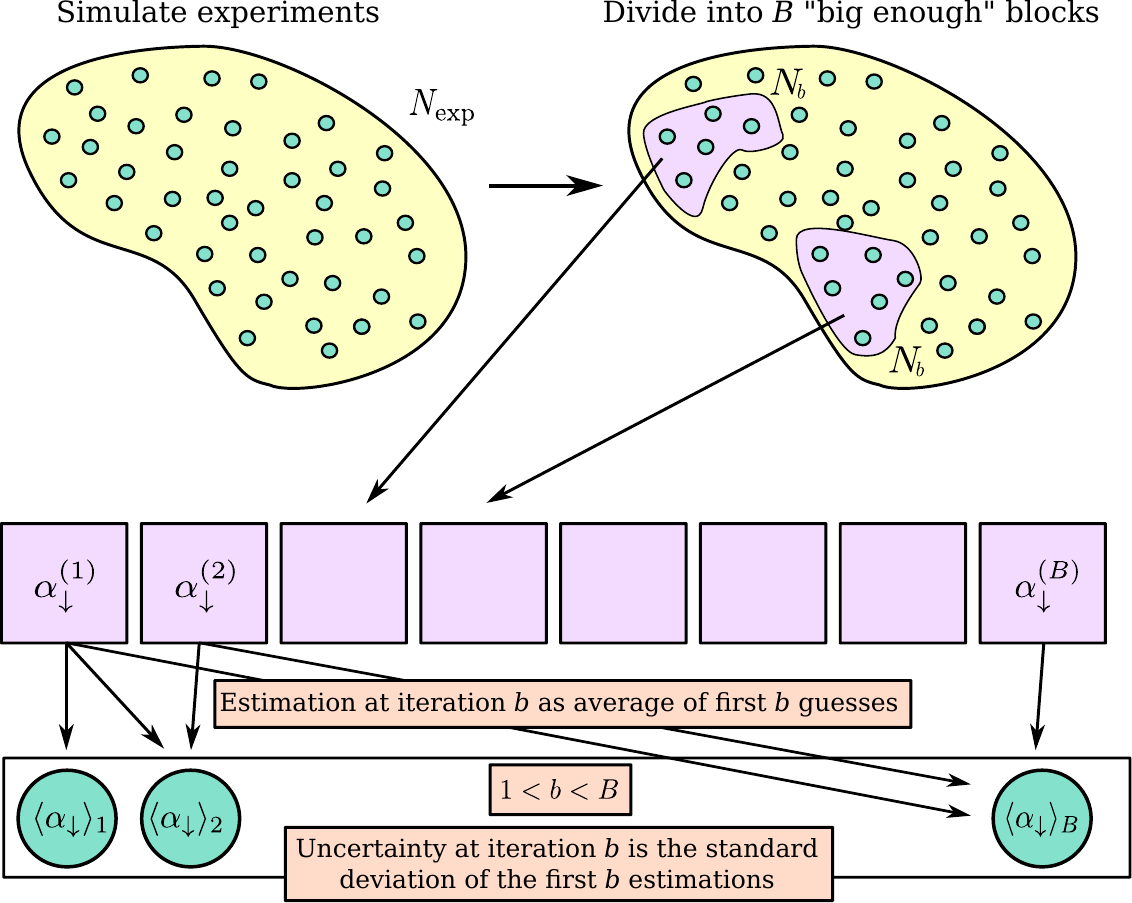}
\caption{\label{fig:blocking_illustration} Illustration of the blocking method applied to the estimation of the lower bound (for upper bound: $\alphadown\to \alphaup$). The $N_\mathrm{exp}$ simulated experiments are divided into $b=1,\ldots,B$ blocks, for each of which the minimum lower bound on $\alphaNP$, $\alphadown^{(b)}$, is computed. The estimations $\lbrace\alphadown^{(b)}\rbrace_{b=1}^B$ are then used to compute an iterative average, the final estimation for the lower bound being $\left\langle \alphadown \right\rangle_{B}$.}
\end{figure}

Since numerical simulations are used in \kifit to estimate the new physics bounds, we need a robust strategy to compute both the estimation and its uncertainty. To do so, we apply a technique inspired by the \emph{blocking average method} (BAM), which is commonly used to compute statistically independent estimations starting from a simulated set of possibly correlated data~\cite{blocking}. 
To facilitate the following discussion, we will refer to the illustration presented in Fig.~\ref{fig:blocking_illustration}.

The BAM involves executing the simulation experiment $N_{\rm exp}$ times, thus collecting $N_{\rm exp}$ raw estimations. These are then divided into $B$ blocks of size $N_b$ (purple subsets in the top right of Fig.~\ref{fig:blocking_illustration}) and each block is used to produce a single estimation of the target variable.
To mitigate eventual correlations among simulated data, it is important to choose the size of the blocks such that the estimation computed from each block can be assumed to be independent from the ones of the other blocks. This is relevant because, if those estimators are independent, their average value and uncertainty stabilises for increasing number of blocks, according to the central limit theorem.

In our case, the raw estimations correspond to confidence intervals of the form  $\left[\alphadownexp{e}, \alphaupexp{e}\right]$, where the superscript refers to the result of the experiment $e$, $1\leq e \leq N_{\rm exp}$.
In Ref.~\cite{blocking}, and more in general in the context of numerical simulations, the estimation for each block is defined as the average value of the data in the block. We use instead a more conservative approach, with the intention of estimating bounds in the least aggressive way possible.
Taking the estimation of the lower bound as an example, we define an estimator as the minimum  of values collected in the block $b$: 
    \begin{align}
        \alphadownblock{b} \equiv \min_{e\in b} \alphadownexp{e}\,.
        \label{eq:block_estimator}
    \end{align}
A similar procedure can be used to compute the upper bound, considering instead the maximum value in the block. Since the $N_\mathrm{exp}$ experiments are considered to be independent (contrary to the time-series simulation in Ref.~\cite{blocking}), we can also assume the estimators in Eq.~\eqref{eq:block_estimator} to be statistically independent.

\begin{figure}[ht]
\includegraphics[width=.9\linewidth]{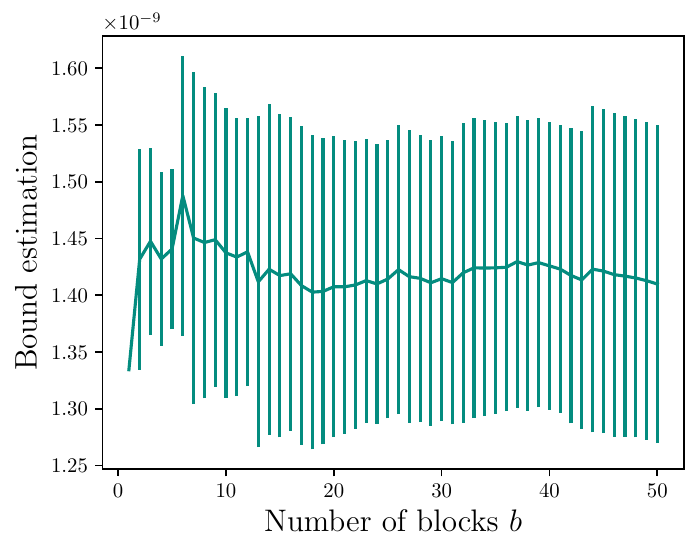}
\caption{
Blocking method applied to the estimation of the upper bound on $\alphaNP/\alphaEM$ after executing the procedure described in Sec.~\ref{sec:consolidation}. These results have been collected
    from $N_{\mathrm{exp}}=500$ experiments divided into $B=50$ blocks.
    }
\label{fig:blocking_plot}
\end{figure}

The estimation of the lower bound and its uncertainty based on $B$ blocks are computed as follows: 
    \begin{align}
            \left\langle \alphadown \right\rangle_{B} & \equiv \frac{1}{B} \sum_{b=1}^B \alphadownblock{b}\,,\\
            \sigma\left[\alphadown\right]_B&\equiv\sqrt{\frac{1}{B-1} 
            \sum_{b=1}^B \left(\alphadownblock{b}- \langle \alphadown \rangle_{b} \right)^2}\,,
    \label{eq:blocking_results_err}
    \end{align}
where $\langle \alphadown \rangle_{b}$ in the second equation corresponds to the estimator defined in the first equation, computed iteratively for any $b \leq B$. A similar approach is used to compute the final estimation and uncertainty for the upper bound. The evolution of $\left\langle \alphadown \right\rangle_{B}$ and $\sigma\left[\alphadown\right]_B$ for increasing number of blocks is shown in Fig.~\ref{fig:blocking_plot}. Ideally, the central value stabilises and the uncertainty decreases for increasing $B$.
The number of blocks $B$ is computed from the number of experiments and the block size, both of which can be set by the user. The ideal values for $N_\mathrm{exp}$ and the block size are ones which ensure the stability of the code. See Appendix~\ref{sec:kifit_general} for a more detailed discussion.

The final confidence intervals for $\alphaNP$ provided by \kifit take all $B$
blocks into consideration. We apply the following conservative definition:
    \begin{align}
        &\left[ \alphaNP^\mathrm{LB}, \alphaNP ^\mathrm{UB}\right]
        =
    \left[  \left\langle \alphadown \right\rangle_{B} -  N\, \sigma\left[\alphadown\right]_B, 
        \left\langle \alphaup \right\rangle_{B} +  N\, \sigma\left[\alphaup\right]_B  \right]\,,
        \label{eq:blocking_LB_UB}
    \end{align}
where $N$ corresponds the number of $\sigma$ of the confidence interval. In Fig.~\ref{fig:experiment_phase}, the uncertainties on the $2\sigma$ confidence
interval are shown as darker teal bands.

Besides the confidence interval, the \kifit code provides a \emph{best}
$\alphaNP$ value (orange star in the upper plot of
Fig.~\ref{fig:experiment_phase}). This corresponds to the median of the best $\alphaNP$ values
$\left\{\bestalphaexp{e}\right\}_{e=1}^{N_\mathrm{exp}}$ (marked by orange dots in the lower plot in Fig.~\ref{fig:experiment_phase}) found in the \textit{experiment phase} (Sec.~\ref{sec:exp}). The uncertainty on the \emph{best} $\alphaNP$ value is estimated by the standard
deviation of the same set of values
$\left\{\bestalphaexp{e}\right\}_{e=1}^{N_\mathrm{exp}}$. 
\\

\begin{figure*}
\includegraphics[width=0.48\linewidth]{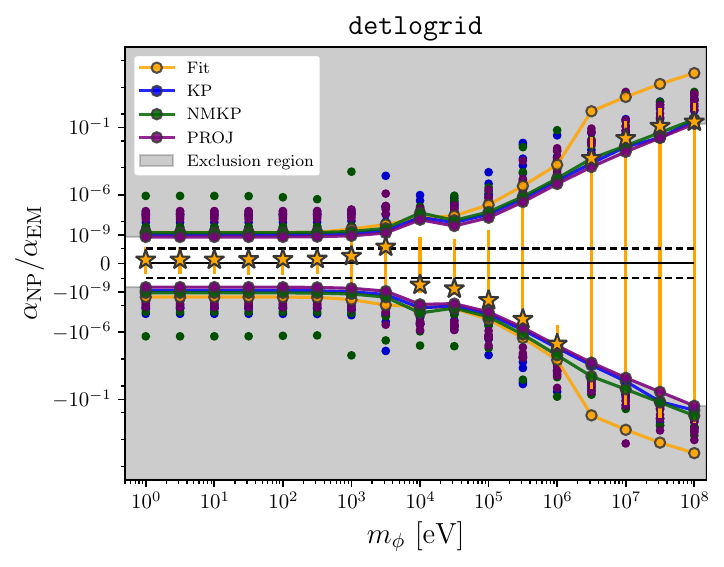} 
\includegraphics[width=0.48\linewidth]{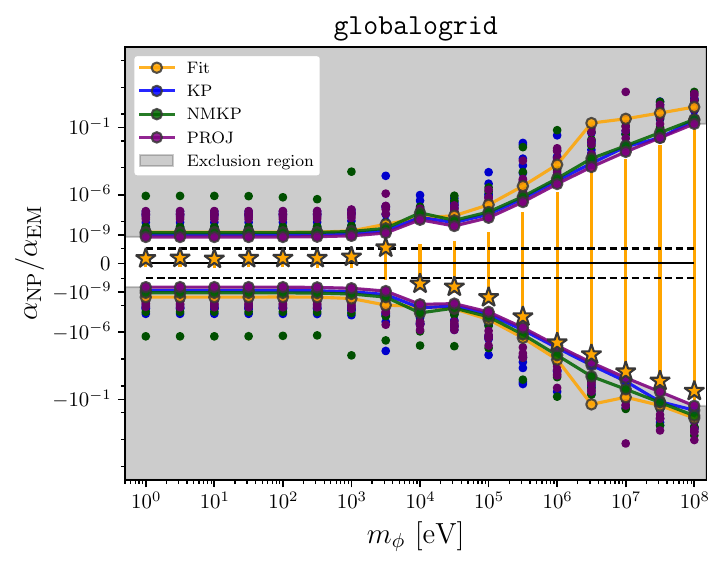}
    \caption{Output of the \kifit code for the \texttt{Ca\_WT\_Aarhus\_PTB\_2024} dataset, showing the $2\sigma$ upper and lower bounds on the new
    physics coupling $\alphaNP/\alphaEM$ as a function of the mass $m_\phi$ of the new
    boson with orange connected round markers. The estimation is repeated for the two different \textit{search phase} approaches (see Sec.~\ref{sec:search}): Left \texttt{detlogrid}, adopting the algebraic-induced logarithmic grid and right \texttt{globalogrid}, using the global logarithmic grid.
    The orange stars and bars indicate the best fit points and the associated uncertainties. 
    The bound obtained via the \kifit algorithm (orange) is compared with
    the algebraic results: minimal KP (Eq.~\eqref{eq:ANP3x2}) in blue, NMKP (Eq.~\eqref{eq:ANP3x3}) in green and projection method (Eq.~\eqref{eq:ANP_proj}) with $n=3$ in purple.
    The solutions for the different
    permutations are illustrated by single markers, whereas the envelopes of the
    most stringent limits are indicated with lines of the same colour.
    Since the three
    algebraic methods lead to very similar results, only the purple markers are
    visible in most cases. 
    For better readability, the regions excluded by any of the plotted bounds  are
    shaded grey.
    The horizontal
    black dashed lines indicate the thresholds between which the plot scale is
    linear.
    }
\label{fig:Ca24}
\end{figure*}

Exclusion plots such as the one shown in Fig.~\ref{fig:Ca24} are obtained by repeating the procedure described in Sections~\ref{sec:build}-\ref{sec:consolidation} for multiple values of the mediator mass $m_\phi$, i.e. for different
values of the $X$ coefficients (see e.g. Eq.~\eqref{eq:Xdef}). 

Fig.~\ref{fig:Ca24} compares the $2\sigma$ bounds 
obtained by the \kifit algorithm (orange) to those obtained using the algebraic methods for subsets of the isotope shift data. The results using the Minimal King Plot formula (Eq.~\eqref{eq:ANP3x2}) are shown in blue, the No-Mass King Plot results (Eq.~\eqref{eq:ANP3x3}) in green and the results using the projection method (Eq.~\eqref{eq:ANP_proj}) with $n=3$ in purple. 
For each of these methods, the most stringent limits on $\alphaNP$ are connected by lines of matching colour. The horizontal black dashed lines indicate the thresholds between which the plot
scale is linear.

As in Fig.~\ref{fig:experiment_phase}, the orange stars and bars represent
the best fit points and their $1\sigma$ uncertainties, as obtained by \kifit. Note that at high-mass values the \kifit sampling becomes more challenging: the points drawn to evaluate the likelihood are more widely spread around the curve that they trace, and consequently the confidence intervals produced by the experiments exhibit greater variance than at low masses. To mitigate this effect, it is advisable to improve the Monte Carlo sampling, for example by increasing the number of sampled points, using variance reduction techniques, or adopting adaptive strategies in which computational effort grows in proportion to the mass under analysis. In this example we have kept the simulation parameters fixed, so what we observe at high masses is, unsurprisingly, larger error bars. Finally, since we employed a very conservative technique to define the bounds estimates, it is also reasonable to see the bounds widen at high masses.

\subsection{Comparison to a Previous King Plot Fit}
As already mentioned in the introduction, the geometric construction used by
\kifit (see Section~\ref{sec:build}) is inspired by the King plot fit proposed
in Ref.~\cite{Frugiuele:2016rii}, but there are some notable differences in the
definition of the log-likelihood and the fit parameters. These differences will
be discussed in this section.

The fit in Ref.~\cite{Frugiuele:2016rii} was developed for the case of similar uncertainties across all transitions and isotope pairs (which was the case for the isotope shift data available at that time) and under the (strong) assumptions of uncorrelated uncertainties. In our notation, the $\chi^2$ function proposed in
Ref.~\cite{Frugiuele:2016rii}, takes the form
\begin{align}
    \chi_a^2 = \sum_{i=1}^m \left(\frac{\dorigai{a}{i}}{\smnuai{a}{i}}\right)^2\,,\quad
    \dorigai{a}{i} = \mnuai{a}{i} - \left(K_{i1} + \alphaNPdEM \ha{a} X_{i1}\right)\,.
    \label{eq:oldchisq}
\end{align}
In order to illustrate the main difference between this definition of $\chi^2$
and Eq.~\eqref{eq:dchi2}, let us consider the 2-dimensional setup illustrated in
Fig.~\ref{fig:kifit_sketch}, which is characterised by one inclination angle
$\phij{21}$. In this case the elements $\dorigai{a}{i}$ take the particularly
simple form
\begin{align}
    \dorigai{a}{1} = \Vert \dveca{a}\Vert \sin \phij{21}\,,\quad\dorigai{a}{2} 
    = \Vert \dveca{a}\Vert \cos \phij{21}\,,
\end{align}
such that Eq.~\eqref{eq:oldchisq} reads
\begin{align}
     \chi_a^2 = \Vert \dveca{a}\Vert^2 
     \left \lbrace 
     \left(\frac{ \sin \phij{21}}{\smnuai{a}{1}}\right)^2 
     + \left(\frac{ \cos \phij{21}}{\smnuai{a}{2}}\right)^2
     \right \rbrace \,,
     \label{eq:old2dchi2}
\end{align}
where $\Vert \dveca{a}\Vert^2 = \Vert \Deltaveca{a}\Vert \cos \phij{21}$.

Eq.~\eqref{eq:old2dchi2} explicitly shows how separately normalising
the components of $\dveca{a}$ by the uncertainties $\smnuai{a}{i}$ distorts the fit:
If the data set under consideration has a direction with significantly smaller
uncertainties, e.g. $\smnuai{a}{1} \gg \smnuai{a}{2}$, the minimisation of
Eq.~\eqref{eq:old2dchi2} will lead to a rotation of the King line, even if this
does not reduce the overall distance of the points to the fitted line. Instead, \kifit minimizes 
$\Vert \dveca{a}\Vert \left(\covdab{ab}\right)^{-1} \Vert \dveca{b} \Vert$, 
where $\covdab{ab}$ is an estimate of the covariance of $\Vert\dveca{a}\Vert$
and $\Vert\dveca{b}\Vert$.
In this way all directions are treated equally, while correlations between the
uncertainties are taken into account.

Note also that Eq.~\eqref{eq:old2dchi2} does not take into account the fact that
a King plot fit cannot capture new physics-induced constant shifts of all data
points (see discussion around Eqs.~\eqref{eq:sdeltaidef}-\eqref{eq:Kprime}) and
will therefore tend to lead to overly aggressive bounds on $\alphaNP$.

Further improvements with respect to the fit proposed in
Ref.~\cite{Frugiuele:2016rii} are: 
the estimation of the uncertainties on the best fit point and the confidence
interval by means of the \emph{blocking average method} (see Sections~\ref{sec:exp} and
\ref{sec:consolidation});
the development of an extensible code framework that allows to combine the
isotope shift data of several elements in a single fit;
an extensive set of numerical tests that ensure the internal consistency of the
code and compare \kifit results in Python with results obtained with Mathematica.\\

In the following section we provide a comparison of the \kifit results with
results obtained by means of the algebraic methods discussed in
Section~\ref{sec:alg}.

\section{Comparison of Fit and Algebraic Methods}
\label{sec:fitvsalg}

In Section~\ref{sec:alg} we reviewed the algebraic methods to extract
bounds on $\alphaNP$ from King plots, whereas in Section~\ref{sec:kifit} we presented
the King plot fit \kifit. Although both methods are based on the King plot formalism, they adopt different strategies when it comes to setting
bounds on $\alphaNP$. This becomes most evident by stating the questions that either method addresses:
\\
\paragraph*{Algebraic methods} 
        \begin{enumerate}
            \item  $\langle\alphaNP\rangle$: Given a data set of dimension as specified in Tab.~\ref{tab:fitvsalg_dims},
                which value of $\alphaNP$ is
                required to perfectly describe the experimental central values
                of the data points in the King plot?
            \item $\sigalphaNP$: How large is the impact of the experimental
                uncertainties on the value of $\alphaNP$?
        \end{enumerate}

\paragraph*{King plot fit} 
        \begin{enumerate}
            \item $\lbrace \Kperpij{j1}, \phij{j1}\rbrace_{j=2}^m$: What is the
                best linear fit, given the central values and uncertainties on
                the data points in the King plot (mass-normalised isotope
                shifts)?
            \item Given the central values and uncertainties on the fit
                parameters and on the data points, what is the likelihood
                (Eq.~\eqref{eq:ll}) associated to the set of sample
                $\alphaNP$ values?
            \item $\left[ \alphaNP ^\mathrm{LB}, \alphaNP
                ^\mathrm{UB}\right]$:
                What are the minimum and maximum $\alphaNP$ values whose $\Delta
                \chi^2$ values are below the critical $\Delta \chi^2$ value
                associated to $2m+1$ degrees of freedom and $N\sigma$?
        \end{enumerate}
In the rest of the section, we will contrast these approaches in more detail,
discuss their respective advantages and disadvantages, present our results and suggest when
to apply which method.

\subsection{Blind Directions}
Since both approaches are based on the King
plot formalism, they inherit the blind directions of the King plot:
As discussed in Section~\ref{sec:alglin}, in particular in
Eqs.~\eqref{eq:ElNuclMatKP} and \eqref{eq:VpredefKP}, the King
plot method is only sensitive to effects that have a component orthogonal to the mass shift
or the field shift, i.e. that lead to a ``misalignment'' both in transition space and in isotope pair space. 
In a standard King plot in transition space such a misalignment manifests itself in a deviation from the King line (Eq.~\eqref{eq:linvecKP}). This is the effect that the
King plot fit looks for. Equivalently, it can be viewed as a component orthogonal to
the plane of King linearity fixed by the basis vectors $\muvec$ and $\drsqvec$
(or equivalently $\mmuvec$ and $\mnuveci{1}$, as in Fig.~\ref{fig:out-of-plane}) in isotope pair space. This is the
effect exploited by the algebraic methods and the nonlinearity decomposition (Eq.~\eqref{eq:lambda_dec}).

The blind directions of the King plot method and their manifestations in the algebraic methods and on the King plot fit \kifit are summarised in Table~\ref{tab:blindirs}: Using the notation of
Eq.~\eqref{eq:VpredefKP}, they can be described by vanishing determinants of the matrix of electronic or nuclear factors or, in \kifit, either by an alignment of the electronic coefficients $X_j/X_1$ and $F_{j1}=F_j/F_1$ leading to a vanishing coefficient $X_{j1}\to 0$, or by a global shift
of the King line, which cannot be distinguished from a redefinition of the
electronic mass shift coefficient $\Kijvec$ (see Fig.~\ref{fig:kifit_sketch}).\\

Another common feature of the fit and the algebraic methods which is inherited from the King plot, is the insensitivity to the experimental uncertainties parallel to the King line:
Whereas this is true by construction for the King plot fit (see Fig.~\ref{fig:kifit_sketch}), it is less obvious in the case of the
algebraic methods. In Appendix~\ref{sec:uncertainty_projections} we check explicitly that the effect of the isotope shift uncertainties parallel to the King line has a negligible impact on $\sigalphaNP$.

 \begin{table}
 \begin{tabular}{c c}
 \hline \hline
Algebraic Methods  & \kifit\\
\hline
     $\det(\mathcal{N})\to 0$ &
     $\ha{a}\to\langle \ha{}\rangle \quad \forall a$    \\
$\det(\mathcal{M})\to 0$ &
$X_{j1}\to 0 \quad \forall j$  \\
 \hline \hline
\end{tabular}
\caption{Blind directions of the King plot method and how they manifest
     themselves in the algebraic methods and in the fit.
     The matrices $\mathcal{M}$ and $\mathcal{N}$ are defined in Eq.~\eqref{eq:VpredefKP} and \eqref{eq:VpredefNMKP} for the cases of the simple King plot and the No-Mass King Plot, respectively, whereas $X_{j1}$ and $\langle\ha{}\rangle$ are defined in Eqs.~\eqref{eq:NPvec} and \eqref{eq:sdeltaidef}, respectively.
     }\label{tab:blindirs}
\end{table}

\begin{table*}
\setlength{\tabcolsep}{10.5pt} 
\centering
\begin{tabular}{ l | c  c  c  }
    \hline \hline
    Data Set & 
    $\left(\alphaNP\pm \sigalphaNP \right)\vert_\mathrm{KP}^\mathrm{(1)}$ &$\left(\alphaNP\pm \sigalphaNP \right)\vert_\mathrm{KP}^\mathrm{MC}$ &     $[\alphaNP^\mathrm{LB}\pm\sigma[\alphadown]_B, \alphaNP^\mathrm{UB}\pm\sigma[\alphaup]_B]$ \\
    \hline
    \texttt{Ca3pointTEST} & 
    $(1.36\pm4.84)\times10^{-10}$ & 
    $(1.36\pm 4.81) \times 10^{-10}$ 
    & $\bigl[-3.51\pm 0.16, 3.59 \pm 0.13\bigr] \times 10^{-11}$
    \\
    \texttt{Ca4pointTEST} &
    $(-0.14\pm 4.50 )\times10^{-11} $ &
    $(-0.14\pm 4.50) \times 10^{-11}$ & 
    $\bigl[-3.44\pm 0.15, 3.29 \pm 0.09\bigr] \times 10^{-11}$ 
    \\
    \texttt{Ca10pointTEST} & 
    $(0.19\pm1.35)\times10^{-10} $ &
    $(0.19\pm 1.31)\times 10^{-10}$ &
    $\bigl[0.32\pm 0.07, 1.65 \pm 0.08\bigr] \times 10^{-10}$  
    \\
    \hline
    \texttt{Ca\_PTB\_2015}  &
    $(0.65\pm1.10)\times10^{-8}$ &
    $(0.65\pm1.10)\times10^{-8}$ & 
    $[-7.04 \pm 0.20, 8.04 \pm 0.50] \times 10^{-10}$ 
    \\
    \texttt{Camin} & 
    $(1.36\pm1.12)\times10^{-9}$ &
    $(1.36\pm 1.1)\times 10^{-9}$ &
    $\bigl[-0.91\pm 0.03, 1.0 \pm 0.05\bigr] \times 10^{-10}$
    \\
    \texttt{Ca24min} &
    $(1.36\pm4.70)\times10^{-10}$ & 
    $(1.36\pm 4.74)\times10^{-10}$ &
    $\bigl[-2.66\pm 0.13, 2.55 \pm 0.07\bigr] \times 10^{-11}$ 
    \\
    \texttt{Ca\_WT\_Aarhus\_PTB\_2024}  &
    $(1.36\pm 4.70)\times10^{-10}$ &
    $(1.36\pm 4.72)\times10^{-10}$ &
    $\bigl[-1.99\pm 0.04, 1.45 \pm 0.05\bigr] \times 10^{-9}$ \\
    \hline \hline
\end{tabular}
\caption{
Estimates of the $1\sigma$ confidence interval for $\alphaNP$ for $m_\phi=1~$eV. 
The first column shows the central value of $\alphaNP$ (Eq.~\eqref{eq:ANP3x2}) and its uncertainty obtained using linear error propagation (Eq.~\eqref{eq:sigalphaNP_alg_estimate}). 
The uncertainties in the second column were estimated using a Monte Carlo approach (see Eq.~\eqref{eq:elemsampling} to Eq.~\eqref{eq:ANP3x2}) with $2000$ samples. 
Both for column 1 and column 2, those subsets of the data were employed that lead to the most stringent bounds on $\alphaNP$. 
The \kifit results are listed in the last column.
These were obtained using a search phase on a \texttt{detlogrid} with \texttt{logrid\_frac}=2, $500$ $\alpha_{\rm NP}$ samples and $200$ input and fit parameter samples per $\alpha_{\rm NP}$ sample. 
The experiment phase consisted of $150$ experiments, with $1000$ $\alpha_{\rm NP}$ samples each and $500$ input and fit parameter samples per $\alphaNP$ sample. The experiments were divided into blocks of size $25$.
} 
\label{tab:fitvsalg}
\end{table*}

\subsection{Geometric Construction \& Form of Data Sets}

The most obvious difference lies in the geometric constructions.
The algebraic
methods compare volumes constructed from a combination of data points and 
predictions (see Fig.~\ref{fig:alg_sketch}), providing one constraint on $\alphaNP$ per data set.
It is for this reason that the data sets used by the algebraic methods must have a fixed size (see Table~\ref{tab:fitvsalg_dims}).

The fit, on the other hand, minimizes the distances $\Vert \dveca{a} \Vert$ of
the data points $\mnuveca{a}$ from the set of King lines
$\lbrace\ell^{(a)}\rbrace_{a=1}^n$, which are determined by a combination of the
best linear fit to data, $\ell^{(0)}$, and the predicted nonlinearity induced by new physics (see
Fig.~\ref{fig:kifit_sketch}). 
Since the fit is based on a log-likelihood, which sums over the data points (Eq.~\eqref{eq:ll}), it can handle data sets of dimension $(n,m)$, with $3 \leq n$ and $2 \leq m$,
but where $n$ and $m$ are independent (see Table~\ref{tab:fitvsalg_dims}), and also combine multiple data sets, enabling global fits to data.
This makes the fit particularly valuable for the search for
new physics, which is expected to couple in the same way to electrons and
neutrons, irrespective of the element.

\subsubsection*{Size of Data Sets}

Whereas the algebraic methods work best for small data sets, the fit relies on the presence of sufficiently large and diverse data sets. When this is not the case, there is a risk of overfitting. 
Overfitting happens when a statistical model is too expressive for a given dataset. The optimal fit parameters are obtained by learning the statistical fluctuations or any other source of error. For example, using a data set with two points only, we would obtain a perfect interpolation and could define a margin for new physics in the form of a finite value for $\sigalphaNP$, but at the same time, we would be performing maximum over-fitting: All
data is used to fix the fit parameters $\lbrace
\Kperpij{j1},\phij{j1}\rbrace_{j=2}^m$ and the bound on the remaining fit
parameter $\alphaNP$ is meaningless. 
Indeed, in order to fit $2(m-1)+1=2m-1$ degrees of freedom with $n$ data points,
each of which contributes an additional scalar constraint (distance
$\Vert\dveca{a}\Vert$), $n\geq 2m-1$ data points are required (see also
Table~\ref{tab:fitvsalg_dims}).

Data sets with just 2 transitions and 2 isotope pairs are not
considered in King plots for precisely this reason.
However, since the statistical uncertainty associated to the spread of the
points is expected to scale as $1/\sqrt{n}$, the fit, which minimises the
distances from the data points to the King line, will significantly
underestimate the bounds on $|\alphaNP|$ in the limit of sparse data.

In Appendix~\ref{sec:sparsity}, we generate mock data starting from a set of linear relations. 
This allows us to study the impact of data sparsity on the fit results in more detail.

\subsubsection*{Numerical Comparisons}

\begin{figure}
\includegraphics[width=\linewidth]{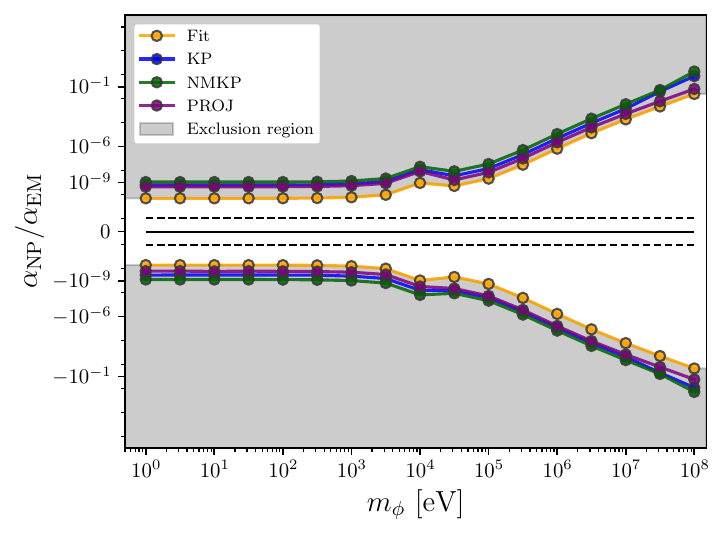}
\caption{Comparison of the fit results with the algebraic results (KP:
    Eq.~\eqref{eq:ANP3x2}, NMKP: Eq.~\eqref{eq:ANP3x3}, PROJ: Eq.~\eqref{eq:ANP_proj}) for the minimal
    dataset (3 isotope pairs, 2 transitions) \texttt{Ca24min}.
For better readability, the regions excluded by any of the plotted bounds  are
    shaded grey, even if these are considered too aggressive in the case of
    \texttt{Ca24min} (see text for discussion).
The black dashed lines indicate the thresholds between which the scaling in the
    plot is linear.}
\label{fig:Ca24min}
\end{figure}

\begin{figure}
    \centering
    \includegraphics[width=\linewidth]{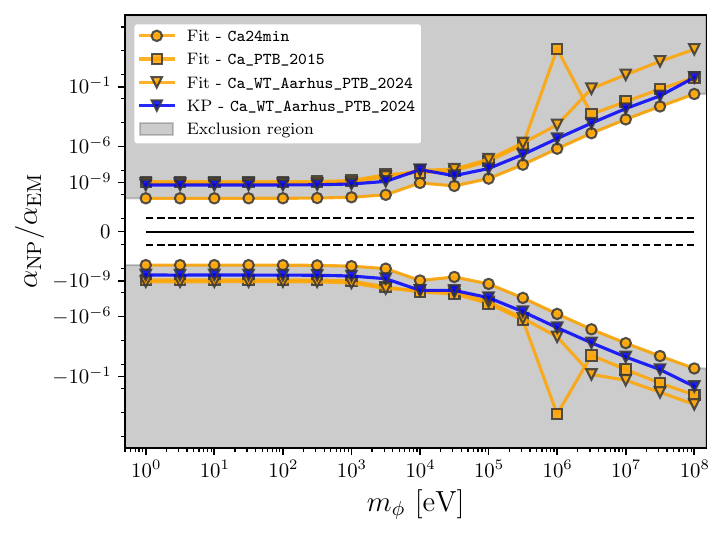}
    \caption{
    Fit results for two data sets of dimensions $(n, m)=(3,2)$ (\texttt{Ca24min},
    \texttt{Ca\_PTB\_2015}) and their combination of dimension $(3,4)$ (\texttt{Ca\_WT\_Aarhus\_PTB\_2024}). The most stringent algebraic bounds (KP: Eq.~\eqref{eq:ANP3x2}) for \texttt{Ca\_WT\_Aarhus\_PTB\_2024} correspond to the algebraic bounds for \texttt{Ca24min} and are shown in blue.
    Whereas the difference between the fit results and the algebraic results are sizeable for the $(3,2)$-dimensional data sets, the agreement is better for \texttt{Ca\_WT\_Aarhus\_PTB\_2024}. 
    }
    \label{fig:Ca}
\end{figure}

To explicitly compare results obtained using the algebraic methods and the fit, we list in Table~\ref{tab:fitvsalg} the values of $\sigalphaNP$ estimated by means of linear error propagation
through Eq.~\eqref{eq:ANP3x2} ($\sigalphaNP\vert _\mathrm{KP}^{(1)}$, first
column) and the Monte Carlo approach outlined in Section~\ref{sec:alglin}, in
particular in Eqs.~\eqref{eq:nsigma_bound} and \eqref{eq:elemsampling}
($\sigalphaNP \vert_\mathrm{KP}^\mathrm{MC}$, second column), as well as the $1\sigma$ confidence
interval determined in \kifit by the BAM, Eq.~\eqref{eq:blocking_LB_UB} (last column).
We observe good agreement between the linear uncertainty propagation and the Monte Carlo approach. In contrast, the \kifit results show some deviations. Notably, for all minimal data sets (2 transitions and 3 isotope pairs), the fit results are roughly an order of magnitude stronger. 
However, we notice that the discrepancy is lifted for larger datasets, such as \texttt{Ca4pointTEST} and \texttt{Ca10pointTEST} (each with 2 transitions and 4 or 10 isotope pairs) and \texttt{Ca\_WT\_Aarhus\_PTB\_2024} (4 transitions and 3 isotope pairs). For a description of these data sets and the relevant references, see Appendix~\ref{sec:data}.

This behaviour can also be observed in the exclusion plot in Fig.~\ref{fig:Ca24min}, which shows the
bounds on $\alphaNP$ for the minimal data set \texttt{Ca24min}. Here the fit bounds (orange) appear to be nearly two orders of magnitude more stringent than
the bounds obtained using the minimal algebraic method (Eq.~\eqref{eq:ANP3x2},
blue) and the projection method (Eq.~\eqref{eq:ANP_proj}, purple). 
Fig.~\ref{fig:Ca} shows the same fit result as Fig.~\ref{fig:Ca24min}, but
complemented with those using the minimal data set \texttt{Ca\_PTB\_2015}, and
\texttt{Ca\_WT\_Aarhus\_PTB\_2024}, which combines \texttt{Ca\_PTB\_2015} and
\texttt{Ca24min} to a data set for 3 isotope pairs and 4
transitions. 
The fit and the algebraic results using \texttt{Ca\_WT\_Aarhus\_PTB\_2024} show significantly better agreement.

\begin{figure}
    \centering
    \includegraphics[width=\linewidth]{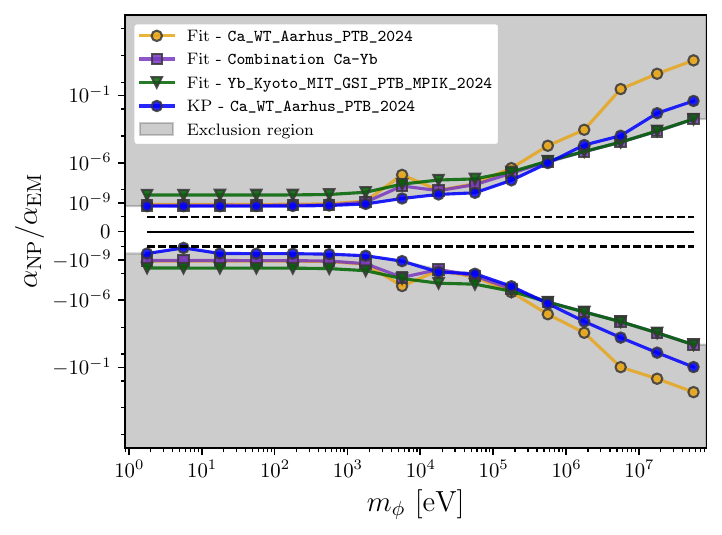}
    \caption{Fit results for a Ca data set, an Yb data set and the combination thereof.
    The (most stringent) algebraic results for the Ca data set are also shown.
    }
    \label{fig:CapYb}
\end{figure}

A similar situation can be observed in Fig.~\ref{fig:CapYb}, which combines
\texttt{Ca\_WT\_Aarhus\_PTB\_2024} with
\texttt{Yb\_Kyoto\_MIT\_GSI\_MPIK\_2024}, a set of Yb isotope shifts for 4
isotope pairs and 5 transitions. The fit result for Yb must be
interpreted with caution, since it is known to contain SM nonlinearities, which are, however, not taken into account by the current version of \kifit.
In Appendix~\ref{sec:sparsity} we discuss the impact of data sparsity in more detail.

In Figs.~\ref{fig:Ca} and \ref{fig:CapYb}, the fit appears to lose sensitivity more quickly than the algebraic methods at the high-mass end.
We trace this behaviour back to the \texttt{Ca\_PTB\_2015} data set, which starts to lose sensitivity comparatively rapidly at around $m_\phi = 10^{-6}~$eV. Since the algebraic bound in Fig.~\ref{fig:Ca} shows the envelope of the bounds, and the algebraic bounds for \texttt{Ca24min} turn out to be more stringent, the behaviour of the \texttt{Ca\_PTB\_2015} bound is not visible in the algebraic results plotted in Figs.~\ref{fig:Ca} and \ref{fig:CapYb}. 
Nonetheless, the robustness of the fit in the high-mass region could be improved by ameliorating the Monte Carlo sampling discussed in Section~\ref{sec:consolidation} and by increasing the resolution in $m_\phi$.
For completeness, we performed a series of higher-resolution  simulations executing the algebraic KP method on the \texttt{Ca\_PTB\_2015} dataset (equivalent to the blue curve in Fig.~\ref{fig:Ca}, but considering eighty mass values in the same range) and we were able to resolve similar spikes to the one shown in the \kifit bound in Fig.~\ref{fig:Ca}.

\section{Conclusions and Outlook}
\label{sec:conclusions}

Over the past few years, several precision isotope shift measurements have been performed, spurring the development of a wide range of King plot-inspired methods to search for new physics and study nuclear structure.
In this work, we systematically compare the available analysis methods, complementing them with the No-Mass Generalised King Plot and the Nuclear Input King Plot formulas, and taking a first step towards a global analysis of isotope shift data with our novel fit framework \kifit.

The key strengths of \kifit originate from its construction based on a log-likelihood: 
While the algebraic methods require data sets with $n=m$ or $n=m+1$ isotope pairs for $m$ transitions, 
the fit is flexible with respect to $n$ and $m$. It thus avoids the omission of data or cherry-picking the subset that yields the most stringent bound on new physics. Under the assumption of uncorrelated measurements, it further allows to combine isotope shifts from different elements.

This is particularly valuable for the search for new bosons whose couplings to atomic nuclei do not introduce element-dependent parameters. 
The Standard Model extension consisting of a new boson coupling linearly to neutrons and electrons, which is commonly investigated using King plots, is particularly predictive.
\kifit can, for the first time, test this model across elements, setting
\textit{global} bounds on the product of the new boson's couplings to neutrons and to electrons, for new boson masses spanning eight orders of magnitude.

The fit does, however, rely on the existence of sufficiently large data sets, since it tends to underestimate the margin for new physics when data is sparse.
For this reason, measuring isotope shifts and nuclear masses in larger arrays of isotopes would not only provide new opportunities to resolve higher-order nuclear effects, but also allow more stable and reliable bounds on new physics to be established.
Example systems with a large number of even stable isotopes are Zn (5 stable isotopes of which 4 are even, see
Ref.~\cite{Roeser:2024_Zn_hyperfine} for recent isotope shift data), 
Sn~\cite{Sn_Goble_1974} (10 stable isotopes of which 7 are even, plus one trace radioisotope with
$\tau_{1/2}\sim 10^5~$y), 
Ba (6 stable isotopes
of which 4 are even, plus one even isotope with a half-life $\tau_{1/2}\sim
10^{21}~$y, measured so far with 100-400\,MHz uncertainty~\cite{Choi_2024}) and
Cd (6 stable isotopes of which 5 are even, plus one even isotope with
$\tau_{1/2}\sim 10^{19}~$y; Ref.~\cite{Hofsaess:2023_CdIS} reached a resolution of few MHz for isotope shifts in two transitions),
plus the hitherto unexploited metastable isotopes in Ca and Yb, amongst others.

Recently, several additional isotope shift measurements were completed: Ref.~\cite{Ishiyama:2025oed} achieved a precision below 10\,Hz in the 431\,nm transition in neutral Yb; Ref.~\cite{Han2025} reduced the uncertainty of isotope shifts in the 369\,nm and 935\,nm transitions in Yb$^+$ to the sub-MHz level; 
Ref~\cite{Diepeveen:2024zax} measured isotope shifts in four metastable state transitions in Yb$^+$ at 30\,MHz and sub-MHz level.
The isotope shift in the fine structure splitting of $P$ states 
in Kr$^+$ was measured at sub-MHz precision in Ref.~\cite{Ando2025}.
Isotope shifts measured at sub-10\,MHz precision in Th$^{3+}$ were reported in Ref.~\cite{Zitzer2025}. 
Ref.~\cite{ZHANG2025} determined isotope shifts in 3 isotope pairs in Rydberg transitions of Sr (including the odd isotope $^{87}$Sr) at a sub-MHz precision. 
Ref.~\cite{Kaklamanakis2025} extended the King plot analysis to isotope shifts in diatomic molecules. 
These developments add to the growing treasure of isotope shift data and pave the path towards high-precision isotope shifts in a multitude of elements and even molecules.

Whereas the current version of the \kifit code is limited to King plots that are linear within uncertainties, King plot nonlinearities could be included in future versions, e.g. by subtracting them as in Ref.~\cite{Wilzewski:2024wap}, provided accurate predictions or complementary experimental input are available. The framework provided by the \kifit code might even be applied to obtain insights into higher-order nuclear effects in a data-driven way, as was suggested in Ref.~\cite{Door:2024qqz}.

Note that all King plot searches for new physics, including \kifit, rely on predictions for the electronic coefficients that enter the new physics term, while the analysis of nonlinear King plots using \kifit-like setups would additionally require predictions for higher-order SM terms. For this reason, it may be beneficial to focus on elements whose dominant King nonlinearities are 
higher-order mass shifts, in which case the nuclear contributions can be
determined from experimental data and the electronic
coefficients can be predicted with comparatively good
precision~\cite{Viatkina:2023qop,Wilzewski:2024wap}. In systems where the second-order mass shift is the only resolvable King nonlinearity, the Nuclear Input King Plot formula presented in this work can be employed to extract the new physics coupling from isotope shift data for just two transitions and four isotope pairs.

In conclusion, we encourage experimentalists to measure isotope shifts across additional transitions and isotopes, even if the initial precision is lower than that of the best current measurements. Additional data points can help eliminate higher-order effects in the isotope shift equations and stabilise the bounds on new physics. They also provide new insights into nuclear structure and can thus be used to refine nuclear models. If new isotope shifts are measured before the corresponding isotope masses, the No-Mass King Plot and its generalised version facilitate a first analysis of the new data and indicate the required precision of the mass measurements to achieve competitive sensitivity to new physics and the nuclear shape.
Ultimately, it is the combination of new measurements, advancements in atomic and nuclear structure theory and global (King plot) analyses that will enhance the sensitivity to new physics and to nuclear effects.\\

\begin{acknowledgments}
    We thank Julian Berengut for providing the $X$ coefficients of Yb that he calculated using the open-source atomic structure code \texttt{AMBiT}~\cite{kahl2019ambit}, and for allowing us to make them publicly available in \kifit.
    FK would like to thank Peter Stangl for fruitful discussions during the early
    stages of this project.
    We thank our collaborators in the Yb$^+$~\cite{Door:2024qqz} and Ca$^+$/Ca$^{14+}$~\cite{Wilzewski:2024wap} King plot analyses for the insightful discussions.
    EF, FK and JR acknowledge funding by the Deutsche Forschungsgemeinschaft (DFG, German Research Foundation) under Germany’s Excellence Strategy -- EXC-2123 QuantumFrontiers -- 390837967.
    AM is partially funded by the DFG via the project B10 of the Collaborative Research Cluster 1227 (DQ-mat) -- Project-ID 274200144 and also acknowledges support by the EXC-2123 QuantumFrontiers.
    EF was funded by CERN in the initial phase of this work and thanks CERN-TH and the CERN QTI for the stimulating environment.
    FK thanks CERN-TH for its hospitality during the early stages of this project and for support via the CERN-TH visitor program.
    AM thanks the Graduate Academy of Leibniz University Hannover for financial support of an extended research stay at CERN during the early stages of the project and CERN-TH for its hospitality.
    MR is supported by CERN through the CERN Quantum Technology Initiative (QTI). 
    This work has been partially funded by the Deutsche Forschungsgemeinschaft
(DFG, German Research Foundation) - 491245950.
    This collaboration came about thanks to the CERN QTI Forum.
\end{acknowledgments}

\bibliography{bibliography.bib}

\appendix
\section{Calculation of Electronic Coefficients in $\text{Ca}^+$}
\label{sec:ambit}

In this section we present some details of the calculation of the electronic new physics coefficients in $\text{Ca}^+$~
\footnote{The $X$ coefficients for Yb used in this paper were calculated in Ref.~\cite{Counts:2020aws, Figueroa:2022mtm,Hur:2022gof} and obtained from J. Berengut via personal communication.}.
Similar to Ref.~\cite{Berengut:2017zuo}, we applied the finite field method. In this approach, the new physics 
potential 
    \begin{equation}
        V_{\mathrm{NP}}(r,m_\phi)=\lambda\frac{1}{4\pi r}e^{- m_\phi r}
        \label{eq:yukawa_ambit}
    \end{equation}    
is directly added to the Hamiltonian in the Dirac equation, including a dimensionless scaling parameter $\lambda$. Then, the eigenenergies of the considered states are calculated by solving the Dirac equation for different values of the parameter $\lambda$. Finally, the $X$ coefficient can be extracted by taking the numerical derivative of the eigenenergies with respect to $\lambda$:
    \begin{equation}
        X_i = \left.\frac{\partial E_i\left(\lambda\right)}{\partial \lambda}\right|_{\lambda=0}.
    \end{equation}
In a similar manner, the FS coefficients $F_i$ are evaluated by calculating the energies for different nuclear radii and taking the numerical derivative of the eigen-energies with respect to $\delta \langle r^2 \rangle$:
    \begin{equation}
        F_i=\left.\frac{\partial E_i\left(\lambda\right)}{\partial \delta \langle r^2 \rangle}\right|_{\delta \langle r^2 \rangle=0}.
    \end{equation}
To obtain the eigenenergies, electronic structure calculations are performed using the \texttt{AMBiT} code, which is based on a combination of configuration interaction (CI) and many-body perturbation theory (MBPT) (see e.g. Refs.~\cite{Dzuba1996,KOZLOV2015,Berengut2016,Torretti2017}) and is thoroughly described in Ref.~\cite{kahl2019ambit}. 
The specifics of the calculations for $\mathrm{Ca}^+$ are overviewed in the appendices of Ref.~\cite{Wilzewski:2024wap}.

\begin{figure}[ht]
        \centering
        \includegraphics[width=0.5\textwidth]{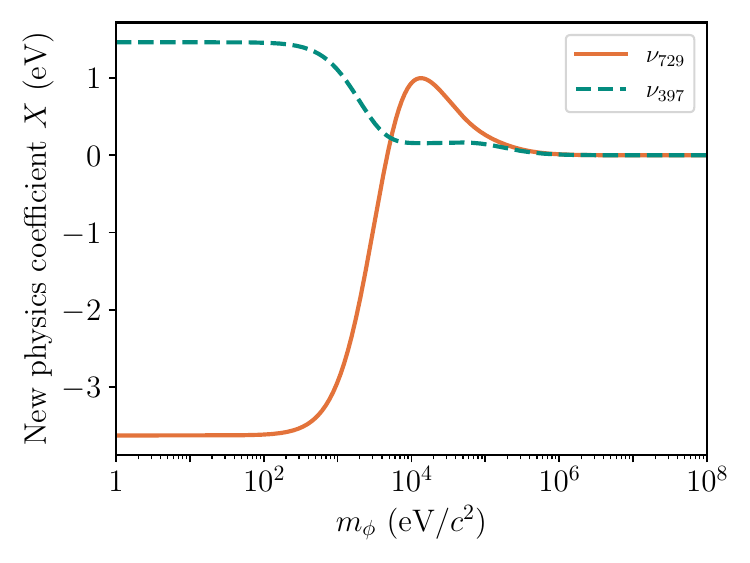}
        \caption{Dependence of the electronic new physics coefficients of the \castodfive~and \castop~transitions in $\mathrm{Ca}^+$ on the mediator mass $m_\phi$. The coefficient of the \castodthree~transition is nearly identical to that of \castodfive~and is therefore not displayed in the figure.}
        \label{fig:X_vs_mass}
    \end{figure}

The results for the $X$ and $F$ coefficients are summarised in Tab.~\ref{tab:el_coeffs} and Fig.~\ref{fig:X_vs_mass} and are in good agreement with reference values from~\cite{Viatkina:2023qop}.\\

As discussed in the main text in the high-mass regime the ratio of the $X$ coefficients approaches the ratio of field shift coefficients, leading to the loss of sensitivity to new physics.
Since even a small discrepancy between the experimental and theoretical values of the field shift ratio $ F_j/F_i $ can significantly impact the high-mass behavior of the bounds on new physics, the $X$ coefficients are rescaled so that in the high-mass limit the ratio $X_j/X_i$ approaches the experimental field shift ratio $ F_j^{\mathrm{exp}}/F_i^{\mathrm{exp}} $, which is obtained from a linear fit to King plot measurements.

Note that the conventions for the new physics coefficients (referred to as $D$ coefficients e.g. in Refs.~\cite{Counts:2020aws,Hur:2022gof}) vary in the literature. For instance, one can include all prefactors of the Yukawa potential of Eq.~\eqref{eq:yukawa_ambit} in the calculation of the new physics coefficients. As a result, when utilising these coefficients to derive constraints on new physics, the obtained bounds are expressed in terms of $y_e y_n$. However, in other conventions a factor of $1/(4\pi\alpha_\mathrm{EM})$ is factored out from the $X$ coefficient, resulting in bounds being expressed in terms of $y_e y_n/(4\pi\alpha_\mathrm{EM}) = \alphaNPdEM$. In this work, the $X$ coefficients are always presented in the latter convention.

\section{Sensitivity to New Physics}
\label{sec:sensitivity_NP}

To gain an understanding of how the choice of transitions can affect the bounds
on the new physics coupling, we compare bounds obtained by
applying Eqs.~\eqref{eq:ANP3x2} and \eqref{eq:nsigma_bound} to subsets of
isotope shift measurements in $\mathrm{Ca}^+$. We fix \castodfive: $3p^6 4s$
$^2S_{1/2}\to 3p^6 3d$ $^2D_{5/2}$ as the reference transition $\nu_1$ and
combine it with either of the following transitions:
    \begin{itemize}
        \item \castop: $3p^64s$ $^2S_{1/2}\to3p^64p$ $^2P_{1/2}$ 
        \item \castodthree: $3p^64s$ $^2S_{1/2}\to3p^63d$ $^2D_{3/2}$
        \item \cadfine: $3p^63d$ $^2D_{3/2}\to3p^63d$ $^2D_{5/2}$
    \end{itemize}
where the subscripts correspond to the transition wavelength in nm, while $\rm DD$ is associated to fine-structure splitting. A comprehensive list of available isotope shift data in Ca can be found in Tab.~\ref{tab:Ca_IS}.

    \begin{figure}
        \centering        \includegraphics[width=\linewidth]{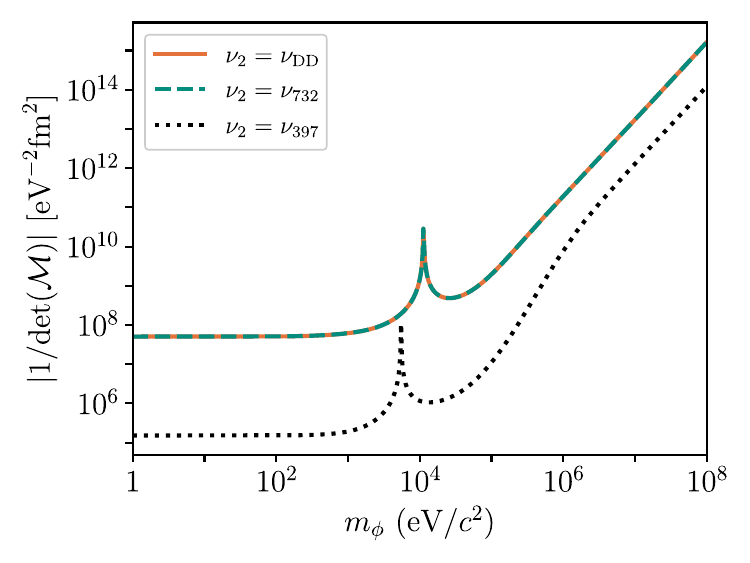}\\
        \caption{Sensitivity of different transition pairs to new physics. The reference transition is fixed to $\nu_1=\nu_{729}$, the second transitions are as specified in the legend.
        The inverse of the electronic dependence of $\Vpred$ (see Eq.~\eqref{eq:VpredefKP}) as a function of the mass $m_\phi$ of the new boson, suggesting the combination $(\nu_{729}$, \castop) is more sensitive than the other two, which are equally sensitive. 
        }
        \label{fig:compare_KP_lin}
    \end{figure}

\subsubsection*{Identifying King Plot Candidates}

The shape of $\Vpred$ can hint at the best transitions for new physics searches: a combination of electronic coefficients that maximises this volume suggests a high sensitivity of the associated transitions to $\alphaNP$. In Tab.~\ref{tab:el_coeffs} we report our calculations of the electronic coefficients of the transitions listed above and in Fig.~\ref{fig:compare_KP_lin} we plot the inverse of the electronic dependence of $\Vpred$:
    \begin{align}
         \frac{1}{\Vpred} \propto \frac{1}{\det(\mathcal{M}(m_\phi))}=\frac{1}{F_1 X_2(m_\phi)-F_2X_1(m_\phi)}\,,
    \end{align}
which determines the
$m_\phi$-dependence of the bounds on $\alphaNP$ (see Eqs.~\eqref{eq:ANP3x2} and
\eqref{eq:VpredefKP}).
The $m_\phi$-independent low-mass behaviour, as well as the loss of sensitivity
at high mediator masses $m_\phi$, are clearly visible. Moreover, we observe
peaks, which are the consequence of accidental cancellations of $\det
(\mathcal{M}(m_\phi))$ at specific values of $m_\phi$.

We identify the combination of $\nu_{729}$ and \castop as the pair of
transitions with the largest value for $\det(\mathcal{M}(m_\phi))$, suggesting a high
sensitivity to new physics. This can be understood as a consequence of the
complementarity of the electronic levels involved in these transitions. In
contrast, the combinations of $\nu_{729}$ with $\nu_{732}$ or \cadfine yield a smaller value for the electronic part of $\Vpred$. Indeed, the
similarity of the $D_{3/2}$ and $D_{5/2}$ states results in similar electronic
coefficients leading to a higher degree of cancellation in $F_1X_2(m_\phi)-F_2 X_1(m_\phi)$.

\subsubsection*{The Perfectly Linear King Plot}
Next, we consider the idealised case of a perfectly linear King plot ($\Vdat=0$). To this end, we generate perfectly linear mock values for the isotope shifts, 
i.e. we use Eq.~\eqref{eq:linIS} to calculate linear isotope shift data based on experimental values for the nuclear radii~\cite{ANGELI201369} and masses~\cite{AME2020II}, as well as on theory calculations of the SM electronic coefficients (reported in Tab.~\ref{tab:el_coeffs}).

\begin{figure}[t]
        \centering
        \includegraphics[width=\linewidth]{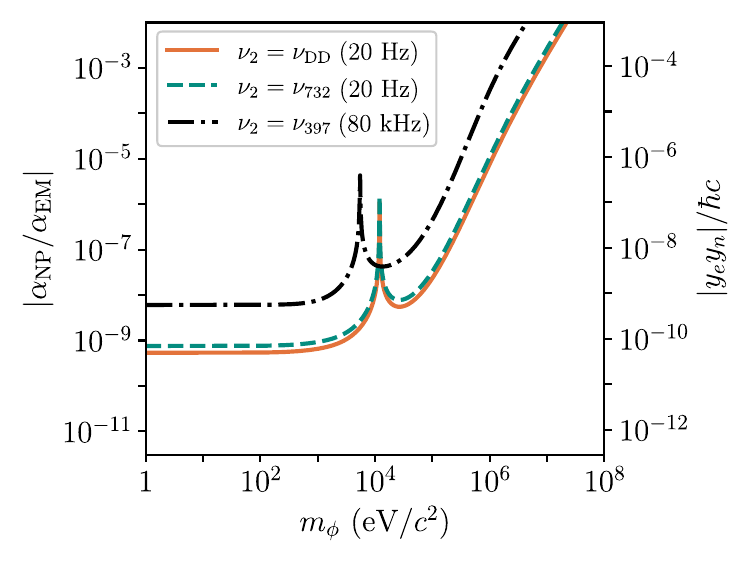}%
        
        \caption{Predicted $2\sigma$
        upper bounds on $|\alphaNP/\alphaEM|$ for linear mock data with uncertainties as reported in the legend.
        Although the combination involving \castop appears to be the most promising based on the behaviour of $|1/\det(\mathcal{M})|$, the bounds on $|\alphaNP/\alphaEM|$ obtained for the other two combinations turn out to be more stringent due to smaller experimental uncertainties. \cadfine wins against \castodthree thanks to the smaller isotope shift.}
        \label{fig:Compare_KP_nonlin_1}
    \end{figure}

Clearly, in this case $\langle\alphaNP\rangle=0$ and the bounds are fully determined by $\sigalphaNP$. 
We can gain some insight into the behaviour of $\sigalphaNP$ by performing linear error propagation on Eq.~\eqref{eq:ANP3x2}:
    \begin{align}
        \sigalphaNP =& \sqrt{\left(\frac{\sigma[\Vdat]}{\Vpred}\right)^2 
            + \left(\frac{\Vdat}{\Vpred^2} \sigma[\Vpred]\right)^2}\,.
        \label{eq:sigalphaNP_Vdat_Vpred}
    \end{align}
with $\Vdat$ and $\Vpred$ as defined in Eq.~\eqref{eq:ANP3x2}.
In linear King plots, only the first term is relevant and $\sigma[\alphaNP]$
is dominated by experimental uncertainties. What determines 
the sensitivity to new physics is thus the interplay of $\Vpred$ and $\sigma[\Vdat]$. Since the latter includes terms of the form $\mnuai{a}{i}\sigma[\mnuai{b}{j}]$, more constraining bounds can be expected from 
comparatively small but precisely measured isotope shifts.
We assign realistic uncertainties to our mock transitions (20\,Hz for \castodfive, \cadfine and \castodthree~\cite{Solaro:2020dxz}, 80\,kHz for \castop~\cite{Gebert:2015_Ca100kHz}) and compare the projections for the $2\sigma$ upper bounds in Fig.~\ref{fig:Compare_KP_nonlin_1}.
Here, we limit the analysis to $\vert\alphaNPdEM\vert$ for simplicity.
Although the combination involving \castop appeared to be the most promising based on the behaviour of its electronic coefficients, the higher experimental uncertainty limits its sensitivity to new physics, relative to the other combinations. \\
    {\setlength{\tabcolsep}{3pt}
    \renewcommand{\arraystretch}{1.5}
    \begin{table*}[]
        \centering
        \begin{tabular}{c|ccccc}
            \hline \hline
            \multirow{2}{*}{Transition ($i$)} & $\Delta E$ & $F_i$ & $K_i$ & $X_i(m_\phi=1\text{ keV})$ &$X_i(m_\phi=10^4\text{ keV})$ \\
             & [eV] & $[\text{eV}/\text{fm}^2]$ & [eV$\cdot$u] & [eV] &  [eV] \\
            \hline
            \textbf{\castop}: $3p^6 4s\;^2S_{1/2}\to 3p^6 4p\;^2P_{1/2}$ & $3.16$ &$-1.19 \times 10^{-6}$ & & $1.11$ & $1.29\times 10^{-4}$ \\
             & $3.12$~\cite{sugar1985atomic} & $-1.18 \times 10^{-6}$~\cite{Viatkina:2023qop} & $-1.59 \times 10^{-3}$~\cite{Viatkina:2023qop} & & \\
             & & & & & \\
            \textbf{\castodthree}: $3p^6 4s\;^2S_{1/2}\to 3p^6 3d\;^2D_{3/2}$ & $1.677$ & $-1.559 \times 10^{-6}$ & & $-2.95$ & $1.71\times 10^{-4}$ \\
             & $1.692$~\cite{sugar1985atomic} & $-1.567 \times 10^{-6}$~\cite{Viatkina:2023qop} & $-1.017 \times 10^{-2}$~\cite{Viatkina:2023qop} & & \\
             & & & & & \\
            \textbf{\castodfive}: $3p^6 4s$ $^2S_{1/2}\to 3p^6 3d$ $^2D_{5/2}$ & $1.685$ & $-1.557 \times 10^{-6}$  & & $-2.93$ & $1.70\times 10^{-4}$ \\
             & $1.700$ \cite{sugar1985atomic} & $-1.565 \times 10^{-6}$~\cite{Viatkina:2023qop} & $-1.012 \times 10^{-2}$~\cite{Viatkina:2023qop} & & \\
            \hline \hline
        \end{tabular}
        \caption{Transition energies ($\Delta E$) and electronic field shift ($F_i$), mass shift ($K_i$) and new physics ($X_i$) coefficients for three transitions in $\text{Ca}^+$. Values of $\Delta E$, $F_i$ and $X_i$ were calculated using the CI + MBPT method implemented in \texttt{AMBiT}~\cite{kahl2019ambit} (see also Appendix~\ref{sec:ambit}). The latter are reported for two different values of the mediator mass $m_\phi$. The results are in good agreement with the transition energies given in Ref.~\cite{sugar1985atomic} and the electronic coefficients of Ref.~\cite{Viatkina:2023qop}. The values for $K_i$ are taken from Ref.~\cite{Viatkina:2023qop}. Values for $\nu_\mathrm{DSIS}$ can be obtained from the difference of \castodfive\, and \castodthree. In Ref.~\cite{Viatkina:2023qop}, the uncertainties on $F_i$ and $K_i$ are estimated to be $5\%$ and $10\%$, respectively. For $X_i$, we estimate the uncertainty to be about $10\%$.}
        \label{tab:el_coeffs}
    \end{table*}
    }
    
    \begin{figure}
        \centering
        \includegraphics[width=\linewidth]{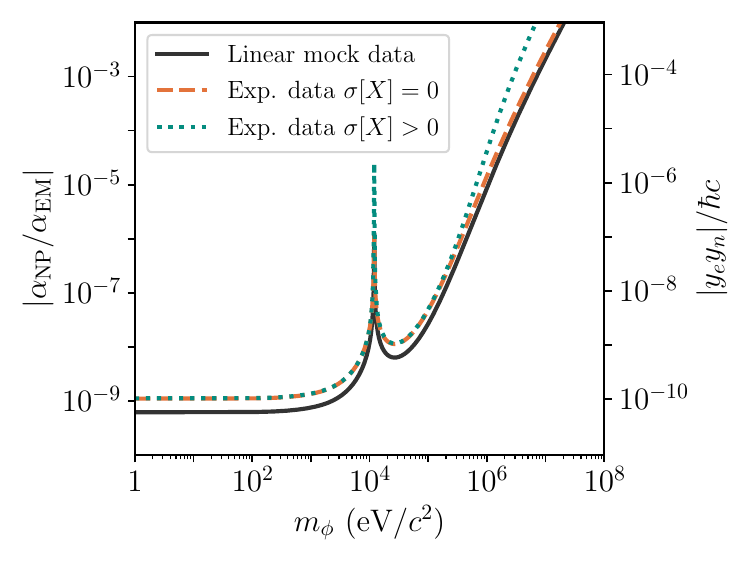}
        \caption{The bounds from the combination of \cadfine and
        \castodfive. The black curve shows the results for perfectly linear
        mock data while the orange dashed and teal dotted curves represent bounds derived
        from the experimental data reported in Refs.~\cite{Solaro:2020dxz,Knollmann2019}, with
        and without including uncertainties on the electronic new physics
        coefficient $X_i$.}
        \label{fig:Compare_KP_nonlin_2}
    \end{figure}
    
\subsubsection*{The Linear King Plot}
Having identified the combination of the \castodfive and \cadfine transitions to generate the most stringent bounds to new physics, we use real data from Refs.~\cite{Knollmann2019,knollmann2023erratum,
Solaro:2020dxz}, which produce a linear King plot within the experimental uncertainties ($\sigma[\Vdat]/\Vdat\lesssim1$). In Fig.~\ref{fig:Compare_KP_nonlin_2}, we compare the resulting bounds with those from perfectly linear mock data.

Since real data predict a non-zero central value for the new physics coupling, the bounds are shifted away from zero. 
This effect is most pronounced in the
low-mass limit, where the King plot method is most sensitive to new physics. Moreover, the second term in Eq.~\eqref{eq:sigalphaNP_Vdat_Vpred} is now non-zero. This has a noticeable impact if the uncertainties on the electronic new physics coefficients $X$, which we take to be $10\%$ of their computed value (see Appendix~\ref{sec:ambit}), are introduced: Since this term scales with $1/\Vpred^2$ and $\Vpred$ becomes small for large values of $m_\phi$, uncertainties on
the $X$ coefficients lead to a loss of sensitivity in the high mass region.
This highlights the importance of considering the uncertainty of $X$ and suggests that more precise $X$ coefficient calculations could improve the bounds, particularly in the high-mass range.

\section{Projection of Uncertainties}
\label{sec:uncertainty_projections}
To better understand how the uncertainties on the measured isotope shifts and atomic masses propagate to the bounds on the new physics coupling, we can
project the experimental uncertainties onto the directions parallel and perpendicular  to the King line and evaluate their impact on the uncertainty on $\alphaNP$:

    \begin{align}
            \Sigma^\alpha_\parallel =& \left(\boldsymbol{\nabla}_{\mnuai{}{}} \alphaNP\right)^\top\boldsymbol{P_\parallel} \boldsymbol{\Sigma}_{\mnuai{}{}} \boldsymbol{P_\parallel} \boldsymbol{\nabla}_{\mnuai{}{}}\alphaNP
            \equiv \left(\sigalphaNP_{\mathrm{KP},\parallel}^{(1)}\right)^2\notag\\
            \Sigma^\alpha_\perp =& 
            \left(\boldsymbol{\nabla}_{\mnuai{}{}} \alphaNP\right)^\top \boldsymbol{P_\perp} \boldsymbol{\Sigma}_{\mnuai{}{}} \boldsymbol{P_\perp} \boldsymbol{\nabla}_{\mnuai{}{}} \alphaNP \equiv \left(\sigalphaNP_{\mathrm{KP},\perp}^{(1)}\right)^2\notag\\
            \Sigma^\alpha_{\parallel\perp} =&
            \left(\boldsymbol{\nabla}_{\mnuai{}{}} \alphaNP\right)^\top \boldsymbol{P_\parallel} \boldsymbol{\Sigma}_{\mnuai{}{}} \boldsymbol{P_\perp} \boldsymbol{\nabla} _{\mnuai{}{}}\alphaNP = (\Sigma^\alpha_{\perp\parallel} )^\top\,,\label{eq:sigalphaNP_alg_perp_estimate}
    \end{align}
Here the superscript (1) indicates that we are applying linear error propagation, whereas
    \begin{align}
        \boldsymbol{\nabla} _{\mnuai{}{}}\alphaNP = \begin{pmatrix}
        \frac{\partial\alphaNP}{\partial \mnuai{1}{1}} & \cdots & \frac{\partial\alphaNP}{\partial \mnuai{1}{m}}\\
        \vdots & \ddots & \vdots\\
        \frac{\partial\alphaNP}{\partial\mnuai{n}{1}} & \cdots & \frac{\partial\alphaNP}{\partial \mnuai{n}{m}}
        \end{pmatrix}\,
        \label{eq:grad_alphaNP_mnuai}
    \end{align}
corresponds to the gradient of $\alphaNP$ with respect to the $n\times m$ mass-normalised isotope shifts $\lbrace \mnuai{a}{i}\rbrace_{1 \leq i \leq m}^{1 \leq a \leq n}$ (points in the King plot), and
    \begin{align}
        \boldsymbol{\Sigma}_{\mnuai{}{}} = \begin{pmatrix}
        \Sigma_{\mnuai{}{}}^{11} & \cdots & \Sigma_{\mnuai{}{}}^{1n}\\
        \vdots      & \ddots & \vdots\\
        \Sigma_{\mnuai{}{}}^{n1} & \cdots & \Sigma_{\mnuai{}{}}^{nn}
        \end{pmatrix}\,
        \label{eq:Sigma_mnuai}
    \end{align}
denotes the $(m \times m \times n \times n)$-dimensional covariance matrix of the set of $n$ mass-normalised isotope shift vectors $\lbrace\mnuveca{a}\rbrace_{a=1}^n$. The entries of $\boldsymbol{\Sigma}_{\mnuai{}{}}$ are $(m\times m)$-dimensional covariance matrices associated to $\mnuveca{a}= (\mnuai{a}{1},\ldots,\mnuai{a}{m})$ and $\mnuveca{b}= (\mnuai{b}{1},\ldots,\mnuai{b}{m})$:
    \begin{align}
        \Sigma_{\mnuai{}{}}^{ab} = 
            \begin{pmatrix}
            \mathrm{Cov}(\mnuai{a}{1}, \mnuai{b}{1}) 
            &
            \cdots 
            &
            \mathrm{Cov}(\mnuai{a}{1}, \mnuai{b}{m})
            \\
            \vdots
            &
            \ddots
            &
            \vdots
            \\
            \mathrm{Cov}(\mnuai{a}{1}, \mnuai{b}{m}) 
            &
            \cdots
            &
            \mathrm{Cov}(\mnuai{a}{m}, \mnuai{b}{m}) 
            \end{pmatrix}\,,
    \label{eq:Sigma_mnuai_ab}
    \end{align}
where
    \begin{align}
    \mathrm{Cov}(\mnuai{a}{i}, &\mnuai{b}{j}) = \sum_{k=1}^m \sum_{c=1}^n \Bigg(
    \frac{\partial \mnuai{a}{i}}{\partial \nuai{c}{k}} \snuai{c}{k}^2 \frac{\partial \mnuai{b}{j}}{\partial \nuai{c}{k}}\notag\\
    & +
    \frac{\partial \mnuai{a}{i}}{\partial \ma{c}} \sma{c}^2 \frac{\partial \mnuai{b}{j}}{\partial \ma{c}}
     +
    \frac{\partial \mnuai{a}{i}}{\partial \map{c}} \smap{c}^2 \frac{\partial \mnuai{b}{j}}{\partial \map{c}}
    \Bigg)\,,
    \label{eq:Sigma_mnuai_abij_analytic}
    \end{align}
assuming the isotope shift and mass measurements are independent. 
The objects $\boldsymbol{P_\parallel}$ and $\boldsymbol{P_\perp}$ in
Eq.~\eqref{eq:sigalphaNP_alg_perp_estimate} are $n \times n \times m\times
m$-dimensional projectors onto the parallel and perpendicular directions,
    \begin{align}
        \boldsymbol{P_\parallel} =& \mathbbm{1}_n\otimes P_\parallel = \mathrm{diag}\left(P_\parallel,\ldots, P_\parallel\right)_n\\
        \boldsymbol{P_\perp} =& \mathbbm{1}_n\otimes P_\perp = \mathrm{diag}\left(P_\perp,\ldots, P_\perp\right)_n\,.
    \end{align}
Here $\otimes$ denotes the tensor product and $P_\parallel$ and $P_\perp$ are
the $m\times m$-dimensional projectors
    \begin{align}
        P_\parallel =&  \eFvec \eFvec^\top \,, & 
        P_\perp =&  \mathbbm{1}_m - \eFvec \eFvec^\top \,,\\
        P_\parallel ^2 =& P_\parallel\,,\quad P_\parallel ^\top = P_\parallel\,,& P_\perp ^2 =& P_\perp\,,\quad P_\perp ^\top = P_\perp\,,
    \end{align}
where $\eFvec$ is the unit vector along the King line (see Eq.~\eqref{eq:eFdef}) and $\mathbbm{1}_m$ is the $(m\times m)$-dimensional identity matrix.

Finally, the impact of the uncertainties perpendicular to the King line on the
estimation of $\sigalphaNP$ can be evaluated by computing
    \begin{align}
        \sqrt{
            \frac{\Sigma^\alpha_\perp}{\Sigma^\alpha}} 
        \equiv 
            \frac{
                \sigalphaNP\vert_{\mathrm{KP},\perp}^{(1)}
            }{
                \sigalphaNP\vert_\mathrm{KP}^{(1)}}\,,
        \label{eq:sigalphaNP_alg_perp_frac_estimate}
    \end{align}
where $\sigalphaNP\vert_{\mathrm{KP}}^{(1)}$ is the full uncertainty
without projectors:
    \begin{align}
        \sigalphaNP\vert_{\mathrm{KP}}^{(1)} \equiv \sqrt{\Sigma_\alpha} \equiv 
        \sqrt{\left(\boldsymbol{\nabla}_{\mnuai{}{}} \alphaNP\right)^\top  \boldsymbol{\Sigma}_{\mnuai{}{}} \boldsymbol{\nabla}_{\mnuai{}{}}\alphaNP}\,.
        \label{eq:sigalphaNP_alg_estimate}
    \end{align}
For all elements in Table~\ref{tab:fitvsalg} we checked that 
$\sigalphaNP\vert_{\mathrm{KP},\perp}^{(1)}/ \sigalphaNP\vert_\mathrm{KP}^{(1)}\approx 1$ 
within numerical uncertainties.

\section{The \kifit Package}
\label{sec:kifit_package}
\kifit is a prototype code and is published strictly for reasons of transparency and reproducibility.
In this section, we briefly explain the structure of the package and give an introductory example of how to use \kifit. We also summarise the \texttt{pytests} implemented in the code.

\subsection{Structure of the \kifit Package}
\label{sec:kifit_general}
\begin{figure}[ht]
  \begin{center}
    \includegraphics[width=\linewidth]{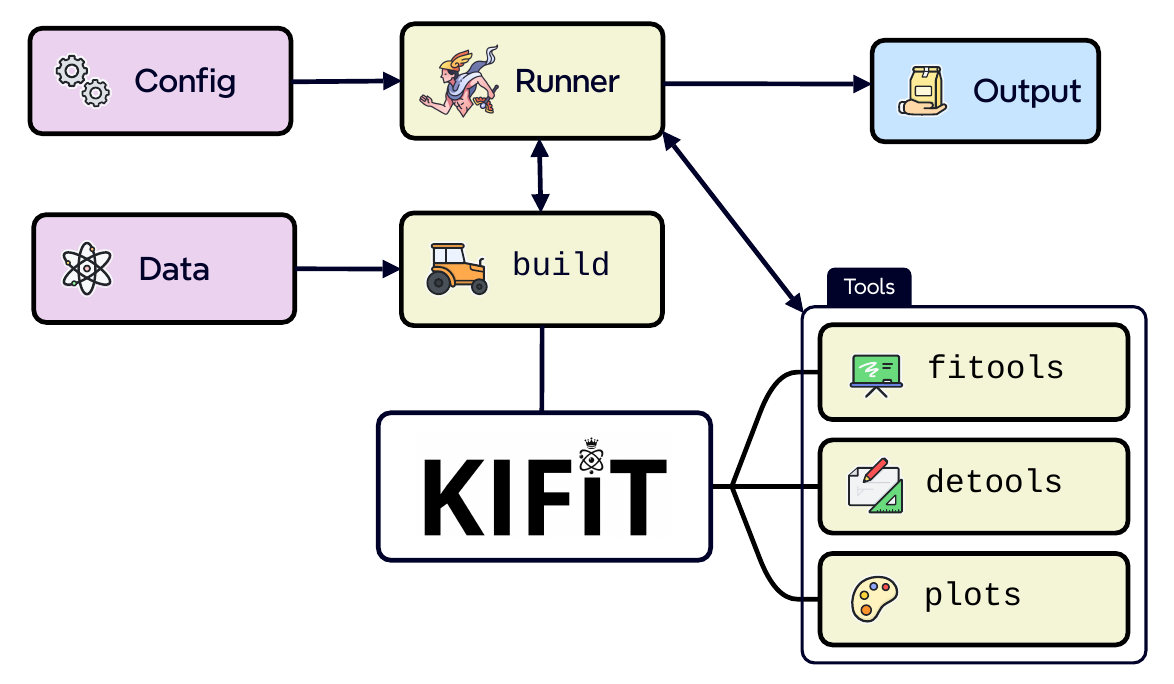}
  \end{center}
  \caption{Schematic structure of \kifit \texttt{0.1.0}.}
    \label{fig:kifit_diagram}
\end{figure}

The source code of \kifit can be divided into three main modules:
\texttt{build}, \texttt{tools} and \texttt{run}, which are represented with
yellow boxes in the diagram of Fig.~\ref{fig:kifit_diagram}.

The operative module is \texttt{run}, where a \texttt{Runner} class makes use of
\texttt{build} and \texttt{tools} to process data and perform either the algebraic methods
and the King plot fit. 

To launch the \kifit scripts, a \texttt{kifit\_data} folder is required, in
which experimental data have to be provided in the same format as in the example
which can be found in the code repository and as described below.
The execution of the code can be
customised by defining a set of initial parameters in an appropriate configuration file (see \href{https://github.com/QTI-TH/kifit}{\texttt{README.md}} file of our repository for more detailed information).

The module \texttt{build} constructs one instance of the class
\texttt{kifit.Elem} for each experimental data set collected in a \texttt{kifit.ElemCollection}. 
The module \texttt{tools} contains two sets of functions which compute the bounds on $\alphaNP$, either using the fit (\texttt{fitools}) or the algebraic methods (\texttt{detools}). 
Finally, the script \texttt{plot} is a compilation of plot scripts that can be used to visualise the fit and determinant results.

\subsection{How to Use \kifit}
\label{sec:kifit_man}

\subsubsection*{Data Preparation}
First, the input data needs to be organised in subfolders of the \texttt{kifit\_data} subfolder and with names corresponding to the element identification (e.g. \texttt{elem}): 
\vspace{0.4cm}
\begin{lstlisting}[language=Python]
kifit/ 
|-- src/ 
|   |- kifit/
|   |   |- ... 
|   |   |- user_elements.py
|   |- kifit_data/
|   |   |- elem/
|   |   |   |- binding_energies_elem.dat
|   |   |   |- isotopes_elem.dat
|   |   |   |- nu_elem.dat
|   |   |   |- sig_nu_elem.dat
|   |   |   |- sig_Xcoeffs_elem.dat
|   |   |   |- Xcoeffs_elem.dat
|   |- tests/
\end{lstlisting}
\vspace{0.4cm}

The \kifit datafiles inside the \texttt{elem} subfolder are to be named as follows:
    \begin{align*}
        \texttt{type\_elem\_ISinstitutions\_massinstitution\_year.dat}
    \end{align*}
where \texttt{type} refers to the type of data saved in the file (for example data files, see the code repository): 
    \begin{itemize}
        \item \texttt{binding\_energies}: electron binding energies, extracted from NIST~\cite{NIST2023}
        \item \texttt{isotopes}: isotope numbers $A, A'$, masses $(\ma{A},\map{A})$, mass uncertainties $(\sma{A}, \smap{A})$.
        \item \texttt{nu}: isotope shift measurements ($\nuai{a}{i}$); (columns: transitions, rows: isotope pairs).
        \item \texttt{sig\_nu}: uncertainties on the isotope shift measurements ($\snuai{a}{i}$).
        \item \texttt{Xcoeffs} and \texttt{sig\_Xcoeffs}: electronic coefficients ($X_i$) and their uncertainties ($\sigma[X_i]$) for the transitions listed in \texttt{nu} datafiles, for different values of the new boson masses $m_\phi$ (each line corresponding to a separate value of $m_\phi$). In the current version of \kifit, the \texttt{sig\_Xcoeffs} files are not employed.
    \end{itemize}
\texttt{ISinstitutions} and \texttt{massinstitution} list the institutions where the isotope shift measurements contained in \texttt{nu} and the isotope masses measurements contained in \texttt{isotopes} datafiles were respectively carried out, while \texttt{year} reports the year of the most recent isotope shift measurements collected in the folder. If no \texttt{massinstitution} is specified, isotope masses measurement are taken from either AME2020~\cite{kondev2017ame2016} or AME2016~\cite{AME2020II}, as reported in the heading of the \texttt{isotopes} datafiles. Similarly, the heading of \texttt{ISinstitutions} datafiles reports, in shorthand notation, the publications from which we extracted the isotope shift measurements, as well as for the electronic transition. Table~\ref{tab:kifit_datafolders} provides a list of the data folders available within \kifit, the shorthand notation adopted to refer to the institutions and the ``dimension'' of the dataset, i.e. number of transitions ($m$) and isotope pairs ($n$). 

{\footnotesize \setlength{\tabcolsep}{6pt}
\renewcommand{\arraystretch}{1.4}
    \begin{table*}[t]
    \centering
        \begin{tabular}{llcc}
        \hline
        \hline
        \textbf{\kifit folder}            & \textbf{Isotope shifts}             & \textbf{Masses}  & $n\times m$ \\
        \hline
        \texttt{Ca\_PTB\_2015            } & Gebert2015~\cite{Gebert:2015_Ca100kHz}  & \cite{kondev2017ame2016} & $3\times2$ \\
        \texttt{Ca\_WT\_Aarhus\_2020     } & Knollmann2019\cite{Knollmann2019}+Solaro2020~\cite{Solaro:2020dxz}  & \cite{kondev2017ame2016} & $4\times2$ \\
        \texttt{Ca\_WT\_Aarhus\_2024     } & Solaro2020~\cite{Solaro:2020dxz}, Knollmann2023~\cite{knollmann2023erratum}, Chang2024~\cite{chang2024systematicfree}  & \cite{kondev2017ame2016} & $4\times2$ \\
        \texttt{Ca\_WT\_Aarhus\_PTB\_2020} & Knollmann2019~\cite{Knollmann2019}, Solaro2020~\cite{Solaro:2020dxz}, Gebert2015~\cite{Gebert:2015_Ca100kHz} & \cite{kondev2017ame2016} & $3\times4$ \\
        \texttt{Ca\_WT\_Aarhus\_PTB\_2024} & Knollmann2023~\cite{knollmann2023erratum}, Chang2024~\cite{chang2024systematicfree}, Gebert2015~\cite{Gebert:2015_Ca100kHz} & \cite{kondev2017ame2016} & $3\times4$ \\
        \texttt{Ca24min                  } & Knollmann2023~\cite{knollmann2023erratum}, Chang2024~\cite{chang2024systematicfree} & \cite{AME2020II} & $3\times2$ \\
        \texttt{Camin                    } & Knollmann2019~\cite{Knollmann2019}, Solaro2020~\cite{Solaro:2020dxz} & \cite{AME2020II} & $3\times2$ \\
        \texttt{Ca\_testdata             } & Knollmann2019~\cite{Knollmann2019}, Solaro2020~\cite{Solaro:2020dxz}, Gebert2015~\cite{Gebert:2015_Ca100kHz} & \cite{AME2020II} & $3\times4$ \\
        \texttt{Ca3pointsTEST            } & *~Knollmann2023~\cite{knollmann2023erratum}, Chang2024~\cite{chang2024systematicfree}  & \cite{AME2020II} & $3\times2$ \\
        \texttt{Ca4pointsTEST            } & *~Knollmann2023~\cite{knollmann2023erratum}, Chang2024~\cite{chang2024systematicfree}                & \cite{AME2020II} & $4\times2$ \\
        \texttt{Ca10pointsTEST           } & mock data                            & \cite{AME2020II} & $10\times2$ \\
        \texttt{Yb\_Kyoto\_MIT\_GSI\_2022} & Ono2022~\cite{ono2022observation}, Counts2020~\cite{Counts:2020aws}, Hur2022~\cite{Hur:2022gof}, Figueroa2022~\cite{Figueroa:2022mtm} & \cite{AME2020II} & $4\times5$ \\
        \texttt{Yb\_Kyoto\_MIT\_GSI\_PTB\_2024} & Ono2022~\cite{ono2022observation}, Door2024~\cite{Door:2024qqz}, Counts2020~\cite{Counts:2020aws}, Figueroa2022~\cite{Figueroa:2022mtm} & \cite{AME2020II} & $4\times5$ \\
        \texttt{Yb\_Kyoto\_MIT\_GSI\_PTB\_MPIK\_2024} & Ono2022~\cite{ono2022observation}, Door2024~\cite{Door:2024qqz}, Counts2020~\cite{Counts:2020aws}, Figueroa2022~\cite{Figueroa:2022mtm} &   \cite{Door:2024qqz} & $4\times5$ \\
        \texttt{Yb\_PTB\_MPIK\_2024} & Door2024~\cite{Door:2024qqz} & \cite{Door:2024qqz} & $4\times2$ \\
        \hline
        \hline
        \end{tabular}
\caption{\label{tab:kifit_datafolders} \kifit data folders with the corresponding references ($2^\mathrm{nd}$ \& $3^\mathrm{rd}$ columns) and dimensions (number of transitions $m$ and isotope pairs $n$, $4^\mathrm{th}$ column). The datafiles marked with an asterisk use artificially modified uncertainties for the isotope shifts and were used as test datafiles. 
The mock data for \texttt{Ca10pointsTEST} was generated using Eq.~\eqref{eq:linIS}, the electronic coefficients from Tab.~\ref{tab:el_coeffs} and the nuclear masses and charge radii from Refs.~\cite{ANGELI201369,AME2020II}, plus small artificial King nonlinearities. 
}
    \end{table*}
}

In order for the element \texttt{elem} to become a ``valid'' \kifit element, its identification (e.g. \texttt{elem}) needs to be added to the list of elements collected in \texttt{kifit/src/user\_elements.py}.

\subsubsection*{Setting up the Simulation Parameters}
The \kifit run can be customised through various hyperparameters. These can be parsed by \kifit directly via the command line, or provided in the form of a JSON file.  An example \kifit configuration could be the following:
\vspace{0.2cm}
\begin{lstlisting}[language=Python]
#                 config.json
{
    # Here we list the elements whose data is to be combined:
    "element_list": ["elem", "extra_elem"],
    # When more than one element is listed above, use globalogrid:
    "search_mode": "globalogrid",
    # Experiments and block sizes for blocking 
    "num_exp": 100,
    "block_size": 20,
    # Sampled points in the MC
    "num_alphasamples_search": 500,
    "num_elemsamples_per_alphasample_search": 500,
    "num_elemsamples_exp": 500,
    "num_alphasamples_exp": 500,
    "min_percentile": 2,
    # The target mass indices 
    "x0_fit": [0],
    "x0_det": [0],
    "gkp_dims": [3],
    # Algebraic methods' params
    "nmgkp_dims": [3],
    "proj_dims": [3],
    "num_det_samples": 5000,
    "num_sigmas": 2,
}
\end{lstlisting}
\vspace{0.4cm}
More parameters can be added to the configuration file, and we recommend referring to the official documentation for a more detailed explanation.

\subsubsection*{Running the Algorithm}

Once the input files are organised and the \kifit configuration is defined, we can write a \texttt{python} script to compute the estimation of the new physics bounds. As shown in Fig.~\ref{fig:kifit_diagram}, we use a \textit{Runner} object (\texttt{Runner.config}) to share information among the \kifit tools. 
\vspace{0.4cm}
\begin{lstlisting}[language=Python]
from kifit.run import Runner
from kifit.config import RunParams

# Set the kifit parameters given in the configuration file
# (Leave the arguments empty if parsing from command line)
params = RunParams(configuration_file="config.json")

# initialize the runner
runner = Runner(params)

# Run
runner.run()

# If needed, generate plots
runner.generate_all_alphaNP_ll_plots()
\end{lstlisting}
\vspace{0.4cm}

\subsubsection*{Collecting results}
The output of \kifit is stored in the \texttt{kifit/results} folder, which can contain two sub-folders: \texttt{output\_data}, containing the numerical results in the form of JSON files, and \texttt{plots}, containing the final plots plus the real-time plots, if requested by the user. Once a \kifit run is completed and the numerical results are collected, the user can exploit the same configuration file as above to process data and generate more plots, using the functions provided in \texttt{plot.py}. (Simply replace \texttt{runner.run()} with a command such as \texttt{runner.generate\_all\_alphaNP\_ll\_plots()}.)

\subsection{Validation of the Algorithm}
\label{sec:kifit_validation}

\begin{figure*}[ht]
\includegraphics[width=0.5\linewidth]{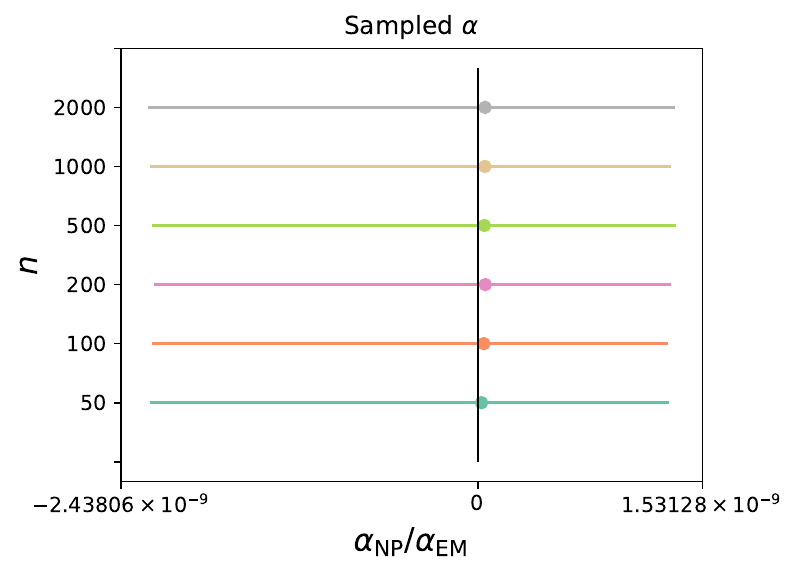}%
\includegraphics[width=0.5\linewidth]{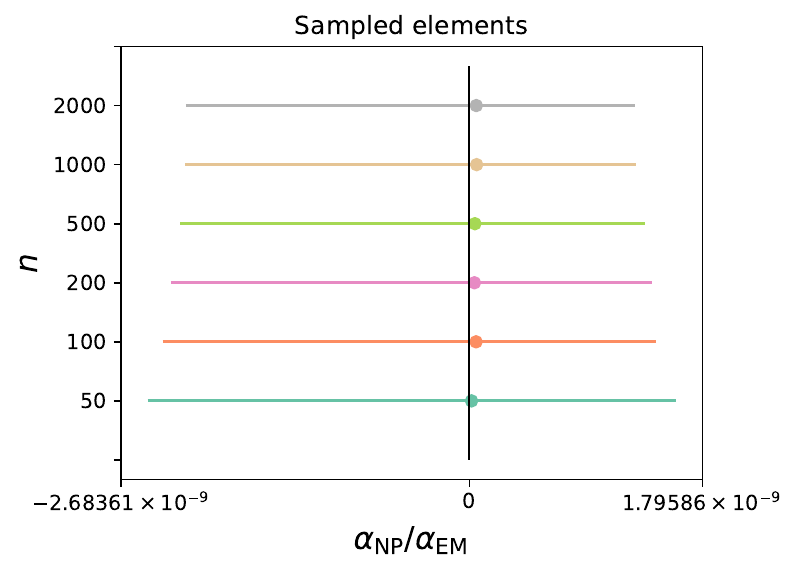}
\includegraphics[width=0.5\linewidth]{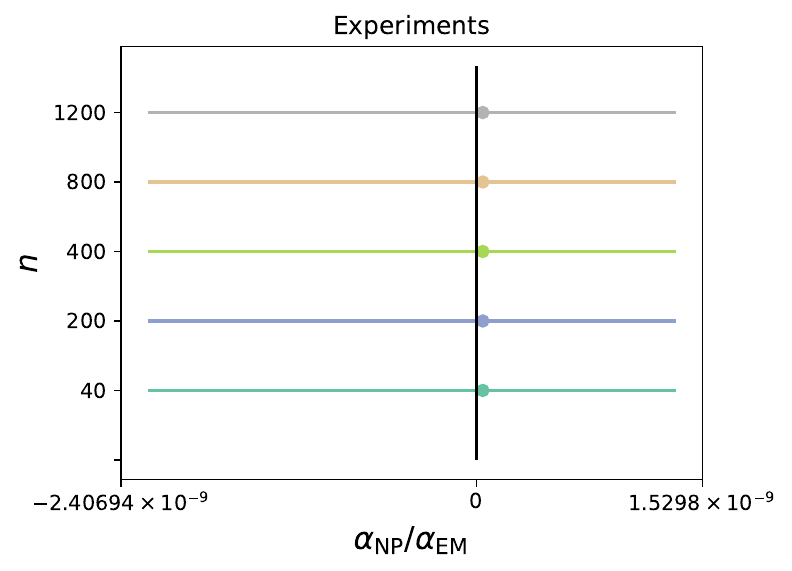}%
\includegraphics[width=0.5\linewidth]{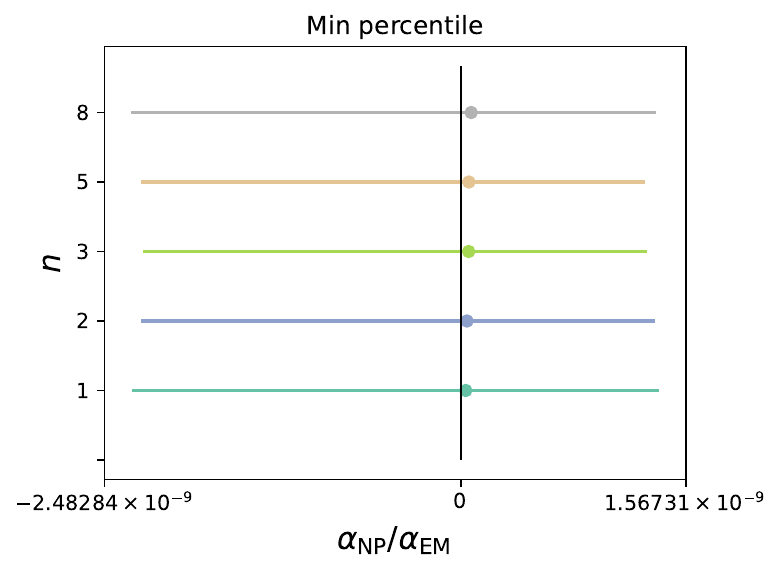}
    \caption{Estimated $\alphaNP/\alphaEM$ values (central dots) and bounds (bars) obtained executing \kifit. The subplots show the impact of varying one of the relevant hyperparameters at a time.}
    \label{fig:validation_sim}
\end{figure*}

We performed a series of validation simulations to highlight the robustness to
the procedure sketched above, and to gain a better understanding of the optimal
hyperparameters required to run a \kifit experiment. 
Table~\ref{tab:default_params} lists the benchmark values, while the bar plots in
Fig.~\ref{fig:validation_sim} illustrate the stability of the code as we vary
one of the hyperparameters after the other.

\renewcommand{\arraystretch}{1.5}
\begin{table}[ht]
\begin{tabular}{cc}
\hline \hline
    \textbf{Hyperparameter} & \textbf{Value} \\
    \hline
    \texttt{num\_elemsaples\_per\_alphasample\_search}  & $500$\\
    \texttt{num\_exp}  & $200$\\
    \texttt{block\_size}  & $20$\\
    \texttt{min\_percentile} & $5$ \\
    \texttt{num\_alphasamples\_exp}  &  $500$ \\
    \texttt{num\_elemsamples\_exp} &  $500$ \\
    \hline
\hline
\end{tabular}
\caption{\label{tab:default_params} Default values of the \kifit hyperparameters set in our validation test. In Fig.~\ref{fig:validation_sim} these values are varied one by one to investigate their impact on the \kifit estimation.}
\end{table}

\begin{figure*}
        \includegraphics[width=0.48\linewidth]{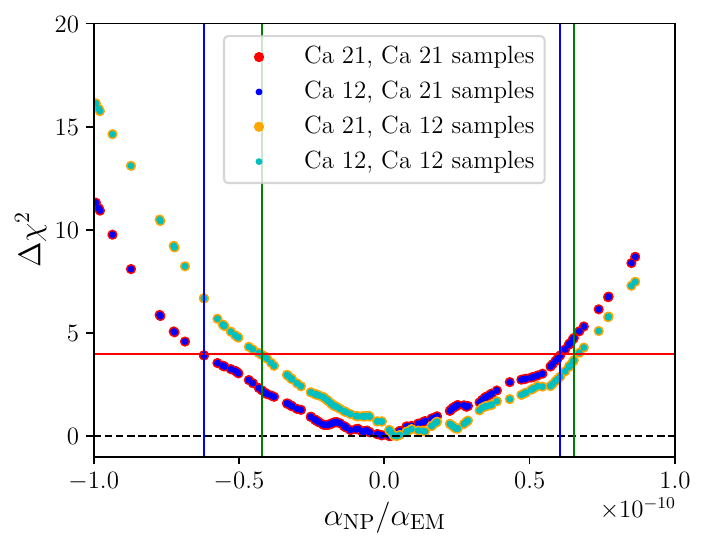}%
         \includegraphics[width=0.48\linewidth]{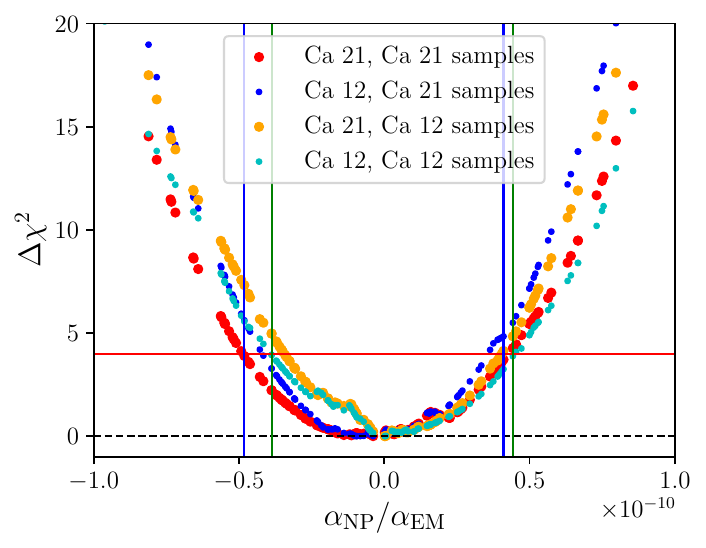}
    \caption{
        Test of the symmetry of the \kifit code with respect to exchanges of
        the transitions.
        \texttt{Ca12} and \texttt{Ca21} correspond to two test data sets that
        differ only in the order of the transitions, and consequently in which
        of the two transitions is used as a reference transition (in
        \texttt{Caij}, $i,j\in \lbrace 1,2\rbrace$, $i$ is considered to be the
        reference transition).
        \texttt{Caij samples} denote the input parameter and fit parameter
        samples obtained for the data set \texttt{Caij}. Based on these samples, 
        the $\Delta \chi^2$ values are computed for each $\alphaNP$ sample,
        once using the \texttt{kifit.Elem} setup provided by the element \texttt{Caij} and once by
        adapting the samples to the conventions of the ``dual'' element
        \texttt{Caji} and using the setup provided by \texttt{Caji}.
        The left (right) figure shows the $(\alphaNP/\alphaEM, \Delta \chi^2)$ values
        obtained adopting the symmetric (asymmetric) definition of $\langle \ha{} \rangle$
        (Eq.~\eqref{eq:symmh} vs Eq.~\eqref{eq:asymmh}).
        }
    \label{fig:llsymmtest}
\end{figure*}

The sensitivity of the code to the choice of the reference transition was
investigated explicitly.
Although the reference transition seems to play a special role in the
construction of the King plot fit (see e.g. Eqs.~\eqref{eq:linISfitdef},
\eqref{eq:NPvec} or Fig.~\ref{fig:kifit_sketch}), the relations between the reference
transition and the other transitions are symmetric and
\begin{align}
    \vert \dai{a}{ji}\vert = \vert \dai{a}{ij}\vert\,,
\end{align}
where we denoted $\dai{a}{j}\equiv \dai{a}{ji}$ (see Eq.~\eqref{eq:ddef}), 
i.e. we explicitly mentioned the reference transition $i$.

This property is particularly easy to observe in the 2-dimensional case: 
\begin{align}
    |\dai{a}{21}|
    =&|\deltai{a}{21}\cos\phij{21}| \notag\\
    =& \left|\alphaNPdEM \left(X_2-\tan\phij{21}X_1\right)
    \left(\ha{a} - \langle \ha{}\rangle^2 \right)\cos \phij{21}\right|\qquad\notag \\
    \overset{\underset{(*)}{}}{=}&
    \left|\alphaNPdEM \left(\tan \phij{12}X_2- X_1\right)
    \left(\ha{a} - \langle \ha{}\rangle^1 \right)\cos \phij{12}\right| \notag\\
    =& |\deltai{a}{12}\cos\phij{12}|
    =|\dai{a}{12}|\,,
\end{align}
where in $(*)$ we assumed that $\langle \ha{}\rangle^2=\langle \ha{}\rangle^1$,
which is the case if we employ the definition given in Eq.~\eqref{eq:symmh}.

The symmetry of the log-likelihood construction can be observed in
Fig.~\ref{fig:llsymmtest}, which shows the outputs of one of
the our numerical tests of \kifit (can be found in the \texttt{kifit/src/tests} folder). Both were produced using the same input data
(central values and uncertainties of the mass and frequency measurements) and
$\alphaNP$ samples, but the upper plot in Fig.~\ref{fig:llsymmtest} shows the $\Delta \chi^2$ values
computed using the manifestly symmetric version of $\sdeltai{a}{i1}$
(Eq.~\eqref{eq:symmh}), while the definition given
in Eq.~\eqref{eq:asymmh} was used in the making of the lower plot in Fig.~\ref{fig:llsymmtest}.
The different colours represent different sets of samples of the input
parameters (transition frequencies and masses, see Eq.~\eqref{eq:elemsampling})
and fit parameters $\Kperpij{21}$ and $\phij{21}$ (see
Eq.~\eqref{eq:fitparamsampling}): The label ``Ca $ji$, Ca $lk$ samples'', means
that the element Ca with the set of transitions $k,l$, where $k$ is the
reference transition, was used to generate samples of the input parameters and
fit parameters, and that the $\Delta \chi^2$ function associated to the element
Ca $ji$, where $i$ is the reference transition, was evaluated. 
As can be seen in Fig.~\ref{fig:llsymmtest}, the manifestly symmetric version of
$\langle \ha{}\rangle$ given in Eq.~\eqref{eq:symmh} leads to a $\Delta \chi^2$
that is invariant under the exchange of transitions ($12\to 21$). The $\Delta
\chi^2$ values vary slightly between different sets of samples (Ca 21 samples
vs. Ca 12 samples), but this is precisely the spread which is captured in the
\emph{experiment phase} (Section~\ref{sec:exp}) and by the \emph{blocking
method} described in Section~\ref{sec:consolidation}.
This spread is present between all four sets of results when using the
definition for $\langle \ha{}\rangle$ given in Eq.~\eqref{eq:asymmh}
(Fig.~\ref{fig:llsymmtest}).

A number of complementary tests were directly implemented as \texttt{pytests} in
the \kifit code. A non-exhaustive list of the tests is given in the following.
For details, please directly check the \texttt{kifit/src/tests} folder.

\subsubsection*{\texttt{test\_build}}
\begin{itemize}
    \item Loading elements, checking dimensions of input data.
    \item Comparison of results of ODR and linear regression in linear fit to
        King plot data.
    \item Numerical cross-checks of \kifit construction against results obtained
        with \texttt{Mathematica}.
    \item Numerical cross-checks of implementation of algebraic methods in
        \kifit against results obtained with \texttt{Mathematica}.
\end{itemize}

\subsubsection*{\texttt{test\_detools}}
\begin{itemize}
    \item Numerical cross-checks of uncertainties computed for algebraic methods
        in \kifit against results obtained with \texttt{Mathematica}.
    \item Plot distribution of $\alphaNP$ values.
\end{itemize}

\subsubsection*{\texttt{test\_fitools}}
\begin{itemize}
    \item Numerical cross-checks of \kifit construction for single point (sample)
        in parameter space against results obtained with \texttt{Mathematica}.
        The procedure is repeated for different values of the fit
        parameters $\lbrace \Kperpij{j1}, \phij{j1}\rbrace _{j=2}^m$, for
        $\alphaNPdEM=0$ and for $\alphaNPdEM=10^{-11}$.
    \item Cross-check of $\Vert\dvec\Vert$, the covariance matrix $\covd$, the
        log-likelihood and various intermediate steps against results obtained
        with \texttt{Mathematica}. 
        The procedure is repeated for different values of the fit
        parameters $\lbrace \Kperpij{j1}, \phij{j1}\rbrace _{j=2}^m$, with and
        without sampling of the fit parameters, for different values of
        $\alphaNP$ and for different sample numbers.
    \item Cross-check of $\covd$ and the log-likelihood for $N_s =10^2, \,10^3,
        \, 10^4,\, 10^5$ samples.
        Check of the condition number of the covariance matrices and, by
        computing the spectral difference, Frobenius norm difference and the
        Kullback-Leibler divergence, comparison to the covariance matrices
        determined in \texttt{Mathematica}.\\
        For $\alphaNP=0$ and $\alphaNP=10^{-11}$, comparison of the distribution of
        log-likelihood values for the $N_s$ samples obtained using \kifit and
        \texttt{Mathematica}.
    \item Check of the numerical accuracy of the inversion of the covariance
        matrix $\covd$.
    \item Tests of the symmetry of the \kifit code with respect to exchange
        of the transitions or the choice of the reference transition (see
        discussion above), varying only the input parameters or varying both the
        input parameters and the fit parameters. 
    \item Comparison of fit and algebraic methods for the test data set
        \texttt{Ca24min\_mod}.
    \item Test of the numerical impact of the regulation of the covariance matrix 
        $\boldsymbol{\Sigma}_{\dvec}^{(\lambda)} = \boldsymbol{\Sigma}_{\dvec} +
        \lambda \mathbf{I}_n$.
    \item Test of \kifit run procedure
\end{itemize}

Finally, \texttt{test\_cache\_update} tests whether the \kifit-internal cache is
working and updated as expected.

\section{The Impact of Data Sparsity on the Fit}
\label{sec:sparsity}

\begin{figure}[!htb]
    \centering    
    \includegraphics[width=\linewidth]{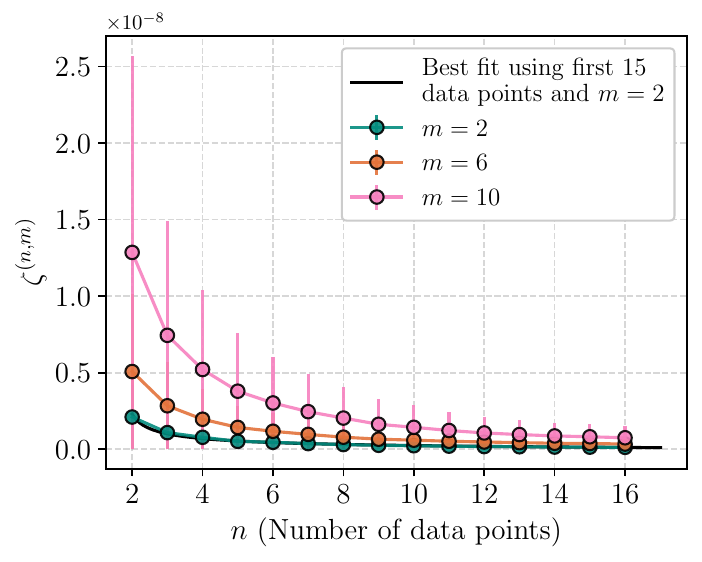} 
    \caption{Illustration of the goodness a fit to $y_j=b_j x + c_j$, $j=1,\ldots,m$ as a function of the number $n$ of mock data points $(x,y_1, \ldots, y_m)$.  $\zeta^{(n)}\equiv \langle \zeta^{(n,s)}\rangle_{(s)}$ with $\zeta^{(n,s)}$ as defined in Eq.~\eqref{eq:zetass} corresponds to the average relative uncertainty on the $2(n-1)$ fit parameters $\lbrace b_j, c_j\rbrace_{j=2}^m$ used to fix the line in $m$-dimensional space. We show the results for the minimal case of $m=2$ and for $m=6$. 
    For each number $n$ of data points, 500 mock data sets with Gaussian uncertainties $\sigma=\sigma_n=10^{-10}$ were generated, leading to relative uncertainties $\sigma_x/x$, $\sigma_y/y$ of the same order.
    In black we show the curve defined by $f(n)=\theta_1 / \sqrt{n-\theta_2} + \theta_3$, with $(\theta_1, \theta_2, \theta_3)\approx(1.5\times 10^{-9}, 1.7, -3\times 10^{-10})$. 
    }
    \label{fig:thetadiff}
\end{figure}

\begin{figure}[!htb]
    \centering%
    \includegraphics[width=\linewidth]{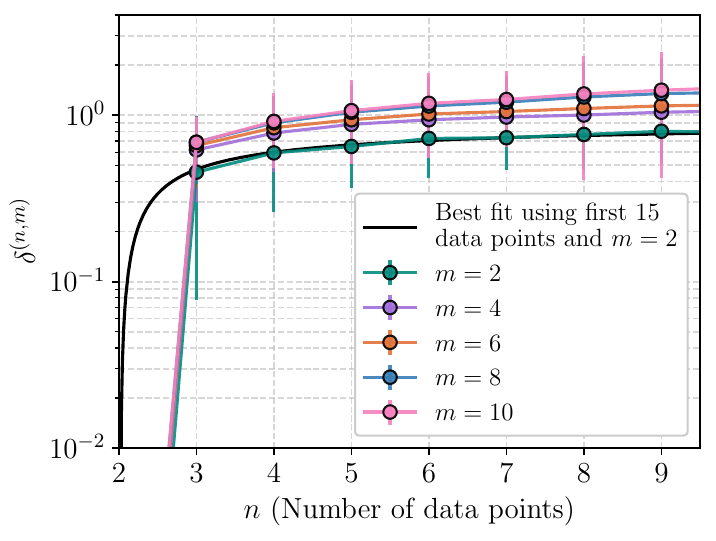}
    \caption{
    $\delta^{(n,m)}$ (see Eq.~\eqref{eq:deltanm}) as a function of the number $n$ of the mock data points.  In black we show the curve defined by $f(n)=\theta_1 / \sqrt{n-\theta_2} + \theta_3$, with $(\theta_1, \theta_2, \theta_3)\approx(-0.6, 1.6, 1)$. 
    $\delta^{(n,m)}$ can be interpreted as the average distance of a point $a$ in an $n$-dimensional data set to the corresponding best-fit line.
    }
    \label{fig:ddiffn}
\end{figure}

In this section we investigate how many data points (isotope pairs) are needed in a given number of dimensions (transitions) to ensure that the initial ODR is not over-fitting the data.
We stress that the purpose of this analysis is to simulate the scaling of the relevant metrics in our estimation procedure with respect to the dimensionality of the King line ($m$) and the number of data points ($n$). The analysis is performed on mock data constructed from $N_s$ samples of $n$ ``data points'' $\lbrace\lbrace(x^{(a,s)}, y_1^{(a,s)},\ldots, y_m^{(a,s)})\rbrace_{a=1}^n\rbrace_{s=1}^{N_s}$ which are normally distributed around a linear relation $y_j=b_j x+ c_j$, $j=1,\ldots, m$ with variance $\sigma_n$. The uncertainties on the mock data points are modelled using Gaussian random variables with variance $\sigma$. 

Concretely, for a given number $n$ of points in $m$ dimensions, we generate the parameters 
\begin{gather}
    b_j, c_j \sim \mathcal{N}(0, 10)\,,\quad j=1,\ldots, m
\end{gather}
which fix the linear relation $y_j=b_j x+ c_j$ and then
generate $N_s=500$ samples ($s=1,\ldots,N_s$) from the distributions
\begin{gather}
    x^{(a,s)} \sim  \mathcal{N}(x^{(a)}, \sigma_n)\,, \quad a=1,\ldots, n\notag \\
    y_j^{(a,s)} \sim  \mathcal{N}(b_j x^{(a)} + c_j, \sigma_n), \quad j=1,\ldots, m \notag\\
    \sigma[x^{(a,s)}], \,\sigma[y^{(a,s)}] \sim  \mathcal{N}(0, \sigma)\,,
    \label{eq:generating_mock_data}
\end{gather}
where we take $\sigma _n = \sigma = 10^{-10}$.
For each sample $\lbrace (x^{(a,s)}, y_1^{(a,s)},\ldots, y_m^{(a,s)})\rbrace_{a=1}^n$ of the $n$ data points in $m$ dimensions, $2m$ fit parameters $\lbrace(b_j^{(n,s)}, c_j^{(n,s)})\rbrace_{j=1}^m$ are determined by means of orthogonal distance regression. 

\begin{figure}[!htb]
    \centering    \includegraphics[width=\linewidth]{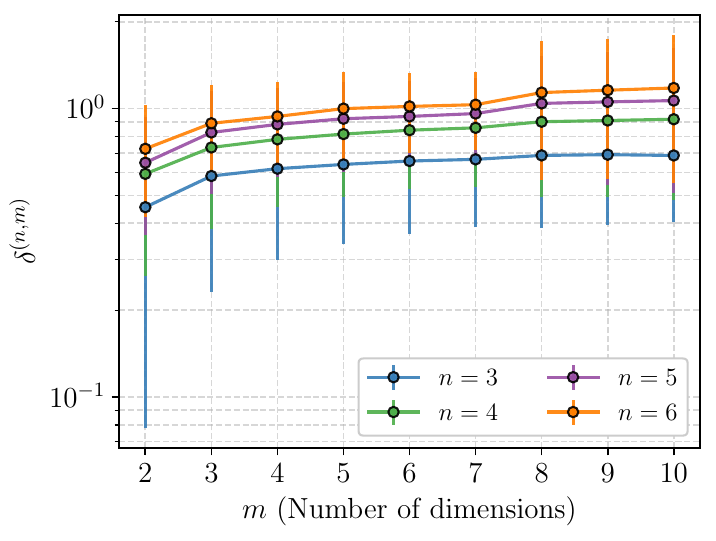}
    \caption{
    $\delta^{(n,m)}$ (see Eq.~\eqref{eq:deltanm}) as a function of the number $m$ of dimensions, for selected sample sizes $n$. 
    $\delta^{(n,m)}$ can be interpreted as the average distance of a point $a$ in an $n$-dimensional data set to the corresponding best-fit line.}
    \label{fig:ddiffm}
\end{figure}

As a first check, we compute the combined fractional differences of the fit parameter samples with respect to the original slopes and intercepts $\lbrace(b_j, c_j)\rbrace_{j=1}^m$, 
\begin{align}
    \zeta^{(n,s)} = \sqrt{\sum_{j=2}^m \left(\frac{(b_j^{(n,s)} - b_j)^2}{b_j^2} + \frac{(c_j^{(n,s)} - c_j)^2}{c_j^2}\right)}\,.
    \label{eq:zetass}
\end{align}
The averages and standard deviations over the samples, $\zeta^{(n)}\equiv \langle \zeta^{(n,s)}\rangle_{(s)}$ and $\sigma[ \zeta^{(n,s)}]_{(s)}$ can be viewed as a measure of the error in the fit parameters due to the finite sample size $n$. 
They are plotted in Fig.~\ref{fig:thetadiff} as a function of the number of points $n$. 
As expected, $\zeta^{(n)}$ approximately follows a relation of the form $f(n)=\theta_1/\sqrt{n-\theta_2} + \theta_3$, which is motivated by the expected asymptotic behaviour of Gaussian uncertainties, by the fact that a linear fit to one point is ill-defined and by the finite number of $n$ values considered for the curve fit. 
The explicit parameters obtained when fitting the first 15 data points for $m=2$ to this curve are $(\theta_1, \theta_2, \theta_3)\approx(1.5\times 10^{-9}, 1.7, -3\times 10^{-10})$.

Since relevant quantity for the log-likelihood in Eq.~\eqref{eq:ll} is the distance of the points from the line, normalised by the respective uncertainties, we apply a definition very similar to that used in \kifit (see Eqs.~\eqref{eq:ddef}, \eqref{eq:eFdef}, \eqref{eq:Deltadef}):
\begin{align}
    \delta^{(a,m,s)} = \frac{1}{\sigma \sqrt{m}} \Vert \boldsymbol{\Delta}^{(a,m,s)} - (\boldsymbol{\Delta}^{(a,m,s)} \cdot \boldsymbol{\hat{e}_f})\boldsymbol{\hat{e}_f}\Vert
    \label{eq:dss}
\end{align}
with 
\begin{align}
  \boldsymbol{\Delta}^{(a,m,s)} =&
  \begin{pmatrix}
      0\\
      y_1^{(a,s)} - (b_1^{(a,s)} x^{(a,s)} + c_1^{(a,s)})\\
      \vdots\\
      y_m^{(a,s)} - (b_m^{(a,s)} x^{(a,s)} + c_m^{(a,s)})
  \end{pmatrix}
\end{align}
and $\boldsymbol{\hat{e}_f} =(1, b_1^{(a,s)},\ldots, b_m^{(a,s)})$,
and a normalisation $\sigma \sqrt{m}$ that captures the asymptotic behaviour of the uncertainties on the sum. We then define
\begin{align}
\begin{split}
    \delta^{(n,m)}\equiv & \left\langle \frac{1}{n}\sum_{a=1}^n \delta^{(a,m,s)}\right\rangle_{(s)}\,,\\
    \sigma[\delta^{(n,m)}]\equiv & \sigma\left[\frac{1}{n}\sum_{a=1}^n \delta^{(a,m,s)}\right]_{(s)}\,,  
    \label{eq:deltanm}
\end{split}
\end{align}
where $\langle .\rangle_{(s)}$ and $\sigma[.]_{(s)}$ denote the average and the standard deviation over the samples $(s)$. In Fig.~\ref{fig:ddiffn}, $\delta^{(n,m)}$ and $\sigma[\delta^{(n,m)}]$ are plotted as a function of the number of points $n$ and for different dimensions $m$.
$\delta^{(n,m)}$ can be viewed as the average distance, normalised by the experimental uncertainties, of a point $a$ in a data set of size $n$ to the best-fit line.  $\delta^{(n,m)}$ increases with $n$ because the uncertainty on the distance to the line receives an additional contribution from the increasing spread of the data points, which for the mock data employed here has variance $\sigma_n$ (see Eq.~\eqref{eq:generating_mock_data}). This residual $n$-dependence is again captured by a function of the form $f(n)=\theta_1 / \sqrt{n-\theta_2} + \theta_3$, this time with $(\theta_1, \theta_2, \theta_3)\approx(-0.6, 1.6, 1)$ for the first 15 points in $m=2$ dimensions. 
Fig.~\ref{fig:ddiffn} shows that increasing the number of data points from 3 by a few can significantly improve the reliability of the fit results.

The $m$-dependence of $\delta^{(n,m)}$ is less pronounced, as can be observed in Fig.~\ref{fig:ddiffm}. Nonetheless, measurements of additional transitions can improve the heterogeneity of the data and thus the reliability of the bounds on $\alphaNP$.

\hspace{3cm}

\section{Isotope Shift and Atomic/Nuclear Mass Data}
\label{sec:data}

In Tables~\ref{tab:Ca_IS} and \ref{tab:Yb_IS}, we collect the
available isotope shift measurements for Ca and Yb. The most recent measurements for the isotope masses can be found in Tables~\ref{tab:Ca_masses} and \ref{tab:Yb_masses}. We mark with an asterisk the
values that are included in the \kifit folders.

%
    {\setlength{\tabcolsep}{5pt}
    \renewcommand{\arraystretch}{1.4}
    \begin{table*}[!htbp]
    \centering
        \begin{tabular}{cccc}
            \hline
            \hline
            Isotope A & $m_A$ [amu]~\cite{AME2016II} & $m_A$ [amu]~\cite{AME2020II} & $m_A/m_{40}$~\cite{Wilzewski:2024wap} \\
            \hline
            40 &  39.962590866(22)& 39.962590851(22)  & 1 \\
            42 &  41.95861783(16)& 41.95861778(16)  & 1.049 961 066 498(15) \\
            44 &  43.9554815(3) & 43.9554815(3)  & 1.099 943 105 797(15)\\
            46 &  45.9536880(24) & 45.9536877(24)  & 1.149 958 773 895(30)\\
            48 &  47.95252290(10) & 47.952522654(19)  & 1.199 990 087 090(40)\\
            \hline
            \hline
        \end{tabular}
    \caption{\label{tab:Ca_nuc_params} Atomic masses of Ca isotopes from Refs.~\cite{AME2016II,AME2020II}. The third column shows the ratios of bare nuclear masses to the mass of isotope $A=40$, as reported in Ref.~\cite{Wilzewski:2024wap}.}
        \label{tab:Ca_masses}
    \end{table*}
    }

   {\setlength{\tabcolsep}{5pt}
    \renewcommand{\arraystretch}{1.4}
    \begin{table*}[h]
    \centering
        \begin{tabular}{cccc}
            \hline
            \hline
            Isotope A & $m_A$ [amu]~\cite{AME2016II} & $m_A$ [amu]~\cite{AME2020II} & $m_A/m_{172}$~\cite{Door:2024qqz} \\
            \hline
            168 &  167.9338891(13) & 167.93389130(10)  & 0.976715921749(4) \\
            170 &  169.934767246(11) & 169.934767243(11)  & 0.988355799258(4)  \\
            172 &  171.936386659(15) & 171.936386654(15)  & 1 \\
            174 &  173.938867548(12)  & 173.938867546(12)   & 1.011649212140(4)\\
            176 &  175.942574709(16) & 175.942574706(16)   & 1.023305557965(4)\\
            \hline
            \hline
        \end{tabular}
    \caption{Atomic masses of Yb isotopes from Refs.~\cite{AME2016II,AME2020II}. The third column shows ratios of bare nuclear masses to the mass of isotope $A=172$, as reported in Ref.~\cite{Door:2024qqz}.}
        \label{tab:Yb_masses}
    \end{table*}
    }

    \begin{table*}[h]
        {\setlength{\tabcolsep}{3pt}
        \renewcommand{\arraystretch}{1.4}
        \centering
        \begin{tabular}{c|c|c|c}
            \hline
            \hline
            \multirow{2}{*}{$(A,A')$} & \Ca{}{+}:\CaStoD{5/2} & \Ca{}{+}:\CaDtoD & \Ca{}{+}:\CaStoP \\
             & [729 nm] & DD  & [397 nm] \\ 
            \hline
            \multirow{3}{*}{(40, 42)} & $^*$2 771 872 467.6(7.6)~\cite{Knollmann2019} & $^*$-3 519 896(24)~\cite{Solaro:2020dxz} & \multirow{2.5}{*}{$^*$425 706 000(94 000)~\cite{Gebert:2015_Ca100kHz}} \\
             & 2 771 873 000(2000)~\cite{Solaro:2020dxz} & $^*$-3 519 910(9.7)~\cite{chang2024systematicfree} & \multirow{2.5}{*}{425 490 000(150 000)~\cite{muller2020collinear}} \\
             & 2 771 872 430.217(27)~\cite{Wilzewski:2024wap} & -3 519 944.6(60)~\cite{Wilzewski:2024wap} & \\
            \multirow{3}{*}{(40, 44)} & $^*$5 340 887 394.6(7.8)~\cite{Knollmann2019} & $^*$-6 792 470(22)~\cite{Solaro:2020dxz} & \multirow{2.5}{*}{$^*$849 534 000(74 000)~\cite{Gebert:2015_Ca100kHz}} \\
             & 5 340 888 000(2000)~\cite{Solaro:2020dxz} & $^*$-6 792 440(6)~\cite{chang2024systematicfree} & \multirow{2.5}{*}{849 000 000(140 000)~\cite{muller2020collinear}} \\
             & 5 340 887 395.288(38)~\cite{Wilzewski:2024wap} & -6 792 440.1(59)~\cite{Wilzewski:2024wap} & \\
            \multirow{3}{*}{(40, 46)} & \multirow{2.5}{*}{$^*$7 768 401 000(2000)~\cite{Solaro:2020dxz}} & $^*$-9 901 524(21)~\cite{Solaro:2020dxz} & \multirow{2.5}{*}{$^*$1 297 610 000(340 000)}~\cite{muller2020collinear} \\
             & \multirow{2.5}{*}{7 768 401 432.916(63)~\cite{Wilzewski:2024wap}} & $^*$-9 901 520(2828.43)~\cite{chang2024systematicfree} &  \\
             & & -9 901 524(21)~\cite{Wilzewski:2024wap} &  \\
            \multirow{3}{*}{(40, 48)} & 9 990 382 525.0(4.9)~\cite{knollmann2023erratum} & -12 746 610(27)~\cite{Solaro:2020dxz} & \multirow{2.5}{*}{1 705 389 000(60 000)~\cite{Gebert:2015_Ca100kHz}} \\
             & 9 990 383 000(2000)~\cite{Solaro:2020dxz} & $^*$-12 746 600(7.5)~\cite{chang2024systematicfree} & \multirow{2.5}{*}{1 705 460 000(140 000)~\cite{muller2020collinear}} \\
             & 9 990 382 526.834(55)~\cite{Wilzewski:2024wap} & -12 746 588.2(57)~\cite{Wilzewski:2024wap} & \\
            \hline
            \hline
            \multirow{2}{*}{$(A,A')$} & \Ca{}{+}:\CaStoD{3/2} & $\mathrm{Ca}^+$: $4^2S_{1/2}\rightarrow4p\;^2P_{3/2}$ & \Ca{}{+}:\CaDtoP \\
             & [732 nm] & [393 nm] & [866 nm] \\
            \hline
            \multirow{3}{*}{(40, 42)} & \multirow{3}{*}{2 775 392 374.8(6.0)~\cite{chang2024systematicfree}} & \multirow{1.75}{*}{426 040 000(150 000)~\cite{muller2020collinear}} & $^*$-2 349 974 000(99 000)~\cite{Gebert:2015_Ca100kHz} \\
             & & \multirow{1.75}{*}{425 932 000 (71 000)~\cite{shi2018unexpectedly}} & 2 366 000 000(59 000 000)~\cite{alt1997shifts} \\
             & & & 2 352 100 000(2 100 000)~\cite{kramida2020atomic} \\
            \multirow{3}{*}{(40, 44)} & \multirow{3}{*}{5 347 679 835.4(5.9)~\cite{chang2024systematicfree}} & \multirow{1.75}{*}{850 090 000(140 000)~\cite{muller2020collinear}} & $^*$-4 498 883 000(80 000)~\cite{Gebert:2015_Ca100kHz} \\
             & & \multirow{1.75}{*}{850 231 000(65 000)~\cite{shi2018unexpectedly}} & 4 509 000 000(24 000 000)~\cite{alt1997shifts} \\
             & & & 4 499 300 000(2 300 000)~\cite{kramida2020atomic} \\
            \multirow{2}{*}{(40, 46)} & \multicolumn{1}{l|}{\multirow{2}{*}{}} & 1 299 070 000(580 000)~\cite{muller2020collinear} & \multirow{2}{*}{} \\
             & \multicolumn{1}{l|}{} & 1 301 000 000(3 600 000)~\cite{garcia2016unexpectedly} &  \\
            \multirow{2}{*}{(40, 48)} & \multirow{2}{*}{10 003 129 115.1(5.7)~\cite{chang2024systematicfree}} & 1 707 580 000(160 000)~\cite{muller2020collinear} & $^*$-8 297 769 000(81 000)~\cite{Gebert:2015_Ca100kHz} \\
             &  & 1 707 945 000(67 000)~\cite{shi2018unexpectedly} & 8 296 700 000(3 200 000)~\cite{kramida2020atomic}  \\
            \hline
            \hline
            \multirow{2}{*}{$(A,A')$} & $\mathrm{Ca}^+$: $^2D_{5/2}\rightarrow^2P_{3/2}$ & $\mathrm{Ca}^+$:$^2D_{3/2}\rightarrow^2P_{3/2}$ & $\mathrm{Ca}^{14+}$: $^3 P_0\rightarrow$ $^3P_1$\\
             & [854 nm] & [850 nm] & [570 nm]\\
            \hline
            \multirow{2}{*}{(40, 42)} & 2 272 000 000(94 000 000)~\cite{alt1997shifts} & 2 359 000 000(64 000 000)~\cite{alt1997shifts} & \multirow{2}{*}{539 088 421.24(12)~\cite{Wilzewski:2024wap}} \\
             & 2 347 600 000(3 900 000)~\cite{kramida2020atomic} & 2 351 450 000(700 000)~\cite{kramida2020atomic}& \\
            \multirow{2}{*}{(40, 44)} & 4 510 000 000(19 000 000)~\cite{alt1997shifts} & 4 538 000 000(27 000 000)~\cite{alt1997shifts}& \multirow{2}{*}{1 030 447 731.64(11)~\cite{Wilzewski:2024wap}} \\
             & 4 489 800 000(3 600 000)~\cite{kramida2020atomic} & 4 497 270 000(900 000)~\cite{kramida2020atomic} & \\
            (40, 46) & 6 470 100 000(2 400 000)~\cite{kramida2020atomic} &  &1 481 135 946.74(14)~\cite{Wilzewski:2024wap}\\
            (40, 48) & 8 277 900 000(5 000 000)~\cite{kramida2020atomic} & 8 295 060 000(1 100 000)~\cite{kramida2020atomic} & 1 894 297 294.53(14)~\cite{Wilzewski:2024wap} \\ 
            \hline
            \hline
        \end{tabular}
        }
    \caption{\label{tab:Ca_IS}
    Isotope shifts of \Ca{}{+} and \Ca{}{14+}, expressed in Hz. Values included in \kifit are marked with an asterisk. Notice that Ref.~\cite{chang2024systematicfree} reports measurements of the \CaStoD{3/2} transition, and the corresponding DD-transition values are derived therein from their measurement of \CaStoD{3/2} and the \CaStoD{5/2} reported in Ref.s~\cite{Knollmann2019, knollmann2023erratum, Solaro:2020dxz}. Similarly, the DD-transition values reported in Ref.~\cite{Wilzewski:2024wap} are derived from their measurement of \CaStoD{5/2} and the \CaStoD{3/2} reported in Ref.\cite{chang2024systematicfree}.} 
    \end{table*}

    \begin{table*}[h]
    \centering
        {\setlength{\tabcolsep}{13pt}
        \renewcommand{\arraystretch}{1.4}
        \begin{tabular}{c|c|c|c}
            \hline
            \hline
            \multirow{2}{*}{$(A,A')$} & $(\alpha)$ Yb$^+$:\YbStoDfive & $(\beta)$ Yb$^+$:\YbStoDthree & $(\gamma)$ Yb$^+$: $^2S_{1/2}\rightarrow^2F_{7/2}$ \\
                & [411 nm] & [436 nm] & [467 nm] \\
            \hline
            \multirow{2}{*}{(168, 170)} & $^*$2 179 098 930(210)~\cite{Counts:2020aws} & \multirow{2}{*}{$^*$2 212 391 850(370)~\cite{Counts:2020aws}} & $^*$-4 438 160 300(500)~\cite{Hur:2022gof} \\
             & $^*$2 179 098 868.0(5.3)~\cite{Door:2024qqz} & & $^*$-4 438 159 671.1(15.7)~\cite{Door:2024qqz} \\
            \multirow{2}{*}{(170, 172)} & $^*$2 044 854 780(340)~\cite{Counts:2020aws} & \multirow{2}{*}{$^*$2 076 421 580(390)~\cite{Counts:2020aws}} & $^*$-4 149 190 380(450)~\cite{Hur:2022gof} \\
                & $^*$2 044 851 281.0(4.9)~\cite{Door:2024qqz} & & $^*$-4 149 190 501.1(15.7)~\cite{Door:2024qqz} \\
            \multirow{2}{*}{(172, 174)} & $^*$1 583 068 420(360)~\cite{Counts:2020aws} & \multirow{2}{*}{$^*$1 609 181 470(220)~\cite{Counts:2020aws}} & $^*$-3 132 321 600(500)~\cite{Hur:2022gof} \\
                & $^*$1 583 064 149.3(4.8)~\cite{Door:2024qqz} & & $^*$-3 132 320 458.1(15.7)~\cite{Door:2024qqz} \\
            \multirow{2}{*}{(174, 176)} & $^*$1 509 055 290(280)~\cite{Counts:2020aws} & \multirow{2}{*}{$^*$1 534 144 060(240)~\cite{Counts:2020aws}} & $^*$-2 976 391 600(480)~\cite{Hur:2022gof} \\
                & $^*$1 509 053 195.8(4.7)~\cite{Door:2024qqz} & & $^*$-2 976 392 045.3(15.7)~\cite{Door:2024qqz} \\
            \hline
            (168, 172) & & & -8 587 352 000(470)~\cite{Hur:2022gof} \\
            (170, 174) & 3 627 922950(500)~\cite{Counts:2020aws} & 3 685 601950(330)~\cite{Counts:2020aws} & -7 281 511 880(450)~\cite{Hur:2022gof} \\
            (172, 176) & & & -6 108 712 930(440)~\cite{Hur:2022gof} \\
            \hline
            \hline
            \multirow{2}{*}{$(A,A')$} & $(\epsilon)$ Yb:\YbStoDtwo & $(\delta)$ Yb:$^1S_0\rightarrow^3P_0$ & Yb:$4f^{14}6s^{2} {}^{1}S_0\longleftrightarrow 4f^{13}5d^6s^2$ \\
                & [361 nm] & [578 nm] & [431 nm] \\
            \hline
            \multirow{2}{*}{(168, 170)} & \multirow{2}{*}{$^*$1 781 785 360(710)~\cite{Figueroa:2022mtm}} & \multirow{2}{*}{$^*$1 358 484 476.2(2.2)$_\mathrm{tot}$~\cite{ono2022observation}} & -1 753 930 000(3000)$_\mathrm{stat}$~\cite{ishiyama2023observation} \\
                & & & -1 753 952 000(26 000)~\cite{kawasaki2024isotope} \\
            \multirow{2}{*}{(170, 172)} & \multirow{2}{*}{$^*$1 672 021 510(300)~\cite{Figueroa:2022mtm}} & \multirow{2}{*}{$^*$1 275 772 006(2.8)$_\mathrm{stat}$~\cite{ono2022observation}} & -1 630 050 000(3000)$_\mathrm{stat}$~\cite{ishiyama2023observation} \\
                & & & -1 630 028 000(26 000)~\cite{kawasaki2024isotope} \\
            \multirow{2}{*}{(172, 174)} & \multirow{2}{*}{$^*$1 294 454 440(240)~\cite{Figueroa:2022mtm}} & \multirow{2}{*}{$^*$992 714 586.6(2.1)$_\mathrm{tot}$~\cite{ono2022observation}} & -1 180 614 000(2000)$_\mathrm{stat}$~\cite{ishiyama2023observation} \\
                & & & -1 180 616 000(25 000)~\cite{kawasaki2024isotope} \\
            \multirow{2}{*}{(174, 176)} & \multirow{2}{*}{$^*$1 233 942 190(310)~\cite{Figueroa:2022mtm}} & \multirow{2}{*}{$^*$946 921 774.9(2.9)$_\mathrm{tot}$~\cite{ono2022observation}} & -1 115 766 000(6000)$_\mathrm{stat}$~\cite{ishiyama2023observation} \\
                & & & -1 115 787 000(24 000)~\cite{kawasaki2024isotope} \\
            \hline
            (168, 172) & 3 453 805 270(83)~\cite{Figueroa:2022mtm} & & \\
            (172, 176) & 2 528 396 500(34)~\cite{Figueroa:2022mtm} & & \\
            \hline
            \hline
            \multirow{2}{*}{$(A,A')$} & $\mathrm{Yb}^{+}$: $6^2S_{1/2}\rightarrow6^2P^o_{1/2}$ & $\mathrm{Yb}^{+}$:$6^2S_{1/2}\rightarrow6^2P^o_{3/2}$ & Yb: $6^1S_0\rightarrow6^3P^o_1$ \\
                & [369 nm] & [329 nm] & [556 nm] \\
            \hline
            (168, 170) & & & -1 368 630 000(500000)~\cite{clark1979optical} \\
            (170, 172) & -1 623 300(800)~\cite{maartensson1994isotope} & -1 459 000(21000)~\cite{berends1992hyperfine} & -1 286 470(500)~\cite{clark1979optical} \\
            (172, 174) & -1 275 300(700)~\cite{maartensson1994isotope} & -1 154 000(11000)~\cite{berends1992hyperfine} & -1 000 280(500)~\cite{clark1979optical} \\
            \hline
            (172, 176) & -2 492 800(1000)~\cite{maartensson1994isotope} & -2 259 000(13000)~\cite{berends1992hyperfine} & -1 955 040(500)~\cite{clark1979optical} \\
            (174, 176) & & & -954 760 000(5000000)~\cite{clark1979optical} \\
            \hline
            \hline 
        \end{tabular}
        }
        {\setlength{\tabcolsep}{3pt}
        \renewcommand{\arraystretch}{1.5}
        \centering
        \begin{tabular}{c|c}
            \multirow{2}{*}{$(A,A')$} & $\mathrm{Yb}^+$:\,$^1S_{0}\rightarrow^1P_{1}$ \\
             & [399 nm] \\
            \hline
            (174, 168) & 1 888 800 000(110 000)~\cite{Kleinert:2016_Yb0} \\
            (174, 172) & 1 190 360 000(490 000)~\cite{Kleinert:2016_Yb0}  \\
            (174, 172) & -250 780 000(330 000)~\cite{Kleinert:2016_Yb0} \\
            (174, 176) & -508 890 000(090 000)~\cite{Kleinert:2016_Yb0} \\
            
            \hline
            \hline
        \end{tabular}
        }
    \caption{\label{tab:Yb_IS}
        Isotope shifts for Yb and Yb$^+$, expressed in Hz.
        Values included in \kifit are marked with an asterisk. Isotopes used in this work are organised in pairs $(A,A')$ with $A'=A+2$. The data for the other isotope pairs can be used
        for cross-checks. 
    }
    \end{table*}

\end{document}